\documentclass[12pt]{article}
\usepackage{amsmath}
\usepackage{graphicx,psfrag,epsf}
\usepackage{booktabs}

\usepackage{amsfonts,amssymb}
\usepackage{algorithm,algpseudocode}
\usepackage{hyperref}
\usepackage{setspace}

\DeclareMathOperator{\Tr}{Tr}

\usepackage{multirow}
\usepackage{rotating}
\usepackage{xcolor}

\usepackage{scalerel,stackengine}
\stackMath
\newcommand\reallywidehat[1]{%
\savestack{\tmpbox}{\stretchto{%
  \scaleto{%
    \scalerel*[\widthof{\ensuremath{#1}}]{\kern.1pt\mathchar"0362\kern.1pt}%
    {\rule{0ex}{\textheight}}
  }{\textheight}%
}{2.4ex}}%
\stackon[-6.9pt]{#1}{\tmpbox}%
}
\usepackage{natbib}
\bibpunct[,]{(}{)}{;}{a}{,}{,}

\usepackage{xparse}
\ExplSyntaxOn
\NewDocumentCommand{\longdash}{ O{2} }
 {
  --\prg_replicate:nn { #1 - 1 } { \negthinspace -- }
 }
\ExplSyntaxOff

\newcommand{\blind}{0}



\newcommand{\model}{Model}

\begin{document}

\def\spacingset#1{\renewcommand{\baselinestretch}%
{#1}\small\normalsize} \spacingset{1}


\if0\blind
{
  \title{\bf Fast Variational Inference for Bayesian Factor Analysis in Single and Multi-Study Settings}
  \author{Blake Hansen\\
    Department of Biostatistics, Brown University\\
    Alejandra Avalos-Pacheco\\
    Applied Statistics Research Unit, TU Wien\\ 
    Harvard-MIT Center for Regulatory Science, Harvard Medical School\\
    Massimiliano Russo\footnote{These authors contributed equally to this work.}\\
    Department of Statistics, The Ohio State University \\
    Roberta De Vito\footnotemark[\value{footnote}]\\
    Department of Biostatistics and Data Science Institute, Brown University}
    \date{}
  \maketitle
} \fi

\if1\blind
{
  \bigskip
  \bigskip
  \bigskip
  \begin{center}
    {\LARGE\bf Title}
\end{center}
  \medskip
} \fi

\bigskip
\begin{abstract}
     Factors models are commonly used to analyze high-dimensional data in both single-study and multi-study settings. Bayesian inference for such models relies on Markov Chain Monte Carlo (MCMC) methods, which scale poorly as the number of studies, observations, or measured variables increase. To address this issue, we propose new variational inference algorithms to approximate the posterior distribution of Bayesian latent factor models using the multiplicative gamma process shrinkage prior. The proposed algorithms provide fast approximate inference at a fraction of the time and memory of MCMC-based implementations while maintaining comparable accuracy in characterizing the data covariance matrix. We conduct extensive simulations to evaluate our proposed algorithms and show their utility in estimating the model for high-dimensional multi-study gene expression data in ovarian cancers. Overall, our proposed approaches enable more efficient and scalable inference for factor models, facilitating their use in high-dimensional settings. An R package VIMSFA implementing our methods is available on GitHub (\href{https://github.com/blhansen/VI-MSFA} {github.com/blhansen/VI-MSFA}).
\end{abstract}

\noindent%

\spacingset{1.45}
\section{Introduction}
\label{sec:intro}

Factor Analysis (FA) models are popular tools for providing low-dimensional  data representations through latent factors. These factors are critical to visualize, denoise and explain patterns of interest of the data, making FA models useful in several fields of application such as biology~\citep{pournara_factor_2007}, finance~\citep{LudvigsonSydneyC.2007Terr}, public policy~\citep{Samartsidis2020ABmf}, and nutrition~\citep{edefonti_nutrient-based_2012, Joo2018}---among many others. 

In high-dimensional settings, FA models shrinking many latent factors' components to zero are particularly helpful. These models allow focusing only on a small set of important variables, making the interpretation of the factors easier~\citep{carvalho_high-dimensional_2008}. Moreover, they can lead to more accurate parameter estimations compared to models that do not include shrinkage or sparsity~\citep{avalos-pacheco_heterogeneous_2022}. In a Bayesian setting, many priors have been proposed to induce shrinkage or sparsity in the latent factors (e.g., \cite{archambeau_sparse_2008, carvalho_high-dimensional_2008, legramanti_bayesian_2020, Cadonna2020TtGU, FruehwirthSchnatter2023Gcsp}). A widely used one is the gamma process shrinkage prior \citep{bhattacharya_sparse_2011, durante_note_2017}  that induces a shrinkage effect increasing with the number of factors. This prior adopts an adaptive approach for automatically choosing the latent dimension (i.e., the number of latent factors), facilitating posterior computation.

In a Bayesian context, inference on the posterior distribution of FA typically relies on Markov-Chain Monte Carlo (MCMC) algorithms~\citep{lopes_bayesian_2004}, which scale poorly to high-dimensional settings \citep{rajaratnam_mcmc-based_2015}. In addition, computing posterior expectations from MCMC can be computationally intensive. Specifically, many parameters are invariant 
to orthogonal transformations and must be post-processed before being averaged for inference, exacerbating time and memory needs. We refer to~\citet{papastamoulis_identifiability_2022} for examples of post-processing algorithms. In high-dimensional settings, MCMC for FA can also show poor mixing \citep{fruehwirth-schnatter_2023}, further increasing the difficulty of obtaining high-quality posterior inference without extensive computational resources.

While some alternatives to MCMC that seek the posterior mode via Expectation-Maximization (EM) algorithms have been developed for Bayesian FA \citep[e.g.,][]{rockova_fast_2016}, fast approximation of the posterior distribution that has the potential to scale inference in high-dimensional settings are still under-explored. Variational Bayes~(VB) is one popular class of approximate Bayesian inference methods which has been successfully applied to approximate the posterior distribution of several models, such as logistic regression \citep{jaakkola97a, durante_conditionally_2019}, probit regression ~\citep{fasano_scalable_2022}, latent Dirichlet allocation \citep{blei_variational_2017},  and network factor models \citep{aliverti_stratified_2022}, among others. VB has also been shown to scale Bayesian inference to high dimensional datasets for models closely related to FA. For example, \cite{ghahramani1999variational}  develop a VB approximation for Bayesian mixtures of factor analysers, \cite{wang_varfa_2020} consider item response theory models with latent factors, while \cite{dang_fitting_2022} propose VB algorithms for structural equation models. However, VB for FA models with shrinkage priors is still under-explored, and to our knowledge, no VB implementations for FA with gamma process shrinkage priors exist. To address this issue and enable the practical use of FA in high-dimensional settings, we present two VB algorithms to approximate the posterior of this model and extend them to settings where multiple data from different sources are available. 
 
 There are many applications where data are collected from multiple sources or studies. These sources of information are then combined to produce more precise estimates and results that are robust to study-specific biases. For example, in biology, different microarray cancer studies can collect the same gene expression, measured in different platforms and/or with different late-stage patients. One standard approach to analyzing these multiple data is to stack them into a single dataset and perform FA~\citep{wang_unifying_2011}.  This approach can lead to misleading conclusions by ignoring potential study-to-study variability arising from both biological and technical sources~\citep{garrett-mayer_cross-study_2008}. Therefore, statistical methods that are able to estimate concurrently common and study-specific signals should be adopted. Recent methods have been proposed to enable FA to integrate multiple sources of information in a single statistical model. These include perturbed factor analysis \citep{roy_perturbed_2020}, Multi-Study Factor Analysis~\citep[MSFA,][]{de_vito_multistudy_2019, de_vito_bayesian_2021}, Multi-Study Factor Regression~\citep{DeVito2023Mfrm}, and Bayesian combinatorial MSFA \citep{grabski_bayesian_2020}.

In this paper, first, we present VB algorithms to conduct fast Bayesian estimation for FA with a continuous response, and second, we extend these algorithms to a multi-study setting. We adopt the MSFA model, which decomposes the covariance matrix of the data in terms of common and study-specific latent factors and extends the gamma process shrinkage prior in this context~\citep{ de_vito_multistudy_2019, de_vito_bayesian_2021}. Compared to current MCMC implementations, we show that VB approximations for FA and MSFA greatly reduce computation time while still providing accurate posterior inference. 

We emphasize that our proposed VB algorithms are tailored for efficient point estimation of the latent factor loadings. This is typically the main goal in FA, which is mainly used as an exploratory tool. In fact, the non-identifiability and the unknown dimensions of the factor loadings make uncertainty quantification of the latent factors themselves challenging.

The structure of this paper is as follows: \S \ref{sec:BFA} defines single-study FA and describes VB estimation of the model parameters through Coordinate Ascent Variational Inference (CAVI) and Stochastic Variational Inference (SVI) algorithms; \S \ref{sec:MSFA} generalizes these algorithms in a multi-study setting;  \S \ref{sec:sims} provides simulation studies benchmarking the performance of the proposed VB algorithms in comparison to previous MCMC based algorithms; \S \ref{sec:Application} applies the proposed algorithms to a dataset containing gene expression values from $1,198$ patients with ovarian cancer across $4$ studies \citep{curatedovarian}; finally, \S \ref{sec:conc} provides a brief discussion. 

\section{Factor Analysis}\label{sec:BFA}
Let $\mathbf{x}_i = (x_{i1}, \ldots, x_{iP}) \in \mathbb R^P$ be a vector of $P$ centered observed variables for individual $i=1,\ldots,N$. FA assumes that $\mathbf{x}_i$  can be modelled as a function of $J \ll P$ latent factors or scores $\mathbf{l}_{i} \in \mathbb{R}^J$, and a corresponding factor loading matrix $\boldsymbol \Lambda \in\mathbb{R}^{P\times J}$ via 
\begin{equation}\label{fa_single_study}
    \mathbf{x}_{i} = \boldsymbol{\Lambda} \mathbf{l}_{i} + \mathbf{e}_{i},
    \hspace{0.5cm}
    i=1,\ldots,N.
\end{equation}
Both the factors  $\{\mathbf{l}_{i}\}$ and the \textit{idiosyncratic errors} $\{\mathbf{e}_{i}\}$ are assumed to be  normally distributed, with $\mathbf{l}_{i} \sim \mathcal{N}_{J}(\boldsymbol{0}, \boldsymbol{I}_{J})$, and  $\mathbf{e}_{i}\sim \mathcal{N}_P(\boldsymbol{0}, \boldsymbol{\Psi})$ where $\boldsymbol{\Psi}=\text{diag}\left(\psi_{1}^2,\ldots,\psi_{P}^2\right)$. 
Factors $\{\mathbf{l}_i\}$ are assumed to be independent from the errors $\{\mathbf{e}_i\}$, and the latent dimension $J \ll P$ is typically unknown. As a consequence, the observed covariance matrix, $\boldsymbol{\Sigma} =  \mathrm{Cov}(\mathbf{x}_i)$, can be expressed as $\boldsymbol{\Sigma}=\boldsymbol{\Lambda}\boldsymbol{\Lambda}^\intercal + \boldsymbol{\Psi}$. Given that $\boldsymbol{\Psi}$ is a diagonal matrix, the non-diagonal elements of $\boldsymbol{\Lambda}\boldsymbol{\Lambda}^\intercal$ represent the pairwise covariances of the $P$ variables, i.e.,
$
\mathrm{Cov}(x_{ip},x_{iq}) = \sum_{j=1}^J\lambda_{pj}\lambda_{qj}
$ for $p\neq q$.

In the following, we will use $\mathbf{X} = \{\mathbf{x}_i\}$ to indicate the $N\times P$ data matrix used for the analysis.

\subsection{Prior Specification}

Following a Bayesian approach to inference for \model~\eqref{fa_single_study},
we use the gamma process shrinkage prior of \citep{bhattacharya_sparse_2011} for each element $\{\lambda_{pj}\}$ of the loading matrix $\boldsymbol \Lambda.$
Indicating with $\Gamma(a,b)$  a gamma distribution with mean $a/b$ and variance $a/b^2$, this prior can be expressed via: 
\begin{gather}
    \lambda_{pj} \mid \omega_{pj}, \tau_{j} \sim \mathcal{N}(0, \omega_{pj}^{-1} \tau_{j}^{-1} ), \quad p=1,\ldots,P, \quad j=1,\ldots\infty, \nonumber \\ 
    \omega_{pj} \sim \Gamma\left(\frac{\nu}{2},\frac{\nu} {2}\right), \quad 
    \tau_{j} =\prod_{l=1}^{j}\delta_{l}, 
    \quad
    \delta_{1} \sim \Gamma(a_{1},1),
   \quad
    \delta_{l} \sim \Gamma(a_{2}, 1), 
    \hspace{0.35cm}
    l \geq 2,
    \label{eq:prior_shrinkage_single_study}
\end{gather}
where $\delta_l \; (l=1,2,\dots)$ are independent, $\tau_j$ is the global shrinkage parameter for column $j$, and $\omega_{pj}$ is the local shrinkage parameter for element $p$ in column $j$. 

In practice,  prior~\eqref{eq:prior_shrinkage_single_study} is truncated to a conservative upper bound $J^*,$ smaller than $P$. We refer to \cite{bhattacharya_sparse_2011} for more details on the choice of $J^*.$

Lastly, an inverse-gamma prior is placed on the diagonal entries of $\boldsymbol \Psi$, which is a common choice in FA \citep{lopes_bayesian_2004, rockova_fast_2016}:
\begin{equation}\label{eq:prior_psi_single_study}
    \psi_{p}^{-2} \sim \Gamma(a^{\psi},b^{\psi}).
\end{equation}

For simplicity, we denote all the model parameters with 
$
 \boldsymbol{\theta} = \left(
    \left\{\boldsymbol \Lambda_{p}\right\}, \left\{\mathbf{l}_{i}\right\}\left\{ \psi_{p}^2 \right\},  \left\{\omega_{pj}\right\}, \left\{\delta_{l} \right\} \right)
$
where $\boldsymbol{\Lambda}_{p} = (\lambda_{1p},\ldots, \lambda_{J^*p})$ is row $p$ of $\boldsymbol{\Lambda}$, and with $\pi(\boldsymbol{\theta})$ the prior \eqref{eq:prior_shrinkage_single_study}--\eqref{eq:prior_psi_single_study}.

\subsection{Variational Inference for Factor Analysis}\label{sec:single_study_VI}

The goal of VB algorithms is to approximate the posterior distribution $p(\boldsymbol{\theta} | \mathbf{X})$, with a density $q^*(\boldsymbol{\theta})$ in a class $\mathcal{Q}$ (see ~\cite{blei_variational_2017} for further details).  Specifically, the VB approximation of the posterior
is defined as the distribution
$q^*(\boldsymbol{\theta}) \in \mathcal{Q}$ closest to $p(\boldsymbol{\theta} 
| \mathbf{X})$ in  Kullback-Leibler (KL) divergence, satisfying the minimization:
\begin{equation*}
    \underset{q(\boldsymbol{\theta}) \in \mathcal{Q}}{\arg\min} \left\{ \text{KL}\left(q(\boldsymbol{\theta})\middle\| p(\boldsymbol{\theta}|\mathbf{X})\right) \right\}=  \underset{q(\boldsymbol{\theta}) \in \mathcal{Q}}{\arg \min} \left\{\int_{\boldsymbol{\Theta}} q(\boldsymbol{\theta}) \log \frac{q(\boldsymbol{\theta})}{p(\boldsymbol{\theta}|\mathbf{X})}d\theta \right\},
\end{equation*}
or equivalently the maximization of the \textit{evidence based lower bound} (ELBO):
\begin{equation}\label{ELBO}
    \underset{q(\boldsymbol{\theta}) \in \mathcal{Q}}{\arg\max}\left\{ \text{ELBO}(q(\boldsymbol{\theta}))\right\}= \underset{q(\boldsymbol{\theta}) \in \mathcal{Q}}{\arg\max}\left\{ \mathbb{E}_{q(\boldsymbol{\theta})}[\log p( \boldsymbol{\theta} , \mathbf{X} )] - \mathbb{E}_{q(\boldsymbol{\theta})}[\log q(\boldsymbol{\theta})]\right\},
\end{equation}
where $p( \boldsymbol{\theta} , \mathbf{X}) = p(\mathbf{X} | \boldsymbol\theta) \pi(\boldsymbol \theta)$ is the joint distribution of the data and $\boldsymbol{\theta},$ and the expectations are with respect to $q(\boldsymbol{\theta})$.
A common approach to select $\mathcal{Q}$ is to use a mean-field variational family, $ \mathcal{Q}^{\text{MF}} = \{ q: q( \boldsymbol \theta ) = \prod_{m=1}^M q(\boldsymbol \theta_m)\},$ that considers a partition of the parameters $\boldsymbol{\theta}$ into $M$ blocks, $(\boldsymbol \theta_1, \ldots, \boldsymbol \theta_M)$ and approximate the posterior with a product of $M$ independent density functions $q(\boldsymbol{\theta}_m),$ referred to as the variational factor for $\boldsymbol{\theta}_m$. When using $\mathcal{Q}^{\text{MF}},$ the variational factors $\{q^*(\boldsymbol \theta_m)\}$ composing the VB approximation of the posterior can obtained from 
\begin{equation}\label{cavi-update}
    q^*_m(\boldsymbol{\theta}_m) \propto \exp \left( \mathbb{E}_{q(\boldsymbol \theta_{-m})}\left[\log(p(\boldsymbol{\theta}_m | \boldsymbol{\theta}_{-m}, \mathbf{X})\right]\right),
\end{equation}
where $\boldsymbol{\theta}_{-m} = (\boldsymbol\theta_1, \ldots,\boldsymbol\theta_{m-1},\boldsymbol \theta_{m+1}, \ldots, \boldsymbol \theta_M)$ is the vector of parameters excluding the $m$-th one (Chapter 10, \citet{bishop_pattern_2006}).

When the conditional distribution $p(\boldsymbol{\theta}_m|\boldsymbol{\theta}_{-m}, \mathbf{X})$ belongs to the exponential family, the corresponding optimal variational factor $q^*(\boldsymbol{\theta}_m)$ also belongs to the same exponential family. Therefore,
the maximization problem~\eqref{ELBO} reduces to learning the value of the parameters characterizing the distribution $q^*_m(\boldsymbol{\theta}_m).$ We will indicate these parameters as $\boldsymbol{\varphi}^*_m$ and the corresponding optimal variational factor as  $q^*(\boldsymbol{\theta}_m;\boldsymbol{\varphi}^*_m).$ Choosing 
$\mathcal{Q}^{\text{MF}}$ in such a way that $\{\pi(\boldsymbol \theta_m)\}$ are conditionally conjugate prior for $\{\boldsymbol \theta_m\}$ enables the development of efficient algorithms 
such as the \textit{Coordinate-Ascent Variational Inference} (CAVI) which learns $\boldsymbol{\varphi}^*$ by iteratively updating each parameter of $q^*(\boldsymbol{\theta}_m;\boldsymbol{\varphi}^*_m)$ conditional to the others variational factors until convergence is reached~(Chapter~10, \citet{bishop_pattern_2006}).

To implement CAVI for FA with priors  \eqref{eq:prior_shrinkage_single_study}--\eqref{eq:prior_psi_single_study}, we propose the following mean-field variational family: 
\begin{equation}\label{mf-approx-single-study}
    \begin{split}
    q(\boldsymbol{\theta}; \boldsymbol{\varphi}) =
    &
    \left[\prod_{l=1}^{J^*} q(\delta_{l};\alpha_{l}^\delta, \beta_{l}^\delta)\right]
    \left[\prod_{p=1}^{P}\prod_{j=1}^{J^*}q(\omega_{pj}; \alpha_{pj}^\omega, \beta_{pj}^\omega) \right] 
    \left[\prod_{i=1}^{N}q(\mathbf{l}_{i}; \boldsymbol{\mu}_{i}^l, \boldsymbol{\Sigma}_{i}^l) \right]
    \times
    \\
    &
    \left[ \prod_{p=1}^{P}q(\boldsymbol{\Lambda}_{p}; \boldsymbol{\mu}_{p}^\Lambda, \boldsymbol{\Sigma}_{p}^\Lambda) \right] 
    \left[\prod_{p=1}^{P}q(\psi_{p}^{-2}; \alpha_{p}^\psi, \beta_{p}^\psi) \right],
    \end{split}
\end{equation}
with variational parameters
$\boldsymbol{\varphi}=\left(\{\alpha_{l}^\delta\}, \{\beta_{l}^\delta\}, \{\alpha_{pj}^\omega\}, \{\beta_{pj}^\omega\}, \{\boldsymbol{\mu}_{i}^l\},\{\boldsymbol{\Sigma}_{i}^l\},\{\boldsymbol{\mu}_{p}^\Lambda\},\{\boldsymbol{\Sigma}_{p}^\Lambda)\},\{\alpha_{p}^\psi\}, \{\beta_{p}^\psi\}\right).$

The mean-field approximation in equation~\eqref{mf-approx-single-study} factorizes the posterior distribution into conditionally conjugate elements, so maximizing the ELBO with respect to $q(\boldsymbol{\theta}_m)$ leads to closed-form expressions for each variational parameter (Supplementary \S F for details). All the steps to obtain the VB posterior approximation are detailed in Algorithm S1 (Supplementary \S A). 

An important aspect of CAVI is that each iteration requires optimizing the \textit{local parameters}
$\boldsymbol{\Sigma}^{l}_{i}$ and $\boldsymbol{\mu}^l_{i},$ relative to the scores $\mathbf{l}_i$ for $i=1,\ldots,N,$ which is computationally expensive in settings with large sample size. Thus, we propose a \textit{Stochastic Variational Inference} (SVI) algorithm, which uses stochastic optimization to reduce the computational cost of each iteration~\citep{hoffman_svi}. The key idea of this algorithm is to compute at each iteration the local parameters only for a small subset of available data and use this subset to approximate the remaining parameters common to all individuals, namely \textit{global parameters}. In the following we use $\mathcal{M}^{\text{G}} \subset \{1, \ldots, M\}$ for the indices of the global parameters. 

As a first step to illustrate the SVI algorithm, we reframe CAVI  as a gradient-based optimization algorithm for the global parameters.  Recall that the mean field approximation is constructed so that the optimal variational factors are distributions in the exponential family, i.e., a function of the natural parameters $\boldsymbol{\eta}\left(\mathbf{X},\boldsymbol{\theta}\right)$ (see \tablename~S1 of Supplementary \S A for details). With this parameterization, the derivative of the ELBO in equation~\eqref{ELBO} with respect to the natural parameters is \citep{hoffman_svi}: 
\begin{equation}\label{elbo-grad}
    \nabla ELBO(q(\boldsymbol{\theta}_m)) = \mathbb{E}_{q(\boldsymbol \theta_{-m})}\left[\boldsymbol{\eta}\left(\mathbf{X},\boldsymbol{\theta}\right)\right] - \boldsymbol{\varphi}_m \quad \mbox{for}\quad m \in \mathcal{M}^{\text{G}}.
\end{equation}
Equating the gradient \eqref{elbo-grad} to $0,$ the solution of~\eqref{ELBO} can be written as $\boldsymbol{\varphi}_m =  \mathbb{E}_{q(\boldsymbol \theta_{-m})}\left[\boldsymbol{\eta}\left(\mathbf{X},\boldsymbol{\theta}\right)\right]$.

The SVI replaces the gradient with an estimate much easier to compute. This is accomplished by
selecting a batch-size parameter $b \in (0,1)$ and drawing a  random sample of the data of dimension $\widetilde N = \lfloor b\times N \rfloor,$ at each iteration of the algorithm. Here $\lfloor x \rfloor,$ indicates the greatest integer less than or equal to $x.$ At iteration $t$, let $\mathcal I(t) \subseteq \{1,\ldots, N\}$ be the subset of sampled observations, and $\widetilde{\mathbf{X}}^t = \{\mathbf{x}_i \mbox{ for } i \in \mathcal I(t)\}$ the corresponding $ \widetilde N  \times P$ data matrix.  SVI proceed by first updating the local parameters $\{\boldsymbol{\Sigma}^{l}_{i}\}$ and $\{\boldsymbol{\mu}^l_{i}\}$ for all $i \in \mathcal I(t),$ and then computing an approximated  gradient for the global parameters via
\begin{equation}\label{noisy-grad}
   \reallywidehat{\nabla ELBO(q(\boldsymbol{\theta}_m))}= \mathbb{E}_{q(\boldsymbol{\theta}_{-m})}\left[\boldsymbol{\eta}(\widetilde{\mathbf{X}}^t,\boldsymbol{\theta})\right] - \boldsymbol{\varphi}_m,
   \quad \mbox{for}\quad m \in \mathcal{M}^{\text{G}},
\end{equation}
where the expectation $\mathbb{E}_{q(\boldsymbol{\theta}_{-m})}\left[\boldsymbol{\eta}(\widetilde{\mathbf{X}}^t,\boldsymbol{\theta})\right]$ is computed giving weight $N/\widetilde N$ to each individual in $\mathcal I(t)$~\citep{hoffman_svi}.  We call $\widehat{\boldsymbol{\varphi}}_m$ the solution of \eqref{noisy-grad}. At each iteration $t$ of the SVI algorithm, the update for the $\{\boldsymbol{\varphi}_m(t)\}$ of the global parameters is obtained as a weighted average of $\widehat{\boldsymbol{\varphi}}_m$ and  $\boldsymbol{\varphi}_m{(t-1)}$ at the previous iteration:
\begin{equation}\label{SVI-updates}
    \boldsymbol{\varphi}_m{(t)} = (1-\rho_t)\boldsymbol{\varphi}_m{(t-1)} + \rho_t\widehat{\boldsymbol{\varphi}}_m,
    \quad \mbox{for}\quad m \in \mathcal{M}^{\text{G}},
\end{equation}
where $\rho_t$ is a \textit{step size} parameter such that $\sum_{t}\rho_t \rightarrow \infty$ and  $\sum_{t}\rho_t^2 < \infty$ \citep{robbins_monro}. The choice of $\rho_t$ can impact convergence. Step sizes that are too large may overweight estimated gradients and lead to unreliable estimates of variational parameter updates for some iterations, while step sizes that are too small may take many iterations to converge. A typical choice is to define $\rho_t = (t+\tau)^{-\kappa}$, where $\kappa\in(0.5,1]$ is the forgetting rate, which controls how quickly the variational parameters change across several iterations, and $\tau>0$ is the delay, which down-weights early iterations \citep{hoffman_svi}. 

To implement SVI for FA, we reparameterize the global parameters in \eqref{mf-approx-single-study} in terms of the natural parameterization of the exponential family (Supplementary \S A \tablename~S1). The details are provided in Algorithm S2 (Supplementary \S A).  

\section{Multi-Study Factor Analysis}\label{sec:MSFA}

We extend the two algorithms presented in \S \ref{sec:BFA} to the setting where there are $S>1$ datasets measuring the same $P$ variables, using the MSFA \model~\citep{de_vito_bayesian_2021}.

We indicate with
 $\mathbf{x}_{si} = (x_{si1},\ldots x_{siP}) \in \mathbb R^P$
the vector of $P$ centered observed  variables for individual $i=1,\ldots,N_s$ in study $s=1,\ldots,S$ and use the model
\begin{equation}\label{multistudy}
    \mathbf x_{si} = \boldsymbol{\Phi} \mathbf{f}_{si} + \boldsymbol{\Lambda}_s \mathbf{l}_{si} + \mathbf{e}_{si}
    \quad
    i=1,\ldots,N_s,
    \quad
    s=1, \ldots, S,
\end{equation}
where $\mathbf{f}_{si} \sim \mathcal{N}_K( \boldsymbol{0}, \boldsymbol{I}_k )$ 
are the $K$-dimensional scores  which correspond to the shared loading matrix $\boldsymbol{\Phi}\in\mathbb{R}^{P\times K},$ and $\mathbf{l}_{si}\sim\mathcal{N}_{J_s}(\boldsymbol{0}, \boldsymbol{I}_{J_s})$ are the $J_s$-dimensional scores with study-specific loading matrices $\boldsymbol{\Lambda}_s\in \mathbb{R}^{P\times J_s}$. Finally, $\mathbf{e}_{is}\sim\mathcal{N}(\boldsymbol{0},\boldsymbol{\Psi}_s)$ are  idiosicratic errors with $\boldsymbol{\Psi}_s = \text{diag} \left(\psi_{s1}^{2},\ldots,\psi_{sp}^{2}\right)$.  \model~\eqref{multistudy} can be seen as a generalization of \model~\eqref{fa_single_study} that incorporates shared latent components, $\boldsymbol{f}_{si}$ and $\boldsymbol{\Phi}$, and study-specific latent components, $\boldsymbol{l}_{si}$ and $\boldsymbol{\Lambda}_s$. 
The covariance matrices of each study $
\boldsymbol{\Sigma}_s = \mathrm{Cov}(\mathbf{x}_{si})$ for $s=1,\ldots,S$ can be decomposed as
$\boldsymbol\Sigma_s= \boldsymbol{\Phi}\boldsymbol{\Phi}^{\intercal} + \boldsymbol{\Lambda}_s\boldsymbol{\Lambda}_s^{\intercal} + \boldsymbol{\Psi}_s$.
The terms $\boldsymbol{\Phi}\boldsymbol{\Phi}^{\intercal}$ and $\boldsymbol{\Lambda}_s\boldsymbol{\Lambda}_s^{\intercal}$ represent the portion of the covariance matrix that is attributable to the common factors and the study-specific factors, respectively.
We refer to the $N_s \times P$ data matrix from study $s$ as $\mathbf{X}_s$.

\subsection{Prior Specification}\label{msfa model spec}
Following~\cite{de_vito_bayesian_2021}, we use the gamma process shrinkage prior~\citep{bhattacharya_sparse_2011} for both the study-specific and the shared loading matrix. The prior for each study-specific loading element $\lambda_{spj}$ is
\begin{gather}
    \lambda_{spj} \mid \omega_{spj}, \tau_{sj} \sim \mathcal{N}\left(0, \omega_{spj}^{-1}\tau_{sj}^{-1}\right), \quad s=1,\ldots, S \quad p=1,\ldots,P, \quad j=1,\ldots,\infty, \nonumber \\
    \omega_{spj} \sim \Gamma\left(\frac{\nu_s}{2},\frac{\nu_s}{2}\right), \quad
    \tau_{sj}=\prod_{l=1}^{j}\delta_{sl}, \quad
    \delta_{s1} \sim \Gamma\left(a_{s1},1\right), \quad
    \delta_{sl} \sim \Gamma\left(a_{s2}, 1\right), 
   \quad
    l \geq 2, \label{prior-lambda_s} 
\end{gather}
where $\delta_{sl} \; (l=1,2, \dots)$ are independent, $\tau_{sj}$ is the global shrinkage parameter for column $j$, and $\omega_{spj}$ is the local shrinkage for the element $p$ in column $j$. 
Similarly, the prior for each shared loading element $\boldsymbol \phi_{pk}$ is
\begin{gather}
    \phi_{pk} \mid \omega_{pk}, \tau_{k} \sim \mathcal{N}(0, \omega_{pk}^{-1}\tau_k^{-1}),  \quad p=1,\ldots,P, \quad k=1,\ldots,\infty, \nonumber \\
    \omega_{pk} \sim \Gamma\left(\frac{\nu} {2},\frac{\nu}{2}\right), \quad 
    \tau_{k}=\prod_{l=1}^{k}\delta_l, \quad \delta_1 \sim \Gamma(a_1,1), \quad  \delta_l \sim \Gamma(a_2, 1), \quad l \geq 2, \label{eq:prior-phi}
\end{gather}
where $\delta_l \; (l=1,2,\dots)$ are independent, $\tau_k$ is the global shrinkage parameter for column $k$ and $\omega_{pk}$ is the local shrinkage parameter for element $p$ in column $k$.
 
These two priors introduce infinitely many factors for both the study-specific and the shared components (i.e.,  $J_s = \infty,$ for $s=1,\ldots,S$, and $K=\infty$). As detailed in \S \eqref{sec:BFA}, we truncate the latent dimensions using upper bounds $K^*$ and $J_s^*,$ to all the studies.
Finally, an inverse-gamma prior is used for the diagonal entries of $\boldsymbol{\Psi}_s$ 
\begin{equation}\label{prior-psi_sp}
    \psi_{sp}^{-2} \sim \Gamma(a^{\psi},b^{\psi}).
\end{equation}

We denote
$
 \boldsymbol{\theta} = \left(
    \left\{ \boldsymbol \Phi_p \right\},  \left\{\boldsymbol \Lambda_{sp}\right\}, \left\{ \psi_{ps}^2 \right\}, \left\{ \omega_{pk} \right\}, \left\{ \delta_l \right\},  \left\{\omega_{spj}\right\}, \left\{\delta_{sl} \right\} \right)
$
as the vector of all model parameters, where $\boldsymbol{\Phi}_p$ is row $p$ of $\boldsymbol{\Phi}$ and $\boldsymbol{\Lambda}_{sp}$ is row $p$ of $\boldsymbol{\Lambda}_{s}$.

\subsection{Variational Inference for Multi-Study Factor Analysis}
The computational cost of \model~\eqref{multistudy} is larger than $S$ times the computational cost of \model~\eqref{fa_single_study} because we need to estimate both the shared and study-specific parameters. Therefore, the benefit of fast VB posterior approximation is even higher than in the single-study setting, especially if several high-dimensional studies are available for analysis. This is increasingly common in many applied settings, such as cancer genomics, where the expression levels of a large set of genes are measured across multiple cancer types. 

To extend CAVI in a multi-study setting, we use the following mean field factorization
\begin{equation}\label{mf-approx-multi}
    \begin{split}
    q(\boldsymbol{\theta}; \boldsymbol{\varphi}) &=
     \left[ \prod_{l=1}^{K^*} q(\delta_l;\alpha_l^\delta, \beta_l^\delta) \right]
    \left[ \prod_{s=1}^S\prod_{l=1}^{J^*_s} q(\delta_{sl};\alpha_{sl}^\delta, \beta_{sl}^\delta)\right]
    \left[ \prod_{p=1}^{P}\prod_{k=1}^{K^*}q(\omega_{pk}; \alpha_{pk}^\omega, \beta_{pk}^\omega)] \right]\\
    &\times
    \left[ \prod_{s=1}^{S}\prod_{p=1}^{P}\prod_{j=1}^{J^*_s}q(\omega_{spj}; \alpha_{spj}^\omega, \beta_{spj}^\omega)] \right] 
    \left[ \prod_{s=1}^{S}\prod_{i=1}^{N_s}q(\boldsymbol{f}_{si}; \boldsymbol{\mu}_{si}^f, \boldsymbol{\Sigma}_{si}^f)] \right]
    \left[ \prod_{s=1}^{S}\prod_{i=1}^{N_s}q(\mathbf{l}_{si}; \boldsymbol{\mu}_{si}^l, \boldsymbol{\Sigma}_{si}^l)] \right] 
    \\
    &\times
    \left[ \prod_{p=1}^{P}q(\boldsymbol{\Phi}_p; \boldsymbol{\mu}_p^\Phi, \boldsymbol{\Sigma}_p^\Phi)] \right]
    \left[ \prod_{s=1}^{S}\prod_{p=1}^{P}q(\boldsymbol{\Lambda}_{sp}; \boldsymbol{\mu}_{sp}^\Lambda, \boldsymbol{\Sigma}_{sp}^\Lambda)] \right] 
    \left[ \prod_{s=1}^{S}\prod_{p=1}^{P}q(\psi_{sp}^{-2}; \alpha_{sp}^\psi, \beta_{sp}^\psi)] \right]
    \end{split}
\end{equation}
where $\boldsymbol{\Phi}_p$ is the $p^{th}$ row of $\boldsymbol{\Phi}$, $\boldsymbol{\Lambda}_{sp}$ is the $p^{th}$ row of $\boldsymbol{\Lambda}_s$, and 
\begin{equation*}
    \begin{split}
        \boldsymbol{\varphi} = 
    &\left(
    \{\alpha_l^\delta\}, \{\beta_l^\delta\},
    \{\alpha_{sl}^\delta\}, \{\beta_{sl}^\delta\},
    \{\alpha_{pk}^\omega\}, \{\beta_{pk}^\omega\},
    \{\alpha_{spj}^\omega\},\{\beta_{spj}^\omega\}, \right. \\
    &\left. 
    \{\boldsymbol{\mu}_{si}^f\}, \{\boldsymbol{\Sigma}_{si}^f\}, 
    \{\boldsymbol{\mu}_{si}^l\}, \{\boldsymbol{\Sigma}_{si}^l\}, 
    \{\boldsymbol{\mu}_p^\Phi\}, \{\boldsymbol{\Sigma}_p^\Phi\},
    \{\boldsymbol{\mu}_{sp}^\Lambda\}, \{\boldsymbol{\Sigma}_{sp}^\Lambda\},
    \{\alpha_{sp}^\psi\}, \{\beta_{sp}^\psi\}
    \right)
    \end{split}
\end{equation*}
is the vector of variational parameters. 
Maximizing the ELBO with equation~\eqref{cavi-update} leads to closed-form expressions for optimal variational parameters for each factor, conditional on the others (see Supplementary \S F for supporting calculations). Our implementation of CAVI for Bayesian MSFA is detailed in Algorithm S3 (Supplementary \S A).

CAVI for MSFA requires computing both shared and study-specific scores for each observation in each study at each iteration. When $\{N_s\}$ are large, this can become very computationally demanding.  Thus, we generalize the SVI algorithm described in \S \ref{sec:BFA}  to the multi-study setting using the natural parameterization in \tablename~S2 (Supplementary \S A). The sub-sampling step of SVI is generalized to a multi-study setting by taking a sub-sample of size $\widetilde N_s =\lfloor b_s \times N_s \rfloor$ from each study. Our implementation of SVI for Bayesian MSFA is presented in Algorithm S4 (Supplementary \S A).

\section{Simulation Study}\label{sec:sims}
To assess the accuracy and computational performance of the CAVI and SVI algorithms for FA (\S \ref{sec:BFA}) 
and MSFA (\S \ref{sec:MSFA}),
we simulated different scenarios with varying numbers of subjects, variables, and studies. We compared the performance of the proposed algorithms (available at \href{https://github.com/blhansen/VI-MSFA}{github.com/blhansen/VI-MSFA}) against several competitors, including the Gibbs Sampler (GS) and Expectation Conditional Maximization (ECM) algorithms (available at \href{https://github.com/rdevito/MSFA}{github.com/rdevito/MSFA}), the principal components based POET algorithm~\citep{fan_large_2013}, and the Automatic Differentiation Variational Inference (ADVI) algorithm~\citep{kucukelbir2017}, implemented in STAN~\citep{carpenter_stan_2017}. The algorithms were evaluated in terms of: 1) computation time measured in seconds, 2) maximum RAM usage in megabytes (Mb), and, 3) estimation accuracy of $\boldsymbol{\Sigma}=\boldsymbol{\Lambda}\boldsymbol{\Lambda}^\intercal+ \boldsymbol{\Psi}$ and $\boldsymbol{\Sigma}_s=\boldsymbol{\Phi}\boldsymbol{\Phi}^\intercal + \boldsymbol{\Lambda}_s\boldsymbol{\Lambda}_s^\intercal + \boldsymbol{\Psi}_s$ in the single and multi-study simulations, respectively. Computation time and RAM usage are monitored using the R package \texttt{peakRAM}~\citep{peakRAM}, while estimation accuracy is evaluated using the RV coefficient between the true, $\boldsymbol{\Sigma}$ or $\boldsymbol{\Sigma_s}$, and the estimated covariance matrices, $\widehat{\boldsymbol{\Sigma}}$ or $\widehat{\boldsymbol{\Sigma}_s}$. The RV coefficient is a measure of similarity between two matrices varying between 0 and 1 defined as 
\begin{equation}\label{eq:RV-def}
    \text{RV}(\mathbf{\boldsymbol{\Sigma}},\widehat{\boldsymbol{\Sigma}}) = \frac{\Tr(\boldsymbol{\Sigma}\widehat{\boldsymbol{\Sigma}}^\intercal\widehat{\boldsymbol{\Sigma}}\boldsymbol{\Sigma}^\intercal)}{\sqrt{\Tr(\boldsymbol{\Sigma}\boldsymbol{\Sigma}^\intercal)^2\Tr(\widehat{\boldsymbol{\Sigma}}\widehat{\boldsymbol{\Sigma}}^\intercal)^2}}.
\end{equation}
An RV coefficient close to 1 (0) indicates strong similarity (dissimilarity) between the two matrices~\citep{robert_unifying_1976}. Additional metrics to evaluate estimation accuracy, including the Frobenius norm and L1 norm of the difference between the estimated covariance matrices and the ground truth, are provided in Supplementary~\S C (\tablename s~S3, S5, S7--S13).

For SVI Algorithms S2 and S4, we set the forgetting rate $\kappa$ and delay parameter $\tau$, which control the step-size schedule for $\rho_t$, to $\kappa=0.75$ and $\tau=1$. We must also specify the batch size parameters $b$, which control the number of points sampled at each iteration to estimate the gradient of the ELBO in Equation \eqref{noisy-grad}. The choice of batch size is non-trivial and consistently affects the accuracy and computational cost of the SVI algorithms. We refer to \cite{tan_stochastic_2017} as a recent work that considers the choice of batch size. To assess the effect of batch size on model performance in our settings, we consider batch sizes corresponding to $5\%,$ $20\%,$ and $50\%$ of the sample size,  $b \in \{0.05, 0.20, 0.50\}$. For convenience, we will refer to SVI algorithms with these batch sizes as SVI-005, SVI-02, and SVI-05, respectively. Parameter initialization for CAVI and SVI is described in Supplementary \S B. Convergence is monitored using the mean squared difference in parameters across iterations. 

For GS, we draw $10,000$ posterior samples after a burn-in of $5,000$ and save in memory the values of the parameters every $5$ iterations, i.e.,  a total of $2,000$ samples.
We then use these samples to compute the averages of  $\boldsymbol{\Phi}\boldsymbol{\Phi}^\intercal$ and $\boldsymbol{\Lambda}_s\boldsymbol{\Lambda}_s^\intercal$ as point estimates of their respective posterior means. Note that $\boldsymbol{\Phi}\boldsymbol{\Phi}^\intercal$ and $\boldsymbol{\Lambda}_s\boldsymbol{\Lambda}_s^\intercal$ are identifiable functionals of the posterior of Model \eqref{multistudy}, therefore posterior means can be estimated directly by averaging the GS draws. 

\subsection{Single-Study Simulations}\label{sec:simulations-single-study}
We consider simulation scenarios for all combinations of $P=\{100,500,5000\}$ and $N=\{100,500, \allowbreak 1000\}$,  generating the data from Model \eqref{fa_single_study} with $J=4$ factors. 
For each scenario, we set each value of $\boldsymbol{\Lambda}$ to zero with probability  $2/3$ or randomly generated it from an $\text{Uniform}(0,1)$ with probability $1/3$. Further simulation studies with sparsity levels of 50\% and 80\% are presented in \tablename s~S4-S7. The values of $\boldsymbol{\Psi}$ are drawn from an $\text{Uniform}(0.1,1)$. Using the generated $\boldsymbol{\Sigma} = \boldsymbol{\Lambda}\boldsymbol{\Lambda}^\intercal + \boldsymbol{\Psi}$, we sample 50  datasets for each simulation scenario by taking $N$ draws from $\mathcal{N}_p(\boldsymbol{0}, \boldsymbol{\Sigma})$. For each dataset, we estimate $\boldsymbol{\theta}$ with $J^*=5$ factors using CAVI, SVI, GS, ECM, ADVI, and POET as previously described. Bayesian methods (CAVI, SVI, GS, and ADVI) use priors \eqref{eq:prior_shrinkage_single_study}-\eqref{eq:prior_psi_single_study} with hyperparameters $\nu=3$, $a_1=2.1$, $a_2=3.1$, $a^\psi=1$, and $b^\psi=0.3$. 

\tablename~\ref{table-single-study-sims} reports, for each simulation scenario, the averages and standard deviations for the computation time, memory usage, and estimation accuracy. Comprehensive results are provided in \figurename s~S1-S4 (Supplementary \S C). Given the computational resources available, we found performing GS, ECM, and ADVI infeasible for some scenarios and therefore, results are not reported (see caption of \tablename~\ref{table-single-study-sims} for details).

As expected, CAVI required much less computational time than GS for every scenario. For example, in the scenario with $P=500$ and $N=1,000$, CAVI required, on average, $68$ seconds compared to  $873$ for the GS. The average computation time of SVI was lower than CAVI in every simulation scenario with $N>100$, while in scenarios with $N=100$ and $P= 100, 500,$ SVI-005 required, on average, more computational time  than CAVI. Note that in these scenarios, SVI-005 only uses $5$ observations to approximate the gradient in~\eqref{noisy-grad}, requiring  many iterations to converge. When the sample size increases, SVI is faster than CAVI. This is particularly evident for scenarios with larger $P$. For example, when $P=5,000$ and $N=1,000$ CAVI required on average approximately $25$ minutes compared to $12$, $5$ and $2$ minutes for SVI-05, SVI-02, and SVI-005 respectively. Additionally, SVI with small batch sizes scales better with increasing $N$. In the scenario with 
$P=5,000$, $N=100$, SVI-005, required an average of $23.65$ seconds. The computational time increases six times when $N=1,000$, while the computational time for SVI-02 and SVI-05 increases by 7 and 8 times, respectively. We note that the frequentist algorithms ECM and POET require very little computational time in simulation scenarios with $P=100$, requiring approximately $3$ seconds of runtime with $N=1000$, while CAVI required $13$ seconds in the same scenario. However, ECM cannot be performed in scenarios with $P\geq N$, and POET does not scale with increasing $P$, requiring approximately $1017$ seconds on average when $P=5000$ and $N=100$, which is approximately 6 times longer than CAVI, which required only $164$ seconds on average. ADVI consistently required the most computational time out of any of the algorithms considered, taking approximately $150$ times longer than CAVI even in the smallest simulation scenario considered (202 seconds vs 1.37 seconds).

\begin{table}[ht!]
\caption{\it \small Average computational cost (in seconds), used RAM (in Mb), and estimation accuracy (RV coefficients) across 50 simulation replicates, mean(sd). Note: we were unable to run GS, ADVI, and POET for some scenarios due to computation time and memory limitations, as the maximum allocation of 24 hours of run time and 128Gb of RAM for each iteration was insufficient. Additionally, we are unable to run ECM in scenarios with $P \geq N$. This is indicated with \longdash[5] in the table.}\label{table-single-study-sims}
\centering
\tiny
\resizebox{\textwidth}{!}{
\begin{tabular}{lr|rrr|rrr|rrr}
\multicolumn{2}{c}{} &
    \multicolumn{3}{c}{Time (Seconds)} & \multicolumn{3}{c}{Memory (Mb)} &
    \multicolumn{3}{c}{Estimation Accuracy $\left(\text{RV}(\widehat{\boldsymbol{\Sigma}}, \boldsymbol{\Sigma})\right)$} \\
method & N & P=100 & P=500 & P=5000 & P=100 & P=500 & P=5000 & P=100 & P=500 & P=5000 \\
\specialrule{\cmidrulewidth}{0pt}{0.5pt}
GS & 100 & 136.58(2.1) & 653.15(7.77) &  \multicolumn{1}{c|}{\longdash[5]} & 209.67(0.1) & 4019.35(0.18) &  \multicolumn{1}{c|}{\longdash[5]} & 0.89(0.04) & 0.86(0.05) &  \multicolumn{1}{c}{\longdash[5]}\\
CAVI & 100 & 1.37(0.04) & 7.46(0.2) & 164.45(13.13) & 23.41(0.04) & 47.84(1.6) & 614.99(61.58) & 0.85(0.04) & 0.76(0.03) & 0.74(0.02)\\
SVI-05 & 100 & 0.91(0.03) & 4.72(0.1) & 87.43(2.21) & 23.2(0) & 48.1(0) & 829.63(57.11) & 0.83(0.04) & 0.73(0.03) & 0.72(0.02)\\
SVI-02 & 100 & 0.53(0.02) & 2.67(0.05) & 45.13(1.23) & 20.7(0) & 47.57(1.87) & 615.2(2.13) & 0.81(0.04) & 0.72(0.03) & 0.7(0.02)\\
SVI-005 & 100 & 3.43(1.25) & 1.71(0.11) & 23.65(0.77) & 17.01(0.06) & 32.49(0.76) & 825.7(69.6) & 0.81(0.05) & 0.63(0.04) & 0.62(0.04)\\
POET & 100 & 0.35(0.02) & 10.59(0.89) & 1017.11(1.5) & 50.22(3.25) & 285.54(30.97) & 24585.98(1134.5) & 0.9(0.02) & 0.89(0.03) & 0.89(0.03)\\
ECM & 100 & \multicolumn{1}{c}{\longdash[5]} & \multicolumn{1}{c}{\longdash[5]} & \multicolumn{1}{c|}{\longdash[5]} & \multicolumn{1}{c}{\longdash[5]} & \multicolumn{1}{c}{\longdash[5]} & \multicolumn{1}{c|}{\longdash[5]} & \multicolumn{1}{c}{\longdash[5]} & \multicolumn{1}{c}{\longdash[5]} & \multicolumn{1}{c}{\longdash[5]} \\
ADVI & 100 & 202.48(120.32) &  \multicolumn{1}{c}{\longdash[5]} &  \multicolumn{1}{c|}{\longdash[5]} & 180.56(21.37) &  \multicolumn{1}{c}{\longdash[5]} &  \multicolumn{1}{c|}{\longdash[5]} & 0.76(0.17) &  \multicolumn{1}{c}{\longdash[5]} &  \multicolumn{1}{c}{\longdash[5]}\\
\specialrule{\cmidrulewidth}{0pt}{0.5pt}
GS & 500 & 157.97(2.48) & 745.66(14.21) & \multicolumn{1}{c|}{\longdash[5]} & 293.63(0.16) & 4133.59(1.02) & \multicolumn{1}{c|}{\longdash[5]} & 0.99(0.01) & 0.98(0.01) & \multicolumn{1}{c}{\longdash[5]}\\
CAVI & 500 & 6.29(0.07) & 34.94(0.5) & 789.25(65.87) & 37.11(0.06) & 46.65(0.33) & 696.27(85.19) & 0.98(0.01) & 0.94(0.01) & 0.93(0.01)\\
SVI-05 & 500 & 3.63(0.05) & 19.78(0.3) & 369.37(24.09) & 37.1(0) & 46.5(0) & 748.79(61.45) & 0.97(0.01) & 0.92(0.01) & 0.91(0.01)\\
SVI-02 & 500 & 1.78(0.04) & 9.62(0.14) & 169.66(10.74) & 32.4(0) & 46.5(0) & 745.16(66.47) & 0.97(0.01) & 0.92(0.01) & 0.91(0.01)\\
SVI-005 & 500 & 0.86(0.04) & 4.49(0.08) & 69.01(4.08) & 24.82(0.13) & 46.23(1.57) & 745.4(72.09) & 0.94(0.01) & 0.9(0.01) & 0.89(0.01)\\
POET & 500 & 1.17(0.09) & 22.86(1.9) & 2596.15(137.93) & 103.6(0.03) & 1195.96(0.07) & 100903.5(1101.22) & 0.98(0.01) & 0.97(0.01) & 0.98(0.01)\\
ECM & 500 & 3.19(0.07) & \multicolumn{1}{c}{\longdash[5]} & \multicolumn{1}{c|}{\longdash[5]} & 37.3(0.03) & \multicolumn{1}{c}{\longdash[5]} & \multicolumn{1}{c|}{\longdash[5]} & 0.98(0.01) & \multicolumn{1}{c}{\longdash[5]} & \multicolumn{1}{c}{\longdash[5]}\\
ADVI & 500 & 988.78(1029.77) & \multicolumn{1}{c}{\longdash[5]} & \multicolumn{1}{c|}{\longdash[5]} & 233.59(59.72) & \multicolumn{1}{c}{\longdash[5]} & \multicolumn{1}{c|}{\longdash[5]} & 0.97(0.02) & \multicolumn{1}{c}{\longdash[5]} & \multicolumn{1}{c}{\longdash[5]}\\
\specialrule{\cmidrulewidth}{0pt}{0.5pt}
GS & 1000 & 182.65(2.25) & 873.28(32.11) & \multicolumn{1}{c|}{\longdash[5]} & 396.72(12.59) & 4275.09(1.39) & \multicolumn{1}{c|}{\longdash[5]} & 0.99(0) & 0.99(0) & \multicolumn{1}{c}{\longdash[5]}\\
CAVI & 1000 & 13.42(2.03) & 68.07(1.38) & 1497.61(50.72) & 36.71(0.06) & 44.65(0.34) & 784.08(106.86) & 0.98(0.01) & 0.97(0.01) & 0.97(0.01)\\
SVI-05 & 1000 & 6.9(0.08) & 38.23(0.66) & 702.83(54.63) & 36.7(0) & 60.7(0) & 747.39(109.88) & 0.98(0.01) & 0.96(0.01) & 0.96(0.01)\\
SVI-02 & 1000 & 3.28(0.05) & 18.01(0.3) & 318.39(24.57) & 36.7(0) & 60.38(2.28) & 761.84(104.72) & 0.97(0.01) & 0.96(0.01) & 0.96(0.01)\\
SVI-005 & 1000 & 1.49(0.04) & 7.89(0.12) & 127.56(8.06) & 33.57(0.17) & 60.42(1.94) & 748.56(130.9) & 0.96(0.01) & 0.94(0.01) & 0.94(0.01)\\
POET & 1000 & 3.18(0.16) & 38.87(3.02) & \multicolumn{1}{c|}{\longdash[5]} & 203(0) & 2385.51(0.07) & \multicolumn{1}{c|}{\longdash[5]} & 0.99(0) & 0.99(0) & \multicolumn{1}{c}{\longdash[5]}\\
ECM & 1000 & 3.08(0.07) & 76.9(0.8) & \multicolumn{1}{c|}{\longdash[5]} & 37(0.03) & 61.04(0.3) & \multicolumn{1}{c|}{\longdash[5]} & 0.99(0) & 0.97(0.01) & \multicolumn{1}{c}{\longdash[5]}\\
ADVI & 1000 & 2011.14(2119.07) & \multicolumn{1}{c}{\longdash[5]} & \multicolumn{1}{c|}{\longdash[5]} & 183.79(15.67) & \multicolumn{1}{c}{\longdash[5]} & \multicolumn{1}{c|}{\longdash[5]} & 0.95(0.09) & \multicolumn{1}{c}{\longdash[5]} & \multicolumn{1}{c}{\longdash[5]}\\
\bottomrule
\end{tabular}}
\end{table}

In terms of memory usage, the computational burden of GS drastically increases with the number of variables. For example, GS required $396$ Mb of RAM memory on average when $N=1,000$ and $P=100$, and $4,275$ Mb when $P=500.$ In other words, when $P=500$ the memory requirements are approximately $11$ times the ones needed for $P=100$. In the same scenarios, the memory requirements of CAVI increase by approximately  $1.2$ times, from $37$ Mb ($P=100$) to $45$ Mb ($P=500$). The ECM and SVI algorithms have memory requirements that are very similar to CAVI. POET and ADVI consistently required more memory than CAVI across all simulation scenarios, with the disparity in their memory requirements increasing in scenarios with $P$.

In terms of accuracy, CAVI and SVI estimate $\boldsymbol{\Sigma}$ with comparable accuracy to the estimates from GS, ECM, and POET in most scenarios. As expected, SVI performance deteriorates in scenarios with a small sample size and batch size. For example, in the scenario with $P=5,000$ and $N=100$, SVI-005 had an average RV of $0.62$ compared to $0.74$ of CAVI. However, the difference in performance between the two algorithms diminishes as the sample size increases. For example, when  $P=5,000$ and $N=1,000$, SVI-005 has an average RV of $0.94$ compared to $0.97$ of CAVI. On average, ADVI had comparable estimation accuracy to CAVI, in terms of RV, but had worse estimation accuracy in terms of the Frobenius and L1 norms (\tablename~S3, Supplementary~\S C).

\subsection{Multi-Study Simulations}
\label{sec:sim}
Analogously to \S \ref{sec:simulations-single-study}, we assess the performance of the proposed VI algorithms in the multi-study setting. We consider simulation scenarios with $S=\{5,10\}$, $P\in\{100,500,5000\}$ and $N_s=\{100,500,1000\}$, generating the data from Model~\eqref{multistudy} with $K=4$ common factors and $J_s=4, s=1,\ldots,S$ study-specific factors. For each scenario, factor matrices $\boldsymbol{\Phi}$ and $\boldsymbol{\Lambda}_s$ are randomly generated setting each values to $0$ with probability $2/3$ or drawing from an $\text{Uniform}(0,1)$ with probability $1/3,$ while the values of 
$\boldsymbol{\Psi}_s, s=1,\ldots,S$ are generated from an $\text{Uniform}(0.1,1).$
Using the generated $\boldsymbol{\Sigma}_s$, we sample $50 \times S$ by taking $N_s$ draws from $\mathcal{N}(\boldsymbol{0}, \boldsymbol{\Phi}\boldsymbol{\Phi}^\intercal + \boldsymbol{\Lambda}_s\boldsymbol{\Lambda}_s^\intercal + \boldsymbol{\Psi}_s)$ for each study. We estimate $\boldsymbol{\theta}$ with $K^*=5$, $J_s^*=5$ using the multi-study versions of CAVI, SVI, GS, ECM, and ADVI as described in \S \ref{sec:sims}. We excluded comparisons with POET in this section since it is not designed to analyze multi-study data. Bayesian methods (CAVI, SVI, GS, and ADVI) use priors \eqref{prior-lambda_s}-\eqref{prior-psi_sp} with hyperparameters $\nu=3=\nu_s=3$, $a_1=a_{s1}=2.1$, $a_2=a_{s2}=3.1$, $a^\psi=1$, and $b^\psi=0.3$ As we found in the single-study simulations, we were unable to run some scenarios with GS, ECM, and ADVI within the limited computational resources available. 

\begin{table}[ht!]
\tiny
\centering
\caption{\it \small Average computation time in seconds, mean(sd), for 50 multi-study simulation replicates. Note: we were unable to run GS and ADVI for some scenarios due to computation time and memory limitations, as the maximum allocation of 24 hours of run time and 128Gb of RAM for each iteration was insufficient. Additionally, we are unable to run ECM in scenarios with $P \geq N_s$. This is indicated with \longdash[5] in the table.
}\label{table-multistudy-time}
\begin{tabular}{lc|rrr|rrr}
\multicolumn{2}{c}{} &
\multicolumn{3}{c}{S = 5} & \multicolumn{3}{c}{S = 10}\\
Method & $\text{N}_s$ & P= 100 & P=500 & P=5000 & P=100 & P=500 & P=5000 \\
\cline{1-8}
GS & 100 & 1742.43(39.39) & 7907.73(254.45) & \multicolumn{1}{c|}{\longdash[5]} & 3288.87(73.37) & 14001.49(857.06) & \multicolumn{1}{c}{\longdash[5]}\\
CAVI & 100 & 14.03(0.3) & 68.38(0.77) & 704.38(10.27) & 26.49(0.32) & 138.1(1.93) & 1434.77(28.46)\\
SVI-05 & 100 & 11.15(0.14) & 54.53(0.56) & 608(5.73) & 22.12(0.3) & 109.7(0.79) & 1216.65(15.54)\\
SVI-02 & 100 & 6.06(0.11) & 30.58(0.33) & 343.61(4.62) & 11.95(0.19) & 59.3(0.67) & 686.83(7.86)\\
SVI-005 & 100 & 5.15(0.42) & 17.86(0.14) & 207.91(3.01) & 7.44(0.46) & 34.2(0.24) & 409.53(8.76)\\
ECM & 100 & \multicolumn{1}{c}{\longdash[5]} & \multicolumn{1}{c}{\longdash[5]} & \multicolumn{1}{c|}{\longdash[5]} & \multicolumn{1}{c}{\longdash[5]} & \multicolumn{1}{c}{\longdash[5]} & \multicolumn{1}{c}{\longdash[5]} \\
ADVI & 100 & 1596(1200.04) & \multicolumn{1}{c}{\longdash[5]} & \multicolumn{1}{c|}{\longdash[5]} & 2851.41(1667.1) & \multicolumn{1}{c}{\longdash[5]} & \multicolumn{1}{c}{\longdash[5]}\\
\specialrule{\cmidrulewidth}{0pt}{0.5pt}
GS & 500 & 2100.87(42.68) & 9534.65(446.68) & \multicolumn{1}{c|}{\longdash[5]} & 3962.02(84.66) & 16861.88(1835.69) & \multicolumn{1}{c}{\longdash[5]}\\
CAVI & 500 & 62.96(0.68) & 316.34(5.56) & 3303.78(66.45) & 126.77(2.1) & 649.44(27.64) & 7001.73(419.42)\\
SVI-05 & 500 & 49.62(3.61) & 232.96(2.9) & 2548.9(63.29) & 94.74(1.3) & 467.26(5.47) & 5055.43(460.32)\\
SVI-02 & 500 & 42.69(2.42) & 111.5(1.18) & 1231.84(13.37) & 64.11(2.54) & 224.64(2.61) & 2475(49.49)\\
SVI-005 & 500 & 39.65(2.1) & 82.77(2.44) & 581.16(9) & 58.95(2.08) & 129.35(1.88) & 1140.81(33.76)\\
ECM & 500 & 1219.83(572.05) & \multicolumn{1}{c}{\longdash[5]} & \multicolumn{1}{c|}{\longdash[5]} & 3063.55(742.01) & \multicolumn{1}{c}{\longdash[5]} & \multicolumn{1}{c}{\longdash[5]}\\
ADVI & 500 & 8642.38(7060.36) & \multicolumn{1}{c}{\longdash[5]} & \multicolumn{1}{c|}{\longdash[5]} & 12078.54(2340.81) & \multicolumn{1}{c}{\longdash[5]} & \multicolumn{1}{c}{\longdash[5]}\\
\specialrule{\cmidrulewidth}{0pt}{0.5pt}
GS & 1000 & 2657.03(56.32) & 11638.49(624.67) & \multicolumn{1}{c|}{\longdash[5]} & 5048.3(99.63) & 21271.9(1256.6) & \multicolumn{1}{c}{\longdash[5]}\\
CAVI & 1000 & 128.28(1.87) & 665.95(11.99) & 6669.16(712.57) & 253.37(5.38) & 1332.73(19.15) & 11674.25(1314.02)\\
SVI-05 & 1000 & 144.7(7.69) & 473.39(7.28) & 5072.69(344.52) & 218.7(12.65) & 965.94(20.17) & 8785.58(808.99)\\
SVI-02 & 1000 & 123.46(5.25) & 259.83(15.78) & 2349.63(45.61) & 181.71(7.47) & 434(7.62) & 4417.48(496.26)\\
SVI-005 & 1000 & 106.77(4.21) & 213.44(5.5) & 1044.76(21.22) & 143.88(4.02) & 329.09(9.48) & 2115.91(92.74)\\
ECM & 1000 & 1023.11(443.41) & 61570.47(19774.57) & \multicolumn{1}{c|}{\longdash[5]} & 4088.28(903.69) & \multicolumn{1}{c}{\longdash[5]}& \multicolumn{1}{c}{\longdash[5]}\\
ADVI & 1000 & 14602.13(13389.31) & \multicolumn{1}{c}{\longdash[5]} & \multicolumn{1}{c|}{\longdash[5]} & 25977.91(17421.21) & \multicolumn{1}{c}{\longdash[5]} & \multicolumn{1}{c}{\longdash[5]}\\
\bottomrule
\end{tabular}
\end{table}

\tablename~\ref{table-multistudy-time} reports the average computational times (refer to \figurename~S5, Supplementary \S C for comprehensive results). CAVI and SVI required much less time on average than GS in every simulation scenario. For example, in the scenario with $S=10$, $P=500$, and $N_s=1,000$, CAVI required an average of approximately $22$ minutes to converge while GS took about $6$ hours, making  CAVI about 16 times faster than GS. In the same scenario, SVI algorithms were even faster, taking approximately $16,$ $7,$ and $6$ minutes on average for SVI-05, SVI-02 and SVI-005 respectively, making SVI algorithms between 22 to 65 times faster than GS. ECM and ADVI require much more computational time than CAVI in all simulation scenarios. 

\begin{table}[ht!]
\tiny
\centering
\caption{\it \small Average peak RAM usage in Mb, mean(sd), for 50 multi-study simulation replicates. Note: we were unable to run GS and ADVI for some scenarios due to computation time and memory limitations, as the maximum allocation of 24 hours of run time and 128Gb of RAM for each iteration was insufficient. Additionally, we are unable to run ECM in scenarios with $P \geq N_s$. This is indicated with \longdash[5] in the table.}\label{table-multistudy-mem}
\begin{tabular}{lc|rrr|rrr}
\multicolumn{2}{c}{} &
\multicolumn{3}{c}{S = 5} & \multicolumn{3}{c}{S = 10}\\
Method & $\text{N}_s$ & P= 100 & P=500 & P=5000 & P=100 & P=500 & P=5000 \\
\cline{1-8}
GS & 100 & 126.24(2.33) & 528.48(22.94) & \multicolumn{1}{c|}{\longdash[5]} & 184.89(1.07) & 856.76(17.8) & \multicolumn{1}{c}{\longdash[5]}\\
CAVI & 100 & 35.81(0.08) & 57.93(1.88) & 3130.33(315.86) & 40.49(6.68) & 69.49(2.16) & 5708.93(710.89)\\
SVI-05 & 100 & 35.51(0.65) & 77.93(7.92) & 3123.61(313.18) & 47.14(3.05) & 82.62(1.27) & 5720.95(701.25)\\
SVI-02 & 100 & 35.51(0.65) & 58.15(0.35) & 2556.88(231.39) & 47.41(2.44) & 82.62(1.27) & 5707.01(703.05)\\
SVI-005 & 100 & 35.51(0.65) & 57.93(1.23) & 2556.78(231.38) & 47.14(3.05) & 82.6(1.39) & 5710.53(697.44)\\
ECM & 100 &  \multicolumn{1}{c}{\longdash[5]} & \multicolumn{1}{c}{\longdash[5]} & \multicolumn{1}{c|}{\longdash[5]} & \multicolumn{1}{c}{\longdash[5]} & \multicolumn{1}{c}{\longdash[5]} & \multicolumn{1}{c}{\longdash[5]} \\
ADVI & 100 & 359.08(23.39) & \multicolumn{1}{c}{\longdash[5]} & \multicolumn{1}{c|}{\longdash[5]} & 610(26.92) & \multicolumn{1}{c}{\longdash[5]} & \multicolumn{1}{c}{\longdash[5]}\\
\specialrule{\cmidrulewidth}{0pt}{0.5pt}
GS & 500 & 211.76(15.62) & 735.71(51.2) & \multicolumn{1}{c|}{\longdash[5]} & 308.18(40.62) & 1170.03(53.33) & \multicolumn{1}{c}{\longdash[5]}\\
CAVI & 500 & 47.66(1.73) & 116.52(4.45) & 3235.65(323.23) & 60.92(1.99) & 155.69(0.1) & 4736.93(650.13)\\
SVI-05 & 500 & 47.83(0.2) & 117.13(1.17) & 2532.19(191.86) & 61.01(2.01) & 155.59(0.08) & 4652.65(633.34)\\
SVI-02 & 500 & 47.6(1.43) & 115.31(7.34) & 2532.29(191.88) & 61.01(2.01) & 155.59(0.08) & 4621.51(501.91)\\
SVI-005 & 500 & 47.56(1.71) & 109.23(13.42) & 2533.01(191.98) & 61.01(2.01) & 155.59(0.08) & 4613.68(499.66)\\
ECM & 500 & 64.31(0.27) & \multicolumn{1}{c}{\longdash[5]} & \multicolumn{1}{c|}{\longdash[5]} & 80.61(0.42) & \multicolumn{1}{c}{\longdash[5]} & \multicolumn{1}{c}{\longdash[5]}\\
ADVI & 500 & 362.45(14.25) & \multicolumn{1}{c}{\longdash[5]} & \multicolumn{1}{c|}{\longdash[5]} & 518.52(28.39) & \multicolumn{1}{c}{\longdash[5]} & \multicolumn{1}{c}{\longdash[5]}\\
\specialrule{\cmidrulewidth}{0pt}{0.5pt}
GS & 1000 & 213.15(13.77) & 1001.85(91.55) & \multicolumn{1}{c|}{\longdash[5]} & 294.72(14.72) & 1557.5(108.83) & \multicolumn{1}{c}{\longdash[5]}\\
CAVI & 1000 & 46.03(0.18) & 168.55(10.13) & 2663.43(258.27) & 87.94(12.32) & 285.26(1.73) & 5080.33(560.21)\\
SVI-05 & 1000 & 60.42(4.8) & 171.02(2.67) & 2674.08(225.38) & 76.06(2.4) & 285.35(1.74) & 5276.57(641.45)\\
SVI-02 & 1000 & 61.06(3.75) & 171.02(2.67) & 2697.99(168.79) & 75.67(3.6) & 274.18(14.38) & 6068.98(888.85)\\
SVI-005 & 1000 & 61.71(2.08) & 169.57(8.52) & 2697.99(168.79) & 75.29(4.46) & 285.35(1.74) & 5901.31(645.49)\\
ECM & 1000 & 62.42(0.27) & 779.87(59.36) & \multicolumn{1}{c|}{\longdash[5]}& 99.76(0.43) & \multicolumn{1}{c}{\longdash[5]} & \multicolumn{1}{c}{\longdash[5]}\\
ADVI & 1000 & 375.78(28.63) & \multicolumn{1}{c}{\longdash[5]} & \multicolumn{1}{c|}{\longdash[5]} & 577.74(82.71) & \multicolumn{1}{c}{\longdash[5]} & \multicolumn{1}{c}{\longdash[5]}\\
\bottomrule
\end{tabular}
\end{table}

\tablename~\ref{table-multistudy-mem} reports the average memory usage (refer to \figurename~S6, Supplementary \S C, for the comprehensive results). The proposed algorithms require much less memory than GS in all the scenarios. For example, GS required 3.5 times as much memory as CAVI in the setting with $S=5$, $N_s=100$, $P=100$ (126 Mb compared to 36 Mb), while  in the setting with $S=10$, $N_s=1000$, $P=500$ GS required about 5.5 times more memory than CAVI (1557 Mb compared to 285 Mb). In all scenarios, CAVI and SVI have comparable memory requirements. ECM had memory requirements to CAVI for scenarios with $P=100$, but appears to scale poorly with increasing $P$, having memory requirements comparable to GS in simulation scenarios with $P=500$. ADVI consistently had the largest memory requirements in every scenario we could complete. 

\begin{table}[ht!]
\tiny
\centering
\caption{\it \small Average estimation accuracy, reported as RV between estimated $\boldsymbol{\Sigma}_s = \boldsymbol{\Phi}\boldsymbol{\Phi}^\intercal + \boldsymbol{\Lambda}_s{\boldsymbol{\Lambda}_s}^\intercal + \boldsymbol{\Psi}_s$ and the simulation truth, mean(sd),
 for 50 multi-study simulation replicates. Note: we were unable to run GS and ADVI for some scenarios due to computation time and memory limitations, as the maximum allocation of 24 hours of run time and 128Gb of RAM for each iteration was insufficient. Additionally, we are unable to run ECM in scenarios with $P \geq N_s$. This is indicated with \longdash[5] in the table.}\label{table-multistudy-rv}
\begin{tabular}{lc|rrr|rrr}
\multicolumn{2}{c}{} &
\multicolumn{3}{c}{S = 5} & \multicolumn{3}{c}{S = 10}\\
Method & $\text{N}_s$ & P= 100 & P=500 & P=5000 & P=100 & P=500 & P=5000 \\
\cline{1-8}
GS & 100 & 0.923(0.025) & 0.894(0.047) & \multicolumn{1}{c|}{\longdash[5]} & 0.927(0.025) & 0.904(0.044) & \multicolumn{1}{c}{\longdash[5]}\\
CAVI & 100 & 0.851(0.032) & 0.842(0.026) & 0.85(0.021) & 0.862(0.031) & 0.855(0.027) & 0.864(0.02)\\
SVI-05 & 100 & 0.835(0.032) & 0.842(0.026) & 0.857(0.022) & 0.843(0.037) & 0.856(0.028) & 0.871(0.021)\\
SVI-02 & 100 & 0.827(0.033) & 0.827(0.027) & 0.842(0.025) & 0.83(0.038) & 0.837(0.03) & 0.849(0.024)\\
SVI-005 & 100 & 0.713(0.049) & 0.676(0.042) & 0.677(0.043) & 0.704(0.047) & 0.679(0.043) & 0.679(0.04)\\
ECM & 100 & \multicolumn{1}{c}{\longdash[5]} & \multicolumn{1}{c}{\longdash[5]} & \multicolumn{1}{c|}{\longdash[5]} & \multicolumn{1}{c}{\longdash[5]} & \multicolumn{1}{c}{\longdash[5]} & \multicolumn{1}{c}{\longdash[5]} \\
ADVI & 100 & 0.798(0.071) & \multicolumn{1}{c}{\longdash[5]} & \multicolumn{1}{c|}{\longdash[5]} & 0.761(0.061) & \multicolumn{1}{c}{\longdash[5]} & \multicolumn{1}{c}{\longdash[5]}\\
\specialrule{\cmidrulewidth}{0pt}{0.5pt}
GS & 500 & 0.986(0.0036) & 0.984(0.0058) & \multicolumn{1}{c|}{\longdash[5]} & 0.982(0.0059) & 0.987(0.0046) & \multicolumn{1}{c}{\longdash[5]}\\
CAVI & 500 & 0.932(0.022) & 0.933(0.03) & 0.956(0.018) & 0.921(0.02) & 0.958(0.013) & 0.966(0.013)\\
SVI-05 & 500 & 0.92(0.025) & 0.927(0.031) & 0.954(0.018) & 0.892(0.032) & 0.952(0.016) & 0.965(0.013)\\
SVI-02 & 500 & 0.925(0.024) & 0.926(0.03) & 0.952(0.018) & 0.896(0.03) & 0.951(0.015) & 0.963(0.013)\\
SVI-005 & 500 & 0.932(0.019) & 0.92(0.028) & 0.935(0.018) & 0.907(0.023) & 0.938(0.015) & 0.943(0.013)\\
ECM & 500 & 0.907(0.012) & \multicolumn{1}{c}{\longdash[5]} & \multicolumn{1}{c|}{\longdash[5]} & 0.907(0.015) & \multicolumn{1}{c}{\longdash[5]} & \multicolumn{1}{c}{\longdash[5]}\\
ADVI & 500 & 0.91(0.045) & \multicolumn{1}{c}{\longdash[5]} & \multicolumn{1}{c|}{\longdash[5]} & 0.906(0.033) & \multicolumn{1}{c}{\longdash[5]} & \multicolumn{1}{c}{\longdash[5]}\\
\specialrule{\cmidrulewidth}{0pt}{0.5pt}
GS & 1000 & 0.992(0.0021) & 0.993(0.0021) & \multicolumn{1}{c|}{\longdash[5]} & 0.993(0.0022) & 0.993(0.0019) & \multicolumn{1}{c}{\longdash[5]}\\
CAVI & 1000 & 0.94(0.029) & 0.949(0.011) & 0.966(0.018) & 0.948(0.0071) & 0.961(0.015) & 0.977(0.011)\\
SVI-05 & 1000 & 0.934(0.028) & 0.944(0.012) & 0.964(0.019) & 0.936(0.012) & 0.955(0.018) & 0.974(0.012)\\
SVI-02 & 1000 & 0.938(0.027) & 0.945(0.012) & 0.963(0.019) & 0.941(0.011) & 0.954(0.018) & 0.974(0.013)\\
SVI-005 & 1000 & 0.946(0.022) & 0.946(0.011) & 0.957(0.018) & 0.948(0.0085) & 0.954(0.016) & 0.966(0.012)\\
ECM & 1000 & 0.929(0.011) & 0.915(0.039) & \multicolumn{1}{c|}{\longdash[5]} & 0.904(0.028) & \multicolumn{1}{c}{\longdash[5]} & \multicolumn{1}{c}{\longdash[5]}\\
ADVI & 1000 & 0.92(0.039) & \multicolumn{1}{c}{\longdash[5]} & \multicolumn{1}{c|}{\longdash[5]} & 0.909(0.031) & \multicolumn{1}{c}{\longdash[5]} & \multicolumn{1}{c}{\longdash[5]}\\
\bottomrule
\end{tabular}
\end{table}

We  report the averages and standard deviations of the RV coefficients between estimated $\boldsymbol{\Sigma}_s$ and the simulation truth (\tablename~\ref{table-multistudy-rv}). We refer to \figurename~S7 for comprehensive results for $\boldsymbol{\Sigma}_s$ and \figurename s~S8--S9 for the RV coefficients of $\boldsymbol{\Phi}$ and $\boldsymbol{\Lambda}_s$, respectively. As before, CAVI estimates $\boldsymbol{\Sigma}_s$ with comparable accuracy to GS, ADVI, and ECM in most scenarios. In scenarios where the number of observations is small, CAVI is less accurate than GS. For example, when $P=100$ and $N_s=100$, CAVI has an average RV of $0.85$ compared to $0.92$ of the GS. When the sample size increases, the average RV of CAVI is closer to GS. For example, when 
$P=500$ and $N_s=1,000$ CAVI as an average RV of $0.96$ compared to $0.99$ of the GS.
When both the sample and batch size are small, the SVI algorithms can present poor performance. 
When $N_s$ is low, the performance of SVI with a small batch size can be inferior to CAVI, especially in high dimensional settings. For example, with $P=5,000$ and $N_s=100$, the average RV of SVI-005 was $0.68$ while CAVI had an average RV of $0.86$. The difference between SVI and CAVI is much smaller with larger batch sizes---the RVs were comparable to CAVI for SVI-02 and SVI-05 across all 18 simulation scenarios. When $N_s>100$, SVI-005 presented an accuracy comparable with CAVI for all the considered values of $P$ and $S$. 


\section{ Case Study: Ovarian Cancer Gene Expression Data}\label{sec:Application}

We apply the proposed algorithms to four high-dimensional datasets containing microarray gene expression data available in the \texttt{R} package \texttt{curatedOvarianData 1.36}
 \citep{curatedovarian}. The curated ovarian cancer studies contain $N_s= \left\{578, 285, 195, 140\right\}$ patient observations. We selected genes with observed variances in the highest 15\% in at least one study, resulting in a final set of $P=3643$ genes. Each dataset was centered and scaled before analysis.

We start by evaluating the ability of MSFA to predict out-of-sample $\mathbf x_{si}$ compared to alternative FA models in a 10-fold cross-validation. In particular, we consider three models:
1) A MSFA as described in \S \ref{sec:MSFA};
2) A FA fitted on a dataset stacking all the $S=4$ studies  (Stacked FA);
3)  Independent FA models for each study (Independent FA).  
For the MSFA model 1) we set  $K^*=10$ and $J^*_s=10,$ and use CAVI and SVI (with $b_s=0.5$ for $s=1,\ldots, 4$). For FA in 2) and 3), we set
$J^*=10$ using CAVI and SVI ($b=0.5$). 
It is important to emphasize that fitting these models with GS, ECM, or ADVI would have required more than 128Gb of RAM to be allocated (the limit set in \S \ref{sec:sims}); therefore, this was considered not feasible.

Predictions are computed as follows:
$$
\textit{MSFA:}\;\widehat{\mathbf{x}}_{si}=\widehat{\boldsymbol{\Phi}}\widehat{\mathbf{f}}_{si} + \widehat{\boldsymbol{\Lambda}}_s\widehat{\mathbf{l}}_{si} \qquad \qquad   \textit{Stacked FA:}\;\widehat{\mathbf{x}}_{si}=\widehat{\boldsymbol{\Phi}}\widehat{\mathbf{f}}_{si}  \qquad \qquad  \textit{Independent FA:}\;\widehat{\mathbf{x}}_{si}=\widehat{\boldsymbol{\Lambda}}_s\widehat{\mathbf{l}}_{si}.
$$
Factor scores for out-of-sample observations are derived by adapting Bartlett's method \citep{bartlett_statistical_1937} for MSFA:
\begin{align}
    \widehat{\mathbf{f}}_{si} &= \left( (\widehat{\boldsymbol{\Phi}}, \widehat{\boldsymbol{\Lambda}}_s)^\intercal \widehat{\boldsymbol{\Psi}}_s^{-1}(\widehat{\boldsymbol{\Phi}}, \widehat{\boldsymbol{\Lambda}}_s)  \right)^{-1} \widehat{\boldsymbol{\Phi}}^\intercal \widehat{\boldsymbol{\Psi}}_s^{-1} \mathbf{x}_{si}, \label{f_hat_eq} \\
    \widehat{\mathbf{l}}_{si} &= \left( (\widehat{\boldsymbol{\Phi}}, \widehat{\boldsymbol{\Lambda}}_s)^\intercal \widehat{\boldsymbol{\Psi}}_s^{-1}(\widehat{\boldsymbol{\Phi}}, \widehat{\boldsymbol{\Lambda}}_s)  \right)^{-1} \widehat{\boldsymbol{\Lambda}}_s^\intercal \widehat{\boldsymbol{\Psi}}_s^{-1} \mathbf{x}_{si}, \label{l_hat_eq}
\end{align}
where $(\widehat{\boldsymbol{\Phi}}, \widehat{\boldsymbol{\Lambda}}_s)$ is obtained by stacking the common and study-specific factor loading matrices, and $\widehat{\boldsymbol{\Phi}}$ and $\widehat{\boldsymbol{\Lambda}}_s$, $\widehat{\Psi}_s$ denote the posterior mode of the respective VI approximation. The Mean Squared Error (MSE) is computed as the average of the prediction errors across all observations in every study:
        $\text{MSE} = \left\{\sum_{s=1}^S N_s\right\}^{-1}\sum_{s=1}^S \sum_{i=1}^{N_s} \sum_{p=1}^P ( x_{sip} - \widehat{x}_{sip})^2.$

\begin{table}[ht!]

\caption{\it \small Computation time (in minutes) and MSE, reported as mean(sd), across the 10 cross-validation folds. Relative MSE is the average MSE divided by the average MSE of MSFA fitted using CAVI (the best-performing algorithm). For independent FA the time include fitting $S=4$ serially.}\label{application-table}
\centering
\begin{tabular}[t]{lllr}
\toprule
Method & Time (minutes) & MSE & Relative MSE\\
\midrule
MSFA (CAVI) & 11.06(0.24) & 2296(53) & 1\\
MSFA (SVI) & 8.83(0.24) & 2315(52) & 1.01\\
Independent FA (CAVI) & 14.41(0.16) & 2435(55) & 1.06\\
Independent FA (SVI) & 6.77(0.036) & 2447(50) & 1.06\\
Stacked FA (CAVI) & 13.58(0.19) & 2531(52) & 1.10\\
Stacked FA (SVI) & 6.07(0.093) & 2527(53) & 1.10\\
\bottomrule
\end{tabular}
\end{table}

Results of the 10-fold cross-validation are reported in \tablename~\ref{application-table}.
Despite the high dimensionality of the dataset, the proposed VB algorithms were able to fit the MSFA model with an average computation time of approximately 11 minutes for CAVI and 9 minutes for SVI. Also, the two considered FA approaches were computed in a small amount of time, approximately 14 minutes for CAVI and 7 minutes for SVI.
In terms of MSE, CAVI and SVI share a similar performance, decreasing the MSE of
 10\% compared to Stacked FA and 6\% compared to Independent FA.

\begin{figure}[ht!]
    \centering
    \includegraphics[width=\textwidth]{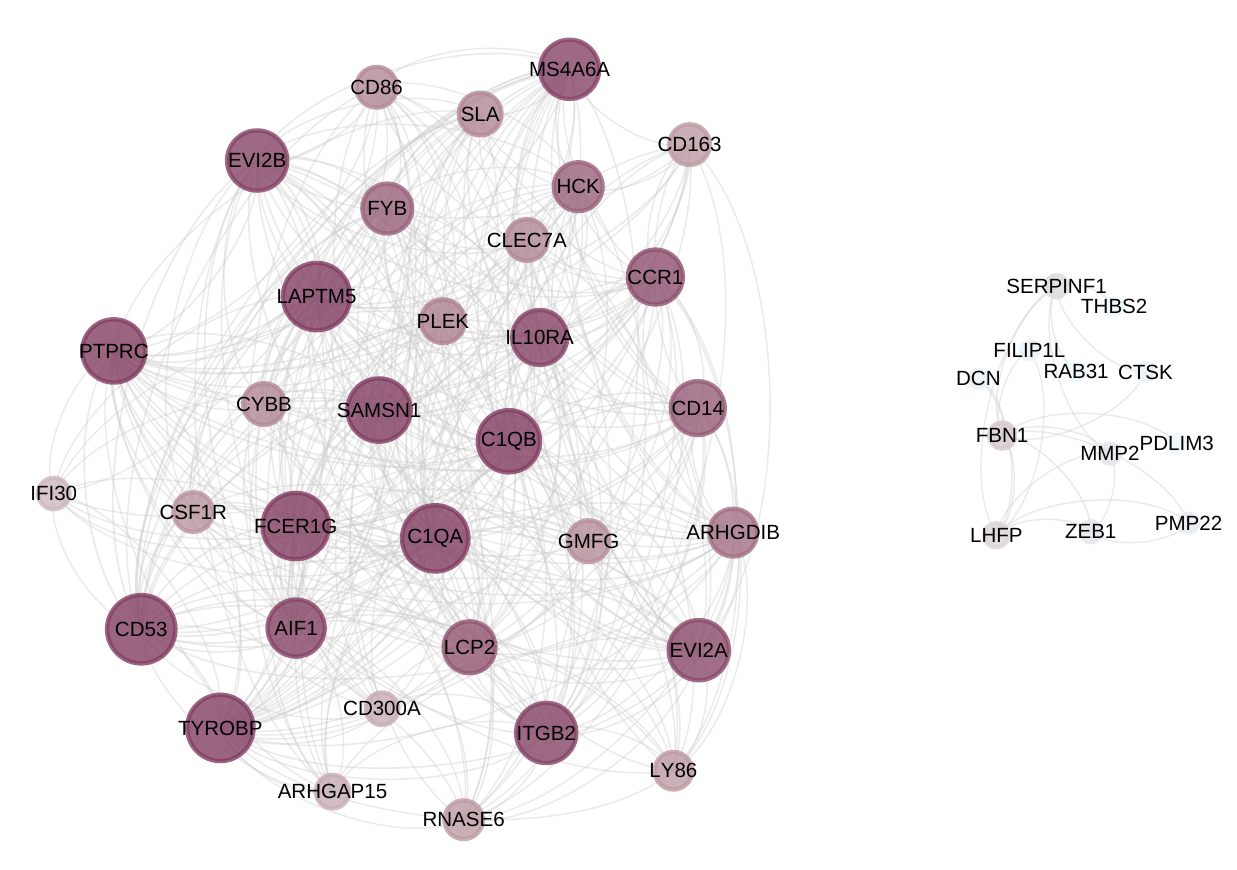}
    \caption{\it \small Common gene co-expression network estimated via $\widehat{\boldsymbol{\Phi}}\widehat{\boldsymbol{\Phi}}^{\intercal}$  across the four ovarian cancer studies. The network is obtained using \texttt{Gephi} via the Fruchterman-Reingold algorithm ~\citep{gephi_paper}. We include edges  if $\widehat{\boldsymbol{\Phi}}\widehat{\boldsymbol{\Phi}}^{\intercal} \geq |0.5|$, and exclude from the figure nodes with no edges. The figure includes 92 nodes and 851 edges. }
    \label{sigma_phi_network}
\end{figure}

We then estimated the common covariance $\boldsymbol{\widehat{\Sigma}}_\Phi=\widehat{\boldsymbol{\Phi}}\widehat{\boldsymbol{\Phi}}^{\intercal}$ with MSFA via CAVI and represented it via a gene co-expression network (Figure~\ref{sigma_phi_network}). A gene co-expression network is an undirected graph where the nodes correspond to genes, and the edges correspond to the degree of co-expression between genes. The number of connections between genes is visualized in the plot by the size of each node, i.e., the bigger, the more connected. 
We identified two different and important clusters. 
 The first cluster contains genes associated with the immune system and cell signaling, such as \emph{CD53}~\citep{dunlock_tetraspanin_2020},  \emph{LAPTM5}~\citep{glowacka_laptm5_2012}, \emph{PTPRC}~\citep{hermiston_cd45_2003}, \emph{TYROBP}~\citep{lanier_immunoreceptor_1998}, \emph{C1QA/C1QB}~\citep{liang_c1qa_2022}. Also, in the first cluster, some genes play a crucial role in cancer, such as \emph{SAMSN1}~\citep{yan_samsn1_2013} and \emph{FCER1G}~\citep{yang_fc_2023}.  The second network includes genes such as \emph{FBN1}  known to promote metastasis in ovarian cancer~\citep{wang_fibrillin-1_2015}, and \emph{SERPINF1} crucial to the prognosis of cancer~\citep{zhang_pan-cancer_2022}.

 The gene network in \figurename~\ref{sigma_phi_network} reveals the ability of the MSFA to discover common signal pathways across different studies, critically helpful in ovarian cancer and generalizable to new context areas. These results further show the  ability of the proposed VB algorithms to provide reliable results in a short computation time.

\section{Discussion}\label{sec:conc}

We have proposed VB algorithms that provide fast Bayesian estimation for FA in both single and multi-study settings. These algorithms provided significant advantages over MCMC-based implementations, requiring substantially less memory and time while achieving comparable performance in characterizing data covariance matrices. Moreover, in the ovarian cancer application case, we showed how these algorithms could help reveal biological pathways in the high-dimensional setting, using computational resources typically available on a laptop rather than a high-performance computing server. 

In Supplementary~\S D, we present additional simulations to evaluate the predictive performance of our VB algorithms compared to GS. We consider scenarios with $S=1,$ sample size $N=100$ and $P=100,500$. These results suggest that CAVI and GS have similar performance in terms of out-of-sample prediction errors and coverage and length of prediction intervals. SVI performed worse than CAVI and GS, achieving inappropriate $95\%$ prediction interval coverage (ranging from $0.92$ to $0.99$) and higher MSE. For example, with $P=500$, SVI algorithms have an MSE ranging from 0.6 to 0.9, compared to 0.3 of GS and CAVI.

While VB algorithms offer numerous advantages, they have some limitations. First, convergence can be sensitive to parameter initialization. This is common to many iterative algorithms, including the EM algorithm used in frequentist MSFA \citep{de_vito_multistudy_2019}. Thus, we have provided effective informative initializations (Supplementary \S B), helpful especially in scenarios with $P>N.$ Second, the performance of SVI algorithms can vary substantially depending on the chosen batch size parameter. Tuning the batch size parameter remains an active area of research \citep{tan_stochastic_2017}. Third, compared to MCMC and GS, VB algorithms tend to underestimate uncertainty. However, when the main focus is on point estimates and exploratory analysis---typically the main goals of FA---the underestimation of uncertainty is less of a concern than in other settings.

A known issue with FA is the non-identifiability of the factor loading matrix, i.e., the factor loadings can be rotated and produce the same covariance structure.  This orthogonal indeterminacy is not a concern if inference only targets an identifiable function of the loading matrix, such as the covariance matrix. Alternatively, the loading matrices in Models~\ref{fa_single_study} and~\ref{multistudy} can be further constrained to be identifiable ~\citep[e.g,][]{lopes_bayesian_2004}. These or alternative constraints can be incorporated in the proposed algorithms by adapting parameter-expansions techniques~\citep[see for example][]{rockova_fast_2016}.  We also refer to \citet{de_vito_multistudy_2019} for two methods to recover factor loading matrices from MSFA, which can be applied to estimates provided by our algorithms.

We have even observed that the impact of the shrinkage priors \eqref{eq:prior_shrinkage_single_study}, and \eqref{prior-lambda_s}--\eqref{eq:prior-phi} on the GS posterior distribution differs from the effect on the approximate VB posterior (\figurename s~S10-S11, Supplementary \S C). When employing the same hyperparameter values, the VB algorithm induces a stronger shrinkage effect compared to GS. Similar findings were reported by \cite{zhao_note_2009} while studying FA under a different shrinkage prior. Further research is required to understand if this behavior is shared when using VB with other popular shrinkage priors, such as the Besov prior~\citep{lassas_discretization-invariant_2009}, the cumulative shrinkage process prior~\citep{legramanti_bayesian_2020}, the triple-gamma prior~\citep{Cadonna2020TtGU}, and non-local shrinkage priors~\citep{avalos-pacheco_heterogeneous_2022}. 

In conclusion, the proposed VB algorithms
enable scaling inference for FA and MSFA to high-dimensional data, enabling new research opportunities in previously inaccessible settings without extensive computational resources.

\section{Acknowledgements}
This work was supported by the US National Institutes of Health, under grants NIGMS/NIH COBRE CBHD P20GM109035, and 5R01CA262710-03.

\section{Supplementary Materials}

\begin{description}

\item[Supplementary A - F:] \S A: Descriptions for CAVI and SVI algorithms for Bayesian FA and Bayesian MSFA as described in \S \ref{sec:BFA}-\ref{sec:MSFA}; \S B: Description of initialization strategies for VB Algorithms; \S C: Additional figures and tables for simulation study; \S D: Prediction accuracy and prediction uncertainty quantification simulations; \S E: Observed factor loading shrinkage per column index; \S F: Derivations supporting coordinate updates in Algorithms 1-4. (pdf)

\item[Supporting code:] VIMSFA: the current version of our R package available at \href{https://github.com/blhansen/VI-MSFA} {github.com/blhansen/VI-MSFA}; example.R: a R script showing how to install and use our algorithms using R version $>4.3.0$; simulations.tar.gz: a zipped directory containing the scripts used to perform the simulation experiments of \S 4.

\end{description}

\bibliography{refs_trunc}

\end{document}


\linespread{2}

\title{\textbf{Supplementary Materials for \\ ``Fast Variational Inference for Bayesian Factor Analysis in Single and Multi-Study Settings.''}}

\date{}

\maketitle

\appendix

\newpage

\section{Variational Inference Algorithms}
In this section, we present the four algorithms outlined in the main manuscript. Convergence of these algorithms can be assessed in several ways, for example monitoring the difference of the ELBO across iterations and stopping the algorithm when this difference is smaller than a pre-specified threshold. In our implementation, we monitor the euclidean distance between the variational parameters in the last and current iteration, divided by the number of variational parameters. Parameter initialization is also important to make the algorithms converge faster, and  potentially avoid  covergence to local optimums; we discuss this in \S B.

\subsection{CAVI for Bayesian Factor Analysis}

\begin{algorithm}[H]
    \setstretch{1.2}
    \tiny
    \begin{flushleft}
    \caption{CAVI for Bayesian FA}\label{alg:CAVIsingle}
    $\bf{Input}$: Data $\mathbf{X}$
    
    $\bf{Initialize}$: Set initial variational parameters $\boldsymbol{\varphi}$:
    \begin{enumerate}
        \item  Fix iteration-independent variational parameters to their optimal value:
            \begin{itemize}
                \item For $l=1,\ldots,J^*,$ $\alpha_{l}^{\delta}=a_{l} + \frac{P}{2}(J^*-l+1)$, \\
                \item For $p=1,\ldots,P$, and $j=1,\ldots,J^*,$ $\alpha_{pj}^{\omega}=\frac{\nu+1}{2}$, \\
                \item For $p=1,\ldots,P,$ $\alpha_{p}^{\psi}= a_{\psi} + \frac{N}{2}$. \\
            \end{itemize}
        \item Initialize the remaining values in $\boldsymbol{\varphi}$ according to Section B.1.
    \end{enumerate}
    \textbf{For $t=1,\ldots$ }
    \begin{enumerate}
            \item For $p=1,\ldots,P$,
            \begin{align*}
                \boldsymbol{\Sigma}^{\Lambda}_{p} &= \left(\boldsymbol{D}^{-1}_{p} +  \frac{\alpha^{\psi}_{p}}{\beta^{\psi}_{p}}\left(\sum_{i=1}^{N} \boldsymbol{\mu}^{l}_{i}\left[\boldsymbol{\mu}^l_{i}\right]^{\intercal} + \boldsymbol{\Sigma}^l_{i}\right)\right)^{-1}, \\
                \boldsymbol{\mu}^{\Lambda}_{p} &= \boldsymbol{\Sigma}^{\Lambda}_{p}\frac{\alpha^{\psi}_{p}}{\beta^{\psi}_{p}}\sum_{i=1}^N x_{ip}\boldsymbol{\mu}_i^l,
            \end{align*}
            where: $\boldsymbol{D}^{-1}_{p}=
            \text{diag}\left(\frac{\alpha^{\omega}_{p1}}{\beta^{\omega}_{p1}}\left( \frac{\alpha^{\delta}_{1}}{\beta^{\delta}_{1}}\right),\ldots,\frac{\alpha^{\omega}_{pJ^*}}{\beta^{\omega}_{pJ^*}}\left( \prod_{l=1}^{J^*} \frac{\alpha^{\delta}_{l}}{\beta^{\delta}_{l}}\right)\right)$. \\
            \item For $p=1,\ldots,P$, 
            \begin{align*}
                \beta^{\psi}_{p}= b^\psi + \frac{1}{2}\sum_{i=1}^N (x_{ip}-\left[\boldsymbol{\mu}_p^\Phi\right]^\intercal\boldsymbol{\mu}_i^f)^2 + \text{Tr}\left\{ \left( \sum_{i=1}^N \boldsymbol{\Sigma}_i^f + \boldsymbol{\mu}_i^f \left[\boldsymbol{\mu}_i^f\right]^\intercal\right)  \boldsymbol{\Sigma}_p^\Phi  \right\} + \left[\boldsymbol{\mu}_p^\Phi\right]^\intercal \left( \sum_{i=1}^N \boldsymbol{\Sigma}_i^f  \right)\boldsymbol{\mu}_p^\Phi.
            \end{align*}
            \item For $i=1,\ldots,N$, 
            \begin{align*}
            \boldsymbol{\Sigma}^{l}_{i}&=\left(\boldsymbol{I}_{J^*} + \sum_{p=1}^P \frac{\alpha^{\psi}_{p}}{\beta^{\psi}_{p}} \left(\boldsymbol{\mu}^{\Lambda}_{p}\left[\boldsymbol{\mu}^{\Lambda}_{p}\right]^{\intercal}+\boldsymbol{\Sigma}^{\Lambda}_{p}\right)\right)^{-1}, \\
            \boldsymbol{\mu}^l_{i} &= \boldsymbol{\Sigma}^l_{i}\left[\mathbf{M}^{\Lambda}\right]^{\intercal}\mathbf{D}^{-1}\mathbf{x}_{i}.
            \end{align*}
            where $\mathbf{M}^{\Lambda}=(\boldsymbol{\mu}_{1}^\Lambda,\ldots,\boldsymbol{\mu}_{P}^\Lambda)^\intercal$ and $\mathbf{D}^{-1}=\text{diag}\left(\frac{\alpha_{1}^\psi}{\beta_{1}^\psi},\ldots,\frac{\alpha_{P}^\psi}{\beta_{P}^\psi}\right)$.
            \item For $p=1,\ldots,P$, and $j=1,\ldots,J^*$,  
                $
                \beta^{\omega}_{pj}=\frac{1}{2} \left(\nu + \left( \prod_{l=1}^j \frac{\alpha^{\delta}_{l}}{\beta^{\delta}_{l}}\right)\left(\left[\mu^{\Lambda}_{pj}\right]^2+\Sigma^{\Lambda}_{pj}\right)\right).
                $
            \item For $l=1,\ldots,J^*$, 
            $\beta_{l}^{\delta}= 1 + \frac{1}{2}\sum_{j=l}^{J^*} \left(\prod_{1\leq r \leq j, r\neq l}\frac{\alpha_{r}^{\delta}}{\delta_{r}^{\xi}}\right)\sum_{p=1}^{P}\frac{ \alpha_{pj}^{\omega}}{ \beta_{pj}^{\omega}}\left(\left[\mu^{\Lambda}_{pj}\right]^2+\Sigma^{\Lambda}_{pj}\right).$
            
    \end{enumerate}
        \textbf{End for}
    \bf{Output:} variational approximation $q^*(\boldsymbol{\theta} ; \boldsymbol{\varphi}^*)$ 
    \end{flushleft}
\end{algorithm}

\subsection{SVI for Bayesian Factor Analysis}

\begin{table}[t!]
    \centering
    \caption{\it \small Reparameterization of $\boldsymbol{\varphi}$ of the mean-field approximation presented in Equation (6) to facilitate calculation of Equations (8)-(9) of the main text. In Algorithm S2, we use both parameterizations to ease of notation of the algorithm.}\label{table:single-study-reparameterization}
    \begin{tabular}{|c|c|}
    \hline
     Natural Parameterization & Original Parameterization \vspace{-0.5cm} \\
     \parbox[c]{5cm}{
         \begin{align*}
            \boldsymbol{\eta}_{p1}^\Lambda &= \left(\boldsymbol{\Sigma}_{p}^\Lambda\right)^{-1} \\
            \boldsymbol{\eta}_{i2}^l &= \left(\boldsymbol{\Sigma}_{p}^\Lambda\right)^{-1}\boldsymbol{\mu}_{p}^\Lambda \\
            \eta_{p}^\psi &= -\beta_{p}^\psi
        \end{align*}}
    &
     \parbox[c]{5cm}{
        \begin{align*}
        \boldsymbol{\Sigma}_{p}^\Lambda &= \left(\boldsymbol{\eta}_{p1}^\Lambda\right)^{-1} \\
        \boldsymbol{\mu}_{p}^\Lambda &= \left(\boldsymbol{\eta}_{p1}^\Lambda\right)^{-1}\boldsymbol{\mu}_{p}^\Lambda \\
        \beta_{p}^\psi &= -\eta_{p}^\psi
        \end{align*}}\\ \hline
    \end{tabular}
    
\end{table}

\begin{algorithm}[H]
  \setstretch{1.2}
    \tiny
    \begin{flushleft}
    \caption{SVI for Bayesian FA}\label{alg:SVIsingle}
    $\bf{Input}$: Data $ \mathbf{X}$
    
    $\bf{Initialize}$: Set initial variational parameters, $\boldsymbol{\varphi}^{(0)},$ according to Section B.1.
    
    \textbf{For $t=1, \ldots$}
    \begin{enumerate}
             \item Calculate step size: $\rho_t=(t+\tau)^{-\kappa}$.
            \item  Create the set $\mathcal{I}(t)$ drawing $\widetilde {N}= \lfloor b\times N \rfloor$ elements without replacement from the set $\{1,\ldots,N\},$ and  define $\widetilde{\mathbf{X}}^t = \{\mathbf{x}_i \mbox{ for } i \in \mathcal I(t)\}.$
            \item For  $i \in \mathcal{I}(t)$ update the local variational parameters
            $\boldsymbol{\Sigma}^{l}_{i}$ and $\boldsymbol{\mu}^l_{i}$ as in Step 3 of algorithm \ref{alg:CAVIsingle}.
            \item Optimize remaining global variational parameters based on $\widetilde{\mathbf{X}}^t,$ with the current step size $\rho_t=(t+\tau)^{-\kappa},$ (refer to Table 1 of the main text for the original parameterization):
            \begin{enumerate}
                \item For $p=1,\ldots,P$:
                \begin{align*}
                    \widehat{\boldsymbol{\eta}}^{\Lambda}_{p1} &= \boldsymbol{D}^{-1}_{p} +  \frac{\alpha^{\psi}_{p}}{\beta^{\psi}_{p}}\left[\frac{N}{\widetilde N}\sum_{i \in \mathcal{I}(t)} \left(\boldsymbol{\mu}^{l}_{i}\left[\boldsymbol{\mu}^l_{i}\right]^{\intercal} + \boldsymbol{\Sigma}^l_{i}\right)\right], \\
                    \widehat{\boldsymbol{\eta}}^{\Lambda}_{p2} &= \frac{\alpha^{\psi}_{p}}{\beta^{\psi}_{p}} \frac{N}{\widetilde N}
                    \sum_{i \in \mathcal{I}(t)} \widetilde{x}^t_{ip}\boldsymbol{\mu}^{l}_{i}, \\
                    \boldsymbol{\eta}^{\Lambda}_{p1}(t) &= (1-\rho_t) \boldsymbol{\eta}^{\Lambda}_{p1}(t-1)  + \rho_t\widehat{\boldsymbol{\eta}}^{\Lambda}_{p1}, \\
                    \boldsymbol{\eta}^{\Lambda}_{p2}(t) &= (1-\rho_t) \boldsymbol{\eta}^{\Lambda}_{p2}(t-1)  + \rho_t \widehat{\boldsymbol{\eta}}^{\Lambda}_{p2}.
                \end{align*}  
                \item For $p=1,\ldots,P$: 
               \begin{align*}
                \widehat{\eta}^{\psi}_{p} &= -b^\psi -\frac{1}{2} \bigg\{\frac{N}{\widetilde N}\sum_{i \in \mathcal I(t)} (x_{ip}-\left[\boldsymbol{\mu}_p^\Phi\right]^\intercal\boldsymbol{\mu}_i^f)^2 \\ 
                & +\text{Tr}\left( \left( \frac{N}{\widetilde N}\sum_{i \in \mathcal{I}(t)} \boldsymbol{\Sigma}_i^f + \boldsymbol{\mu}_i^f \left[\boldsymbol{\mu}_i^f\right]^\intercal\right) \boldsymbol{\Sigma}_p^\Phi \right) +  \left[\boldsymbol{\mu}_p^\Phi\right]^\intercal \left( \frac{N}{\widetilde N}\sum_{i \in \mathcal{I}(t)} \boldsymbol{\Sigma}_i^f \right)\boldsymbol{\Sigma}_p^\Phi\bigg\},
            \end{align*}
                $$
                    \eta^{\psi}_{p}(t) = (1-\rho_t) \eta^{\psi}_{p}(t-1) + \rho_t \widehat{\eta}^{\psi}_{p}.
                $$
              \end{enumerate}
      \item Update parameters which do not depend on local variables according to algorithm~\ref{alg:CAVIsingle} steps 4-5.
    \end{enumerate}
    \textbf{End for.}
    \bf{Output:} variational approximation $q^*(\boldsymbol{\theta} ; \boldsymbol{\varphi}^*)$ 
    \end{flushleft}
\end{algorithm}

\FloatBarrier

\subsection{CAVI for Bayesian Multi-Study Factor Analysis}

\begin{breakablealgorithm}
 \setstretch{1.2}
    \tiny
    \begin{flushleft}
    \caption{CAVI for Bayesian MSFA}\label{alg:CAVI}
    $\bf{Input}$: = Data $\{ \mathbf{X}_s \}$
    
    $\bf{Initialize}$: Set initial variational parameters, $\boldsymbol{\varphi}_0$:
    \begin{enumerate}
        \item  Fix iteration-independent variational parameters:
            \begin{itemize}
                \item For $l=1,\ldots,K^*,$ $\alpha_{l}^{\delta}=a_l + \frac{P}{2}(K^*-l+1)$.
                \item For $s=1,\ldots,S$ and $l=1,\ldots,J_s^*,$ $\alpha_{sl}^{\delta}=a_{sj} + \frac{P}{2}(J_s-l+1)$. 
                \item For $p=1,\ldots,P$ and $k=1,\ldots,K^*,$ $\alpha_{pk}^{\omega}=\frac{\nu+1}{2}$. 
                \item For $s=1,\ldots,S$, $p=1,\ldots,P$, and $j=1,\ldots,J^*,$$\alpha_{spj}^{\omega}=\frac{\nu_s+1}{2}$. 
                \item For $s=1,\ldots,S$ and $p=1,\ldots,P,$  $\alpha_{sp}^{\psi}= a_{\psi} + \frac{N_s}{2}$.
            \end{itemize}
        \item Initialize the remaining values in  $\boldsymbol{\varphi}$ according to Section B.2.
    \end{enumerate}
        
    \textbf{For $t=1,\ldots$}
    \begin{enumerate}
        \item For $s=1,\ldots,S$ and $p=1,\ldots,P$, 
            \begin{align*}
                \boldsymbol{\Sigma}^{\Lambda}_{sp} &= \left(\boldsymbol{D}^{-1}_{sp} +  \frac{\alpha^{\psi}_{sp}}{\beta^{\psi}_{sp}}\left(\sum_{i=1}^{N_s} \boldsymbol{\mu}^{l}_{si}\left[\boldsymbol{\mu}^l_{si}\right]^{\intercal} + \boldsymbol{\Sigma}^l_{si}\right)\right)^{-1}, \\
                \boldsymbol{\mu}^{\Lambda}_{sp} &= \boldsymbol{\Sigma}^{\Lambda}_{sp}\frac{\alpha^{\psi}_{sp}}{\beta^{\psi}_{sp}}\sum_{i=1}^{N_s}\left(x_{sip}-\left[\boldsymbol{\mu}^\Phi_{p}\right]^\intercal\boldsymbol{\mu}^f_{si}\right)\boldsymbol{\mu}^l_{si},
            \end{align*}
            where $\boldsymbol{D}^{-1}_{sp}=\text{diag}\left(\frac{\alpha^{\omega}_{sp1}}{\beta^{\omega}_{sp1}}\left( \frac{\alpha^{\delta}_{s1}}{\beta^{\delta}_{s1}}\right),\ldots,\frac{\alpha^{\omega}_{spJ^*_s}}{\beta^{\omega}_{spJ^*_s}}\left( \prod_{l=1}^{J^*_s} \frac{\alpha^{\delta}_{sl}}{\beta^{\delta}_{sl}}\right)\right)$.

            \item For $p=1,\ldots,P$, 
            \begin{align*}
                \boldsymbol \Sigma^{\Phi}_{p}&=\left(\boldsymbol{D}^{-1}_p + \sum_{s=1}^S \frac{\alpha^{\psi}_{sp}}{\beta^{\psi}_{sp}}\sum_{i=1}^{N_s} \boldsymbol{\mu}^f_{si}\left[\boldsymbol{\mu}^f_{si}\right]^{\intercal} + \boldsymbol{\Sigma}^f_{si}\right)^{-1}, \\
                \boldsymbol{\mu}^{\Phi}_{p} &= \boldsymbol{\Sigma}^{\Phi}_{p}\sum_{s=1}^S \frac{\alpha^{\psi}_{sp}}{\beta^{\psi}_{sp}}\sum_{i=1}^{N_s} \left(x_{sip} - \left[\boldsymbol{\mu}^{\Lambda}_{sp}\right]^\intercal \boldsymbol{\mu}_{si}^l \right)\boldsymbol{\mu}^f_{si},
            \end{align*}
            where $\boldsymbol{D}^{-1}_p=\text{diag}\left(\frac{\alpha^{\omega}_{p1}}{\beta^{\omega}_{p1}}\left(\frac{\alpha^{\delta}_{1}}{\beta^{\delta}_{1}}\right),\ldots,\frac{\alpha^{\omega}_{pK}}{\beta^{\omega}_{pK}}\left(\prod_{l=1}^K\frac{\alpha^{\delta}_{l}}{\beta^{\delta}_{l}}\right)\right)$.

            \item For $s=1,\ldots,S$ and $p=1,\ldots,P,$
            \begin{align*}
               \beta^{\psi}_{sp}=b_{\psi} + &\frac{1}{2} 
                 \left\{\sum_{i=1}^{N_s}  \left( x_{sip} - \left[\boldsymbol{\mu}_p^\Phi\right]^\intercal\boldsymbol{\mu}_{si}^f - \left[\boldsymbol{\mu}_{sp}^\Lambda\right]^\intercal \boldsymbol{\mu}_{si}^l\right)^2 + \left[\boldsymbol{\mu}^{\Phi}_{p}\right]^\intercal  \left( \sum_{i=1}^{N_s}  \boldsymbol{\Sigma}^f_{si}\right)  \boldsymbol{\mu}^{\Phi}_{p} + \left[\boldsymbol{\mu}^{\Lambda}_{sp}\right]^\intercal\left( \sum_{i=1}^{N_s}  \boldsymbol{\Sigma}^l_{si}\right)\boldsymbol{\mu}^{\Lambda}_{sp} + \right. \\
                & \qquad \left.  \text{Tr}\left( \left( \sum_{i=1}^{N_s}\boldsymbol{\mu}^f_{si}\left[\boldsymbol{\mu}^f_{si}\right]^\intercal  + \boldsymbol{\Sigma}^f_{si}\right)\boldsymbol{\Sigma}^{\Phi}_{p} \right) +  \text{Tr}\left( \left( \sum_{i=1}^{N_s} \boldsymbol{\mu}^l_{si} \left[\boldsymbol{\mu}^l_{si}\right]^\intercal + \boldsymbol{\Sigma}^l_{si}\right)\boldsymbol{\Sigma}^{\Lambda}_{sp} \right)\right\}.
            \end{align*}

            \item For $s=1,\ldots,S$ and $i=1,\ldots,N_s$, 
            \begin{align*}
                 \boldsymbol{\Sigma}^{l}_{si}&=\left(\boldsymbol{I}_{J_s} + \sum_{p=1}^P \frac{\alpha^{\psi}_{sp}}{\beta^{\psi}_{sp}} \left(\boldsymbol{\mu}^{\Lambda}_{sp}\left[\boldsymbol{\mu}^{\Lambda}_{sp}\right]^{\intercal}+\boldsymbol{\Sigma}^{\Lambda}_{sp}\right)\right)^{-1} \\
                 \boldsymbol{\mu}^l_{si} &= \boldsymbol{\Sigma}^l_{si}\left[\mathbf{M}^{\Lambda}_s\right]^{\intercal}\boldsymbol{D}^{-1}_s (\mathbf{x}_{si}- \mathbf{M}^{\Phi}\boldsymbol{\mu}_{si}^f) \\
                 \boldsymbol{\Sigma}^f_{si}&=\left(\boldsymbol{I}_k + \sum_{p=1}^P \frac{\alpha^{\psi}_{sp}}{\beta^{\psi}_{sp}} \left( \boldsymbol{\mu}^{\Phi}_{p} \left[\boldsymbol{\mu}^{\Phi}_{p}\right]^{\intercal}+\boldsymbol{\Sigma}^{\Phi}_{p}\right)\right)^{-1} \\
                \boldsymbol{\mu}^{f}_{si} &= \boldsymbol{\Sigma}^f_{si} \left[\mathbf{M}^{\Phi}\right]^{\intercal} \boldsymbol{D}^{-1}_s \left(\mathbf{x}_{si} - \mathbf{M}^{\Lambda}_s\boldsymbol{\mu}^{l}_{si}\right)
            \end{align*}
            where $\mathbf{M}^{\Phi}=\left(\boldsymbol{\mu}^{\Phi}_{1},\ldots,\boldsymbol{\mu}^{\Phi}_{P}\right)^\intercal$, $\mathbf{M}^{\Lambda}_s=\left(\boldsymbol{\mu}^{\Lambda}_{s1},\ldots,\boldsymbol{\mu}^{\Lambda}_{sP}\right)^{\intercal}$, $\boldsymbol{D}^{-1}_s=\text{diag}\left(\frac{\alpha^{\psi}_{s1}}{\beta^{\psi}_{1s}},\ldots,\frac{\alpha^{\psi}_{sP}}{\beta^{\psi}_{sP}}\right)$. 

            \item For $p=1,\ldots,P$ and $k=1,\ldots,K^*$,
             $\beta_{pk}^{\omega}=\frac{1}{2}\left(\nu + \left( \prod_{l=1}^{k} \frac{\alpha_{l}^{\delta}}{\beta_{l}^{\delta}} \right) \left(\left[\mu^{\Phi}_{pk}\right]^2+\Sigma^{\Phi}_{pk}\right)\right).$
            \item For $s=1,\ldots,S$, $p=1,\ldots,P$, and $j=1,\ldots,J^*$, $\beta^{\omega}_{spj}=\frac{1}{2} \left(\nu_s + \left( \prod_{l=1}^j \frac{\alpha^{\delta}_{sl}}{\beta^{\delta}_{sl}}\right)\left(\left[\mu^{\Lambda}_{spj}\right]^2+\Sigma^{\Lambda}_{spj}\right)\right).$

            \item For $l=1,\ldots,K^*$, 
            $\beta_{l}^{\delta}= 1 + \frac{1}{2}\sum_{k=l}^{K^*} \left(\prod_{1\leq r \leq k, r\neq l}{\frac{\alpha_{r}^{\delta}}{\beta_{r}^{\delta}}}\right)\sum_{p=1}^{P}\frac{\alpha_{pk}^{\omega}}{ \beta_{pk}^{\omega}}\left(\left[\mu^{\Phi}_{pk}\right]^2+\Sigma_{pk}^{\Phi}\right).$
            \item For $s=1,\ldots,S,$ $l=1,\ldots,J^*,$  $\beta_{sl}^{\delta}= 1 + \frac{1}{2}\sum_{j=l}^{J^*} \left(\prod_{1\leq r \leq j, r\neq l}\frac{\alpha_{sr}^{\xi}}{\beta_{sr}^{\xi}}\right)\sum_{p=1}^{P}\frac{ \alpha_{spj}^{\upsilon}}{ \beta_{spj}^{\upsilon}}\left(\left[\mu^{\Lambda}_{spj}\right]^2+\Sigma^{\Lambda}_{spj}\right).$
    \end{enumerate}
    \textbf{End For.}
    \bf{Output:} variational approximation $q^*(\boldsymbol{\theta};\boldsymbol{\varphi}^*)$ 
    \end{flushleft}
\end{breakablealgorithm}

\FloatBarrier

\subsection{SVI for Bayesian Multi-Study Factor Analysis}

\begin{table}[h!]
    \centering
    \caption{\it \small Reparameterization of $\boldsymbol{\varphi}$ of the mean-field approximation presented in Equation (14) to facilitate calculation of Equations (8)-(9) for multi-study SVI.}
    \begin{tabular}{|c|c|}
    \hline
    Natural Parameterization & Original Parameterization   \vspace{-0.5cm} \\
         \parbox[c]{2cm}{
         \begin{align*}
             \boldsymbol{\eta}^\Phi_{p1} &= \left(\boldsymbol{\Sigma}_p^\Phi\right)^{-1} \\
            \boldsymbol{\eta}^\Phi_{p2} &= \left(\boldsymbol{\Sigma}_p^\Phi\right)^{-1}\boldsymbol{\mu}_p^\Phi \\
            \boldsymbol{\eta}^\Lambda_{sp1} &= \left(\boldsymbol{\Sigma}_{sp}^\Lambda\right)^{-1} \\
            \boldsymbol{\eta}^\Lambda_{sp2} &= \left(\boldsymbol{\Sigma}_{sp}^\Lambda\right)^{-1}\boldsymbol{\mu}_{sp}^\Lambda, \\
            \eta_{sp}^\psi &= -\beta_{sp}^\psi
         \end{align*}
         }
         &
         \parbox[c]{2cm}{
            \begin{align*}
                \boldsymbol{\Sigma}_p^\Phi &= \left(\boldsymbol{\eta}^\Phi_{p1}\right)^{-1} \\
                \boldsymbol{\mu}_p^\Phi &= \left(\boldsymbol{\eta}^\Phi_{p1}\right)^{-1}\boldsymbol{\eta}^\Phi_{p2} \\
                \boldsymbol{\Sigma}_{sp}^\Lambda &= \left(\boldsymbol{\eta}^\Lambda_{sp1}\right)^{-1} \\
                \boldsymbol{\mu}_{sp}^\Lambda &= \left(\boldsymbol{\eta}^\Lambda_{sp1}\right)^{-1}\boldsymbol{\eta}^\Lambda_{sp2}, \\
                \beta_{sp}^\psi &= -\eta_{sp}^\psi
            \end{align*}
         } \\
         \hline
    \end{tabular}
    
    \label{table:multi-study-reparameterization}
\end{table}

\begin{breakablealgorithm}
    \tiny
    \begin{flushleft}
    \caption{SVI for Bayesian MSFA} \label{alg:SVI}
    $\bf{Input}$: Data $\{ \mathbf{X}_{s} \}$
    
    $\bf{Initialize}$: Set initial variational parameters according to Section B.2.
    
    \textbf{For $t=1,\ldots$ until convergence}:
    \begin{enumerate}
        \item Calculate step size: $\rho_t=(t+\tau)^{-\kappa}$.
        \item For $ s=1, \ldots, S $: create the set $\mathcal{I}_s(t)$ drawing $\widetilde{N}_s = \lfloor b_s \times N_s \rfloor$ elements without replacement from the set $\{1,\ldots,N_s\}$ and define $\widetilde{\mathbf{X}}_s^t = \{\mathbf{x}_{si} \mbox{ for } i \in \mathcal{I}_s(t)\}$.
        \item For $s=1,\ldots,S$, $i \in \mathcal{I}_s(t)$ compute local variational parameters for each observation in  $\widetilde{\mathbf{x}}_{s}^t$:
        \begin{enumerate}
            \item  Calculate $\boldsymbol{\Sigma}^f_{si}$, $\boldsymbol{\mu}^{f}_{si}$, $\boldsymbol{\Sigma}^l_{si}$ and $\boldsymbol{\mu}^{l}_{si}$ according to Algorithm~\ref{alg:CAVI} Step 4.
        \end{enumerate}
        \item Optimize remaining variational parameters using $\widetilde{\mathbf{X}}^t_{s}$ and update them according to the current step size:
            \begin{enumerate}
            \item For $s=1,\ldots,S$ and $p=1,\ldots,P$, Set:
            \begin{align*}
                \widehat{\boldsymbol{\eta}}^{\Lambda}_{sp1} &= \boldsymbol{D}^{-1}_{sp} +  \left(\frac{\alpha^{\psi}_{sp}}{\beta^{\psi}_{sp}}\frac{N_s}{\widetilde{N}_s}\sum_{i \in \mathcal{I}_s(t)} \boldsymbol{\mu}^{l}_{si}\left[\boldsymbol{\mu}^l_{si}\right]^{\intercal} + \boldsymbol{\Sigma}^l_{si}\right), \\
                \widehat{\boldsymbol{\eta}}^{\Lambda}_{sp2} &=  \frac{\alpha^{\psi}_{sp}}{\beta^{\psi}_{sp}} \frac{N_s}{\widetilde{N}_s}\sum_{i \in \mathcal{I}_s(t)}\left(x_{sip} - \left[\boldsymbol{\mu}_p^\Phi\right]^\intercal\boldsymbol{\mu}_{si}^f\right)\boldsymbol{\mu}_{si}^l, \\
                \boldsymbol{\eta}^{\Lambda}_{p1}(t)&= (1-\rho_t)\boldsymbol{\eta}^{\Lambda}_{p1} + (\rho_t) \hat{\boldsymbol{\eta}}^{\Lambda}_{p1}, \\
                \boldsymbol{\eta}^{\Lambda}_{p2}(t)&= (1-\rho_t)\boldsymbol{\eta}^{\Lambda}_{p2} + (\rho_t) \hat{\boldsymbol{\eta}}^{\Lambda}_{p2}.
            \end{align*}
            where $\boldsymbol{D}^{-1}_{sp}=\text{diag}\left(\frac{\alpha^{\omega}_{sp1}}{\beta^{\omega}_{sp1}}\left( \frac{\alpha^{\delta}_{s1}}{\beta^{\delta}_{s1}}\right),\ldots,\frac{\alpha^{\omega}_{spJ^*_s}}{\beta^{\omega}_{spJ^*_s}}\left( \prod_{l=1}^{J^*_s} \frac{\alpha^{\delta}_{sl}}{\beta^{\delta}_{sl}}\right)\right)$.
            \item For $p=1,\ldots,P$:
            \begin{align*}
                \hat{\boldsymbol{\eta}}^{\Phi}_{p1} &= \boldsymbol{D}^{-1}_p + \left(\sum_{s=1}^S \frac{\alpha^{\psi}_{sp}}{\beta^{\psi}_{sp}}\frac{N_s}{\widetilde{N}_s}\sum_{i \in \mathcal{I}_s(t)} \boldsymbol{\mu}^f_{si}\left[\boldsymbol{\mu}^f_{si}\right]^{\intercal} + \boldsymbol{\Sigma}^f_{si}
                \right), \\
                \widehat{\boldsymbol{\eta}}^{\Phi}_{p2} &= \sum_{s=1}^S \frac{\alpha^{\psi}_{sp}}{\beta^{\psi}_{sp}}\frac{N_s}{\widetilde{N}_s}
                \sum_{i \in \mathcal{I}_s(t)}\left(x_{sip} - \left[\boldsymbol{\mu}_{sp}^{\Lambda}\right]^\intercal \boldsymbol{\mu}_{si}^l \right)\boldsymbol{\mu}_{si}^f, \\
                \boldsymbol{\eta}^{\Phi}_{p1}(t)&= (1-\rho_t)\boldsymbol{\eta}^{\Phi}_{p1} + (\rho_t) \hat{\boldsymbol{\eta}}^{\Phi}_{p1}, \\
                \boldsymbol{\eta}^{\Phi}_{p2}(t)&= (1-\rho_t)\boldsymbol{\eta}^{\Phi}_{p2} + (\rho_t) \hat{\boldsymbol{\eta}}^{\Phi}_{p2}.
            \end{align*}
            where $\boldsymbol{D}^{-1}_p=\text{diag}\left(\frac{\alpha^{\omega}_{p1}}{\beta^{\omega}_{p1}}\left(\frac{\alpha^{\delta}_{1}}{\beta^{\delta}_{1}}\right),\ldots,\frac{\alpha^{\omega}_{pK}}{\beta^{\omega}_{pK}}\left(\prod_{l=1}^K\frac{\alpha^{\delta}_{l}}{\beta^{\delta}_{l}}\right)\right)$.
            
            \item For $s=1,\ldots,S$ and $p=1,\ldots,P$: 
            \begin{align*}
                \widehat{\eta}^{\psi}_{sp}=-b_{\psi} - \frac{1}{2} 
                 &\left\{\frac{N_s}{\widetilde{N}_s}\sum_{i \in \mathcal{I}_s(t)} \left( x_{sip} - \left[\boldsymbol{\mu}_p^\Phi\right]^\intercal\boldsymbol{\mu}_{si}^f - \left[\boldsymbol{\mu}_{sp}^\Lambda\right]^\intercal \boldsymbol{\mu}_{si}^l\right)^2 + \right. \\
                 & \left. \left[\boldsymbol{\mu}^{\Phi}_{p}\right]^\intercal \left( \frac{N_s}{\widetilde{N}_s}\sum_{i \in \mathcal{I}_s(t)} \boldsymbol{\Sigma}^f_{si} \right)  \boldsymbol{\mu}^{\Phi}_{p} + \left[\boldsymbol{\mu}^{\Lambda}_{sp}\right]^\intercal\left(\frac{N_s}{\widetilde{N}_s}\sum_{i \in \mathcal{I}_s(t)}\boldsymbol{\Sigma}^l_{si}\right)\boldsymbol{\mu}^{\Lambda}_{sp} + \right. \\ 
                & \left. \text{Tr}\left( \boldsymbol{\Sigma}^{\Phi}_{p}\left(\frac{N_s}{\widetilde{N}_s}\sum_{i \in \mathcal{I}_s(t)}\boldsymbol{\mu}^f_{si}\left[\boldsymbol{\mu}^f_{si}\right]^\intercal  \boldsymbol{\Sigma}^f_{si}\right) \right) +  \text{Tr}\left(\boldsymbol{\Sigma}^{\Lambda}_{sp} \left( \frac{N_s}{\widetilde{N}_s}\sum_{i \in \mathcal{I}_s(t)} \boldsymbol{\mu}^l_{si} \left[\boldsymbol{\mu}^l_{si}\right]^\intercal + \boldsymbol{\Sigma}^l_{si}\right) \right)\right\}, \\
                \eta^{\psi}_{sp}(t) &= (1-\rho_t)\eta^{\psi}_{sp} + (\rho_t)\widehat{\eta}^{\psi}_{sp}.
            \end{align*}
            \item Let $\boldsymbol{\varphi}=\boldsymbol{\varphi}(t)$ for updated parameters in Steps 4a-c.
        \end{enumerate}
        \item Update parameters which do not depend on local variables according to Algorithm \ref{alg:CAVI} Steps 5-6.
        \item If convergence met, end loop.
    \end{enumerate}
    \bf{Output:} variational approximation $q^*(\boldsymbol{\theta};\boldsymbol{\varphi}^*)$ 
    \end{flushleft}
\end{breakablealgorithm}

\FloatBarrier

\section{Initialization of parameters}
In the paper, we propose four variational Bayes algorithms which approximate the posterior distribution with a mean field factorization. These algorithms require an initial guess for the parameters that are sequentially updated. Informative initializations can speed up convergence and, sometimes, avoid converging to local optimums. We rely on sparse principal components analysis to provide an informative initial guess of the parameters of  FA and MSFA. Note that the optimal value of some parameters, such as $\alpha_{p}^\Psi$ in Algorithm 1, can be obtained analytically. These parameters are directly set to their optimal value and omitted in the following.

\subsection{Initialization for Factor Analysis}
\begin{enumerate}
	\item \textit{Initial values for $\boldsymbol{\mu}^\Lambda_p$ and $\boldsymbol{\mu}_{i}^l$}: 
We decompose the data matrix $\mathbf{X}$ via $\mathbf{X} =   \mathbf{\widetilde l} {\boldsymbol{\widetilde \Lambda}}^\intercal $, with the constraint that $\text{Cov}(\mathbf{\widetilde l})=\mathbf{I}_{J^*}.$ 
Here the loadings matrix $ \boldsymbol{\widetilde \Lambda}$ of dimensions $P \times J^*$ matrix and the factor scores $ \mathbf{\widetilde l}$ of dimensions $N \times J^*$ are estimated using an unconstrained sparse principal components analysis with $J^*$ components~\citep{erichson_sparse_2020} and appropriately scaled. Finally, set $\boldsymbol{\mu}^\Lambda_{pj} = {\widetilde \lambda_{pj}},$ where  $\widetilde \lambda_{pj}$ is the element $(p,j)$  of $ \boldsymbol{\widetilde \Lambda}$ and  $\boldsymbol{\mu}_{ij}^l = \widetilde l_{ij},$ with $\widetilde l_{ij}$ being the element $(i,j)$ of $ \mathbf{\widetilde l}$.
    
    \item \textit{Initial values for $\beta_{p}^\Psi$}: 
    We set $\beta_{p}^\Psi = \alpha^\Psi_{p}\times \widetilde\psi_{p}^{2}$, where
    $(\widetilde\psi_{1}^{2}, \ldots,\widetilde\psi_{1}^{2}) = \mbox{diag}(\mathbf X \mathbf {X^{\intercal}} - \boldsymbol{\Lambda}^{*}{\boldsymbol{\Lambda}^{*}}^\intercal)$. 
    \item \textit{Initial values for $\alpha^\omega_{pj}$ and $\beta^\omega_{pj}$}: We set there paramters equal to the prior: $\alpha^\omega_{pj}={\nu}/{2}$ and $\beta^\omega_{pj}={\nu}/{2}$. 
    \item \textit{Initial values for $\alpha^\delta_{l}$ and $\beta^\delta_{l}$}:
    We these parameters equal to the prior: $\alpha^\delta_{1}=a_1$ for $l=1$, $\alpha^\delta_{l}=a_2$ for $ 2 \leq l \leq J^*$, and $\beta^\delta_l=1$ for $l =1,\cdots, J^*$. 

    \item \textit{Inital values for $\boldsymbol{\Sigma}_p^\Lambda$}:
    We set the initial $\boldsymbol{\Sigma}_p^\Lambda=\mathbf{D}_p^{-1}$ where \newline $\mathbf{D}_p=\text{diag}\left(\frac{\alpha_{p1}^\omega}{\beta_{p1}^\omega}\frac{\alpha_1^\delta}{\beta_1^\delta},\cdots,\frac{\alpha_{pJ^*}^\omega}{\beta_{pJ^*}^\omega}\prod_{l=1}^{J^*}\frac{\alpha_l^\delta}{\beta_l^\delta}\right)$.

    \item \textit{Initial values for $\boldsymbol{\Sigma}_{i}^l$}:
    We set the initial $\boldsymbol{\Sigma}_p^\Lambda = \left(\boldsymbol{I}_{J^*} + \sum_{p=1}^P \frac{\alpha^{\psi}_{p}}{\beta^{\psi}_{p}} \left(\boldsymbol{\mu}^{\Lambda}_{p}\left[\boldsymbol{\mu}^{\Lambda}_{p}\right]^{\intercal}+\boldsymbol{\Sigma}^{\Lambda}_{p}\right)\right)^{-1}$.
\end{enumerate}

\subsection{Initialization for Multi-study Factor Analysis}
\begin{enumerate}
	\item \textit{Initial values for $\boldsymbol{\mu}^\Phi_p$ and $\boldsymbol{\mu}_{si}^f$}: 
    Let  $N=\sum_{s=1}^S N_s$, and
    $\mathbf{Z}$ be the $N\times P$ dimensional matrix obtained by stacking the study-specific data matrices $X_s$ for $s=1,\ldots,S,$  we decompose $\mathbf{Z}$ via $\mathbf{Z} = 
    \mathbf{\widetilde f} {\boldsymbol{\widetilde \Phi}}^\intercal,$
  with the constraint that $\text{Cov}(\mathbf{\widetilde f})=\mathbf{I}_{K^*}.$   The  common loadings matrix $\boldsymbol{\widetilde \Phi}$ of dimensions $P \times K^*$  and factor scores $\mathbf{\widetilde f}$ of dimensions $N \times K^*$ are estimated via an uncostrained sparse principal components analysis~\citep{erichson_sparse_2020}  with $K^*$ components and appropriately scaled. 
  Finally,  we set ${\mu}^\Phi_{pk} = {\widetilde \phi_{pk}}$, where ${\widetilde \phi}_{pk}$ is the element $(p,k)$ of $\boldsymbol{\widetilde \Phi}^*$ and  ${\mu}_{sik}^f = \widetilde f_{sik},$ where with  $\widetilde f_{sik}$ we refer to the element of the stacked matrix $\boldsymbol{f}^*$ corresponding to subject $i$ in study $s$ and dimension $k$.

    \item \textit{Initial values for $\boldsymbol{\mu}_{sp}^\Lambda$ and $\boldsymbol{\mu}_{si}^l$}:
    We define the study-specific residuals $\widetilde{\mathbf{E}}_s$:
     $$\widetilde{\mathbf{E}}_s = \mathbf{X}_s \left(\mathbf{I}_{P} - \boldsymbol{\Phi}^{(0)} \left({\boldsymbol{\Phi}^{(0)}}^\intercal \boldsymbol{\Phi}^{(0)} \right)^{+}{\boldsymbol{\Phi}^{(0)}}^\intercal\right),$$
    where $M^+$ indicates the Moore-Penrose inverse of $M$. These are the sparse Principal Componenets Analysis residuals proposed by ~\cite{camacho_all_2020}, which subtract the common part of the variability, $\boldsymbol{\Phi}^{(0)}{\boldsymbol{\Phi}^{(0)}}^\intercal$ , used in Step 1 from the data. For $s=1,\ldots S,$ we decompose $\widetilde{\mathbf{E}}_s$ via $\widetilde{\mathbf{E}}_s=\mathbf{\widetilde l}_{s}   {\boldsymbol{\widetilde \Lambda}_s}^\intercal$  with the constraint that $\text{Cov}(\mathbf{\widetilde l}_{s}) = \mathbf{I}_{J_s^*}.$ 
   The study-specific factor loadings $\boldsymbol{\widetilde \Lambda}_s$  of dimensions $P \times J_s^*,$ and the study-specific scores $\mathbf{\widetilde l}_{s}$  of dimensions $N_s \times J_s^*$ are estimated via an unconstrained sparse principal components analysis~\citep{erichson_sparse_2020}  with $J_s^*$ components and appropriately scaled. Finally, we set ${\mu}_{spj} = \widetilde \lambda_{spj}, $ where $\widetilde \lambda_{spj},$ is the $(p,j)$ element of
    $\boldsymbol{\widetilde \Lambda}_s$ and ${\mu}_{si}^l = \widetilde l_{sij},$ where $\widetilde l_{sij}$ is the element $(i,j)$ of $\mathbf{\widetilde l}_s$.
    
    \item \textit{Initial values for $\alpha_{sp}^\Psi$, $\beta_{sp}^\Psi$}:
    For $s=1,\ldots,S,$ let   
     $(\widetilde \psi^{2}_{s1}, \ldots, \widetilde \psi^{2}_{sP}) = \mbox{diag}( 
 \mathbf X_s \mathbf {X_s^{\intercal}} - \boldsymbol{\widetilde \Phi}{\boldsymbol{\widetilde \Phi}}^\intercal - \boldsymbol{\widetilde \Lambda}_s{\boldsymbol{\widetilde \Lambda}_s}^\intercal),$ we set $\beta_{sp}^\Psi = \alpha^\Psi_{sp} \times \widetilde \psi_{sp}^{2}.$ 

    \item \textit{Initial values for $\alpha_{pk}^\omega$, $\beta_{pk}^\omega$}
    We set initial variational parameters for $\omega_{pk}$ equal to the prior: $\alpha_{pk}^\omega={\nu}/{2}$ and $\beta_{pk}^\omega={\nu}/{2}$.

    \item \textit{Initial values for $\alpha_{spj}^\omega$, $\beta_{spj}^\omega$}:
We set these parameters equal to the prior: $\alpha_{spj}^\omega={\nu_s}/{2}$ and $\beta_{spj}^\omega={\nu_s}/{2}$.

    \item \textit{Initial values for $\alpha_l^\delta$, $\beta_l^\delta$}:
    We set these parameters equal to the prior: $\alpha_1^\delta = a_1$ for $l=1$, $\alpha_l=a_2$ for $2 \leq l \leq K^*$, $\beta_l=1$ for $l=1,\cdots,K^*$.

    \item \textit{Initial values for $\alpha_{sl}^\delta$, $\beta_{sl}^\delta$}:
    We set these parameters  equal to the prior: $\alpha_{s1}^\delta = a_{s1}$ for $l=1$, $\alpha^\delta_{sl}=a_{s2}$ for $2 \leq l \leq J_s^*$, $\beta^\delta_{sl}=1$ for $l=1,\cdots,J_s^*$.

    \item \textit{Initial values for $\boldsymbol{\Sigma_{p}^\Phi}$}:
    We set   $\boldsymbol{\Sigma}_p^\Phi=\mathbf{D}_p^{-1}$ where $\mathbf{D}_p=\text{diag}\left(\frac{\alpha_{p1}^\omega}{\beta_{p1}^\omega}\frac{\alpha_1^\delta}{\beta_1^\delta},\cdots,\frac{\alpha_{pK^*}^\omega}{\beta_{pK^*}^\omega}\prod_{l=1}^{K^*}\frac{\alpha_l^\delta}{\beta_l^\delta}\right)$.

    \item \textit{Initial values for $\boldsymbol{\Sigma}_{sp}^\Lambda$}:
    We set $\boldsymbol{\Sigma}_{sp}^\Lambda=\mathbf{D}_{sp}^{-1}$ where $\mathbf{D}_{sp}=\text{diag}\left(\frac{\alpha_{sp1}^\omega}{\beta_{sp1}^\omega}\frac{\alpha_{s1}^\delta}{\beta_{s1}^\delta},\cdots,\frac{\alpha_{spJ^*}^\omega}{\beta_{spJ^*}^\omega}\prod_{l=1}^{J^*}\frac{\alpha_{sl}^\delta}{\beta_{sl}^\delta}\right)$.

    \item \textit{Initial values for $\boldsymbol{\Sigma}_{si}^f$}:
    We set  $\boldsymbol{\Sigma}^f_{si}=\left(\boldsymbol{I}_k + \sum_{p=1}^P \frac{\alpha^{\psi}_{sp}}{\beta^{\psi}_{sp}} \left( \boldsymbol{\mu}^{\Phi}_{p} \left[\boldsymbol{\mu}^{\Phi}_{p}\right]^{\intercal}+\boldsymbol{\Sigma}^{\Phi}_{p}\right)\right)^{-1}$.

    \item \textit{Inital values for $\boldsymbol{\Sigma}_{si}^l$}:
    We set $\boldsymbol{\Sigma}^{l}_{si}=\left(\boldsymbol{I}_{J_s} + \sum_{p=1}^P \frac{\alpha^{\psi}_{sp}}{\beta^{\psi}_{sp}} \left(\boldsymbol{\mu}^{\Lambda}_{sp}\left[\boldsymbol{\mu}^{\Lambda}_{sp}\right]^{\intercal}+\boldsymbol{\Sigma}^{\Lambda}_{sp}\right)\right)^{-1}$.
    
\end{enumerate}

\newpage

\section{Additional Simulation Study Results}

In this section, we provide additional information to benchmark the performance of the proposed VB algorithms (available at \href{https://github.com/blhansen/VI-MSFA}{github.com/blhansen/VI-MSFA}) against other FA and covariance estimation algorithms as described in \S 4 of the main manuscript. We evaluate the accuracy of the algorithms to correctly estimated the covariance based on the Frobenius norm (\tablename s~S3, S8--S10), defined as $\Vert \boldsymbol{\Sigma}-\widehat{\boldsymbol{\Sigma}}\Vert_2$, and L1 norm (\tablename s~S3, S11--13), defined as $\Vert \boldsymbol{\Sigma}-\widehat{\boldsymbol{\Sigma}}\Vert_1$. Additionally, we provide simulations with 50\% and 80\% levels of sparsity in the true factor loadings matrix. In \figurename s~S1--S9, we visualize the results of the simulation study described in \S 4 of the main manuscript. 

\subsection{Factor Analysis}

\begin{table}[h!]
\caption{\it \small Estimation accuracy (Frobenius norm and L1 norm), mean(sd), across 50 simulation replicates described in \S 4.1 of the main text. Note: we were unable to run GS, ADVI, and POET for some scenarios due to computation time and memory limitations, as the maximum allocation of 24 hours of run time and 128Gb of RAM for each iteration was insufficient. Additionally, we are unable to run ECM in scenarios with $P \geq N$. This is indicated with \longdash[5] in the table.}\label{single_study_extended_sims}
\centering
\tiny
\begin{tabular}{lr|lll|lll}
\multicolumn{2}{c}{} &
    \multicolumn{3}{c}{Frobenius Norm} &
    \multicolumn{3}{c}{L1 Norm} \\
method & N & P=100 & P=500 & P=5000 & P=100 & P=500 & P=5000 \\
\specialrule{\cmidrulewidth}{0pt}{0pt}
GS & 100 & 7.76(1) & 42.27(6) & \multicolumn{1}{c|}{\longdash[5]} & 13.38(2.4) & 82.73(16.16) & \multicolumn{1}{c}{\longdash[5]}\\
CAVI & 100 & 9.32(1) & 54.51(3) & 591.25(35) & 15.8(2.64) & 105.32(14.43) & 1293.68(116.97)\\
SVI-05 & 100 & 10.23(1) & 60.41(3) & 656.18(39) & 17.2(2.75) & 117.22(14.59) & 1438.19(133.67)\\
SVI-02 & 100 & 11.87(1) & 70.1(4) & 755.49(48) & 20.98(3.57) & 138.95(14.85) & 1757.61(236.83)\\
SVI-005 & 100 & 27.02(3) & 145.33(15) & 1577.6(136) & 56.23(10.89) & 313.69(59.56) & 4039.96(560.84)\\
POET & 100 & 8.76(1) & 41.99(3) & 437.71(42) & 14.27(2.75) & 77.19(12.27) & 919.12(158.45)\\
ECM & 100 & \multicolumn{1}{c}{\longdash[5]} & \multicolumn{1}{c}{\longdash[5]} & \multicolumn{1}{c|}{\longdash[5]} & \multicolumn{1}{c}{\longdash[5]} & \multicolumn{1}{c}{\longdash[5]} & \multicolumn{1}{c}{\longdash[5]} \\
ADVI & 100 & 309.03(489) & \multicolumn{1}{c}{\longdash[5]} & \multicolumn{1}{c|}{\longdash[5]} & 548.91(829.37) & \multicolumn{1}{c}{\longdash[5]} & \multicolumn{1}{c}{\longdash[5]}\\
\specialrule{\cmidrulewidth}{0pt}{0pt}
GS & 500 & 3.57(0) & 18.45(2) & \multicolumn{1}{c|}{\longdash[5]} & 5.85(0.96) & 33.34(6.23) & \multicolumn{1}{c}{\longdash[5]}\\
CAVI & 500 & 4.52(0) & 25.62(2) & 256.78(13) & 7.47(1.17) & 48.67(5.92) & 584.62(58.2)\\
SVI-05 & 500 & 4.98(0) & 28.96(2) & 294.23(12) & 8.24(1.16) & 56.01(6.05) & 682.42(60.33)\\
SVI-02 & 500 & 5.52(0) & 31.43(2) & 317.45(13) & 8.93(1.35) & 57.24(6.28) & 694.62(54.81)\\
SVI-005 & 500 & 8.26(1) & 47.25(3) & 482.57(25) & 13.57(2.29) & 87.08(10.04) & 1062.94(116.4)\\
POET & 500 & 4.28(0) & 19.72(2) & 186.11(14) & 7.14(1.54) & 35.33(6.53) & 388.53(54.47)\\
ECM & 500 & 4.03(0) & \multicolumn{1}{c}{\longdash[5]} & \multicolumn{1}{c|}{\longdash[5]} & 7.1(1.28) & \multicolumn{1}{c}{\longdash[5]} & \multicolumn{1}{c}{\longdash[5]}\\
ADVI & 500 & 59.06(70) & \multicolumn{1}{c}{\longdash[5]} & \multicolumn{1}{c|}{\longdash[5]} & 99.25(117.6) & \multicolumn{1}{c}{\longdash[5]} & \multicolumn{1}{c}{\longdash[5]}\\
\specialrule{\cmidrulewidth}{0pt}{0pt}
GS & 1000 & 2.68(0) & 12.81(1) & \multicolumn{1}{c|}{\longdash[5]} & 4.19(0.72) & 22.99(3.36) & \multicolumn{1}{c}{\longdash[5]}\\
CAVI & 1000 & 3.54(0) & 18.13(1) & 175.53(16) & 5.86(0.94) & 35.86(4.63) & 397.51(52.98)\\
SVI-05 & 1000 & 3.96(0) & 20.53(1) & 200.16(15) & 6.64(0.97) & 41.44(4.75) & 468.96(57.89)\\
SVI-02 & 1000 & 4.31(0) & 22.22(1) & 217.43(14) & 6.95(0.9) & 42.92(4.85) & 481.41(54.2)\\
SVI-005 & 1000 & 6(0) & 31.25(1) & 311.39(12) & 9.22(1.16) & 56.57(6.01) & 651.99(59.91)\\
POET & 1000 & 3.31(0) & 13.97(1) &  \multicolumn{1}{c|}{\longdash[5]} & 5.44(1.07) & 25.19(3.75) &  \multicolumn{1}{c}{\longdash[5]}\\
ECM & 1000 & 2.93(0) & 19.4(3) & \multicolumn{1}{c|}{\longdash[5]} & 4.64(0.97) & 89.97(40.76) & \multicolumn{1}{c}{\longdash[5]}\\
ADVI & 1000 & 56.15(62) & \multicolumn{1}{c}{\longdash[5]} & \multicolumn{1}{c|}{\longdash[5]} & 86.38(94.39) & \multicolumn{1}{c}{\longdash[5]} & \multicolumn{1}{c}{\longdash[5]}\\
\bottomrule
\end{tabular}
\end{table}

\begin{table}[h!]
\caption{\it \small Simulation study of \S 4.1 with sparsity level of 50\% in true Factor Loading matrices. Average computational cost (in seconds), used RAM (in Mb) and estimation accuracy (RV coefficients) across 50 simulation replicates, mean(sd). Note: we were unable to run GS, ADVI, and POET for some scenarios due to computation time and memory limitations, as the maximum allocation of 24 hours of run time and 128Gb of RAM for each iteration was insufficient. Additionally, we are unable to run ECM in scenarios with $P \geq N$. This is indicated with \longdash[5] in the table.}\label{table-single-study-sims-50-sparse}
\centering
\tiny
\resizebox{\textwidth}{!}{
\begin{tabular}{lr|rrr|rrr|rrr}
\multicolumn{2}{c}{} &
    \multicolumn{3}{c}{Time (Seconds)} & \multicolumn{3}{c}{Memory (Mb)} & \multicolumn{3}{c}{RV Coefficient} \\
method & N & P=100 & P=500 & P=5000 & P=100 & P=500 & P=5000 & P=100 & P=500 & P=5000 \\
\specialrule{\cmidrulewidth}{0pt}{0.5pt}
GS & 100 & 134.22(1.87) & 642.09(9.16) & \multicolumn{1}{c|}{\longdash[5]} & 209.67(0.09) & 4019.33(0.19) & \multicolumn{1}{c|}{\longdash[5]} & 0.92(0.02) & 0.9(0.03) & \multicolumn{1}{c}{\longdash[5]}\\
CAVI & 100 & 1.37(0.04) & 8.1(0.22) & 153.89(7.74) & 23.41(0.06) & 47.48(1.58) & 681.48(102.02) & 0.81(0.04) & 0.78(0.03) & 0.76(0.02)\\
SVI-05 & 100 & 0.87(0.03) & 4.83(0.15) & 84.83(2.76) & 23.2(0) & 47.6(0) & 797.92(39.34) & 0.78(0.03) & 0.75(0.03) & 0.72(0.02)\\
SVI-02 & 100 & 0.51(0) & 2.71(0.07) & 43.35(1.52) & 20.7(0) & 47.01(1.86) & 824.36(62.48) & 0.77(0.03) & 0.74(0.03) & 0.72(0.02)\\
SVI-005 & 100 & 4.43(0.8) & 1.68(0.07) & 22.6(0.79) & 17.11(0.04) & 32.39(0.75) & 656.4(39.43) & 0.77(0.04) & 0.67(0.04) & 0.65(0.03)\\
POET & 100 & 0.36(0.02) & 9.39(0.1) & 1070.8(21.11) & 50.22(3.25) & 285.53(31.03) & 24585.96(1134.49) & 0.91(0.03) & 0.91(0.03) & 0.91(0.02)\\
ECM & 100 & \multicolumn{1}{c}{\longdash[5]} & \multicolumn{1}{c}{\longdash[5]} & \multicolumn{1}{c|}{\longdash[5]} & \multicolumn{1}{c}{\longdash[5]} & \multicolumn{1}{c}{\longdash[5]} & \multicolumn{1}{c|}{\longdash[5]} & \multicolumn{1}{c}{\longdash[5]} & \multicolumn{1}{c}{\longdash[5]} & \multicolumn{1}{c}{\longdash[5]} \\
ADVI & 100 & 196.09(173.23) & \multicolumn{1}{c}{\longdash[5]}& \multicolumn{1}{c|}{\longdash[5]} & 180.57(21.36) & \multicolumn{1}{c}{\longdash[5]} & \multicolumn{1}{c|}{\longdash[5]} & 0.81(0.11) & \multicolumn{1}{c}{\longdash[5]} & \multicolumn{1}{c}{\longdash[5]}\\
\specialrule{\cmidrulewidth}{0pt}{0.5pt}
GS & 500 & 151.83(3.45) & 743.81(9.66) & \multicolumn{1}{c|}{\longdash[5]} & 293.64(0.16) & 4133.68(1.06) & \multicolumn{1}{c|}{\longdash[5]} & 0.98(0.01) & 0.98(0.01) & \multicolumn{1}{c}{\longdash[5]}\\
CAVI & 500 & 6.47(0.08) & 35.46(1.9) & 685.42(30.67) & 37.11(0.06) & 46.15(0.34) & 834.97(84.65) & 0.97(0.01) & 0.95(0.01) & 0.93(0.01)\\
SVI-05 & 500 & 3.74(0.07) & 20.25(0.36) & 345.16(27) & 37(0) & 46.1(0) & 843.07(107.52) & 0.96(0.01) & 0.94(0.01) & 0.91(0.01)\\
SVI-02 & 500 & 1.84(0.04) & 9.72(0.15) & 160.42(11.33) & 32.3(0) & 46.1(0) & 862.58(117.53) & 0.96(0.01) & 0.93(0.01) & 0.91(0.01)\\
SVI-005 & 500 & 0.89(0.05) & 4.54(0.08) & 66.89(3.67) & 24.72(0.14) & 45.69(1.56) & 812.91(109.64) & 0.94(0.01) & 0.92(0.01) & 0.89(0.01)\\
POET & 500 & 1.21(0.08) & 20.4(0.13) & 2819.49(105.58) & 103.51(0.04) & 1196(0.03) & 100854.73(1061.55) & 0.98(0.01) & 0.98(0.01) & 0.98(0.01)\\
ECM & 500 & 3.21(0.07) & \multicolumn{1}{c}{\longdash[5]} & \multicolumn{1}{c|}{\longdash[5]} & 37.29(0.06) & \multicolumn{1}{c}{\longdash[5]} & \multicolumn{1}{c|}{\longdash[5]} & 0.98(0.01) & \multicolumn{1}{c}{\longdash[5]} & \multicolumn{1}{c}{\longdash[5]}\\
ADVI & 500 & 1002.54(953.73) & \multicolumn{1}{c}{\longdash[5]} & \multicolumn{1}{c|}{\longdash[5]} & 193.7(48.32) &\multicolumn{1}{c}{\longdash[5]} & \multicolumn{1}{c|}{\longdash[5]} & 0.97(0.06) & \multicolumn{1}{c}{\longdash[5]} & \multicolumn{1}{c}{\longdash[5]}\\
\specialrule{\cmidrulewidth}{0pt}{0.5pt}
GS & 1000 & 181.91(2.42) & 866.59(23.88) & \multicolumn{1}{c|}{\longdash[5]} & 396.83(12.59) & 4275.14(1.37) & \multicolumn{1}{c|}{\longdash[5]} & 0.99(0) & 0.99(0) & \multicolumn{1}{c}{\longdash[5]}\\
CAVI & 1000 & 15.16(0.61) & 69.81(2.19) & 1490.86(50.03) & 36.71(0.06) & 44.25(0.34) & 872.5(122.64) & 0.98(0.01) & 0.98(0.01) & 0.97(0.01)\\
SVI-05 & 1000 & 7.31(0.15) & 39.02(0.8) & 702.72(61.81) & 36.6(0) & 60.3(0) & 846.89(124.74) & 0.97(0.01) & 0.97(0.01) & 0.96(0.01)\\
SVI-02 & 1000 & 3.45(0.06) & 18.16(0.3) & 319.72(21.62) & 36.6(0) & 59.98(2.28) & 844.4(96.93) & 0.97(0.01) & 0.97(0.01) & 0.96(0.01)\\
SVI-005 & 1000 & 1.57(0.08) & 8.12(0.17) & 126.49(7.39) & 33.47(0.18) & 60.03(1.94) & 838.5(105.43) & 0.96(0.01) & 0.96(0.01) & 0.95(0.01)\\
POET & 1000 & 3.46(0.11) & 35.82(0.21) & \multicolumn{1}{c|}{\longdash[5]}& 203(0.01) & 2385.51(0.06) & \multicolumn{1}{c|}{\longdash[5]} & 0.99(0) & 0.99(0) & \multicolumn{1}{c}{\longdash[5]}\\
ECM & 1000 & 3.08(0.06) & 74.03(1.61) & \multicolumn{1}{c|}{\longdash[5]} & 36.91(0.04) & 60.64(0.3) & \multicolumn{1}{c|}{\longdash[5]} & 0.99(0) & 0.97(0.01) & \multicolumn{1}{c}{\longdash[5]}\\
ADVI & 1000 & 1506.61(1405.29) & \multicolumn{1}{c}{\longdash[5]} & \multicolumn{1}{c|}{\longdash[5]} & 180.27(21.42) & \multicolumn{1}{c}{\longdash[5]} & \multicolumn{1}{c|}{\longdash[5]} & 0.97(0.03) & \multicolumn{1}{c}{\longdash[5]} & \multicolumn{1}{c}{\longdash[5]}\\
\bottomrule
\end{tabular}}
\end{table}

\begin{table}[h!]
\caption{\it \small  Estimation accuracy: Frobenius norm, L1 Norm, reported as mean(sd) across 50 simulation replicates for simulation study of \S 4.1 with sparsity level of 50\% in true Factor Loading matrices. Note: we were unable to run GS, ADVI, and POET for some scenarios due to computation time and memory limitations, as the maximum allocation of 24 hours of run time and 128Gb of RAM for each iteration was insufficient. Additionally, we are unable to run ECM in scenarios with $P \geq N$. This is indicated with \longdash[5] in the table.}\label{single_study_extended_sims}
\centering
\tiny
\begin{tabular}{lr|lll|lll}
\multicolumn{2}{c}{} &
    \multicolumn{3}{c}{Frobenius Norm} &
    \multicolumn{3}{c}{L1 Norm} \\
method & N & P=100 & P=500 & P=5000 & P=100 & P=500 & P=5000 \\
\specialrule{\cmidrulewidth}{0pt}{0pt}
GS & 100 & 10.77(1) & 53.57(6) & \multicolumn{1}{c|}{\longdash[5]} & 16.19(2.38) & 91.14(14.02) & \multicolumn{1}{c}{\longdash[5]}\\
CAVI & 100 & 15.18(1) & 79.14(5) & 815.78(39) & 23.11(2.67) & 145.38(12.49) & 1695.73(120.97)\\
SVI-05 & 100 & 16.85(1) & 87.76(4) & 909.99(40) & 25.73(2.48) & 161.04(11.27) & 1881.96(125.05)\\
SVI-02 & 100 & 18.32(1) & 95.48(5) & 991.45(52) & 27.36(2.74) & 168.6(14.76) & 2030.55(185.6)\\
SVI-005 & 100 & 30.85(3) & 164.84(15) & 1760.47(166) & 55.73(9.57) & 328.73(51.78) & 4142.47(569.32)\\
POET & 100 & 12.09(1) & 56.52(5) & 566.17(59) & 18.45(2.64) & 98.66(14.22) & 1159.9(231.95)\\
ECM & 100 & \multicolumn{1}{c}{\longdash[5]} & \multicolumn{1}{c}{\longdash[5]} & \multicolumn{1}{c|}{\longdash[5]} & \multicolumn{1}{c}{\longdash[5]} & \multicolumn{1}{c}{\longdash[5]} & \multicolumn{1}{c}{\longdash[5]} \\
ADVI & 100 & 414.19(481) & \multicolumn{1}{c}{\longdash[5]} & \multicolumn{1}{c|}{\longdash[5]} & 659.69(797.02) & \multicolumn{1}{c}{\longdash[5]} & \multicolumn{1}{c}{\longdash[5]}\\
\specialrule{\cmidrulewidth}{0pt}{0pt}
GS & 500 & 4.57(0) & 25.35(3) & \multicolumn{1}{c|}{\longdash[5]} & 7.12(1.11) & 44.69(6.04) & \multicolumn{1}{c}{\longdash[5]}\\
CAVI & 500 & 5.97(1) & 35.91(3) & 380(25) & 9.36(1.28) & 68.49(7.6) & 837.46(87.16)\\
SVI-05 & 500 & 6.58(1) & 41.38(2) & 443.04(23) & 10.35(1.21) & 80.88(8.12) & 1007.78(94.52)\\
SVI-02 & 500 & 7.11(0) & 43.67(2) & 463.7(23) & 11.03(1.25) & 81.49(7.88) & 1012.11(91.87)\\
SVI-005 & 500 & 9.77(1) & 58.08(3) & 608.59(25) & 14.99(1.74) & 99.65(9.99) & 1220.6(95.45)\\
POET & 500 & 5.1(0) & 25.52(2) & 254.85(25) & 8.26(1.08) & 44.13(5.94) & 500.7(76.68)\\
ECM & 500 & 4.75(0) & \multicolumn{1}{c}{\longdash[5]} & \multicolumn{1}{c|}{\longdash[5]} & 7.46(1.06) & \multicolumn{1}{c}{\longdash[5]} & \multicolumn{1}{c}{\longdash[5]}\\
ADVI & 500 & 68.09(84) &\multicolumn{1}{c}{\longdash[5]} & \multicolumn{1}{c|}{\longdash[5]} & 105.23(130.38) & \multicolumn{1}{c}{\longdash[5]} & \multicolumn{1}{c}{\longdash[5]}\\
\specialrule{\cmidrulewidth}{0pt}{0pt}
GS & 1000 & 3.89(0) & 17.76(2) & \multicolumn{1}{c|}{\longdash[5]} & 5.93(0.95) & 31.26(4.73) & \multicolumn{1}{c}{\longdash[5]}\\
CAVI & 1000 & 5.8(1) & 22.84(2) & 253.82(24) & 9.76(1.34) & 43.01(5.72) & 577.23(80.6)\\
SVI-05 & 1000 & 6.39(0) & 26.17(2) & 297.76(23) & 11.03(1.3) & 50.71(5.58) & 709.21(82.24)\\
SVI-02 & 1000 & 6.71(0) & 27.92(2) & 314.77(22) & 11.2(1.31) & 51.62(4.9) & 710.3(81.01)\\
SVI-005 & 1000 & 8.33(1) & 37.44(2) & 408.52(19) & 12.88(1.58) & 61.18(6.37) & 794.99(60.09)\\
POET & 1000 & 4.33(0) & 17.66(2) &  \multicolumn{1}{c|}{\longdash[5]} & 6.69(1.18) & 31.22(5.93) &  \multicolumn{1}{c}{\longdash[5]}\\
ECM & 1000 & 4.14(1) & 24.27(3) & \multicolumn{1}{c|}{\longdash[5]} & 6.76(3.11) & 85.25(19.69) & \multicolumn{1}{c}{\longdash[5]}\\
ADVI & 1000 & 49.26(74) & \multicolumn{1}{c}{\longdash[5]} & \multicolumn{1}{c|}{\longdash[5]} & 73.32(107.34) & \multicolumn{1}{c}{\longdash[5]} & \multicolumn{1}{c}{\longdash[5]}\\
\bottomrule
\end{tabular}
\end{table}

\begin{table}[h!]
\caption{\it \small Simulation study of \S 4.1 with sparsity level of 80\% in true Factor Loading matrices. Average computational cost (in seconds), used RAM (in Mb) and estimation accuracy (RV coefficients) across 50 simulation replicates, mean(sd). Note: we were unable to run GS, ADVI, and POET for some scenarios due to computation time and memory limitations, as the maximum allocation of 24 hours of run time and 128Gb of RAM for each iteration was insufficient. Additionally, we are unable to run ECM in scenarios with $P \geq N$. This is indicated with \longdash[5] in the table.}\label{table-single-study-sims-50-sparse}
\centering
\tiny
\resizebox{\textwidth}{!}{
\begin{tabular}{lr|rrr|rrr|rrr}
\multicolumn{2}{c}{} &
    \multicolumn{3}{c}{Time (Seconds)} & \multicolumn{3}{c}{Memory (Mb)} & \multicolumn{3}{c}{RV Coefficient} \\
method & N & P=100 & P=500 & P=5000 & P=100 & P=500 & P=5000 & P=100 & P=500 & P=5000 \\
\specialrule{\cmidrulewidth}{0pt}{0.5pt}
GS & 100 & 136.28(1.8) & 648(7.78) & \multicolumn{1}{c|}{\longdash[5]} & 209.65(0.1) & 4019.33(0.19) & \multicolumn{1}{c|}{\longdash[5]} & 0.89(0.04) & 0.75(0.08) & \multicolumn{1}{c}{\longdash[5]}\\
CAVI & 100 & 1.37(0.04) & 7.73(0.34) & 161.82(11.09) & 23.41(0.04) & 48.14(1.59) & 654.16(100.72) & 0.83(0.04) & 0.74(0.03) & 0.7(0.03)\\
SVI-05 & 100 & 0.87(0.03) & 4.69(0.09) & 86.4(2.74) & 23.2(0) & 47.86(2.67) & 881.59(94.07) & 0.81(0.04) & 0.72(0.03) & 0.69(0.03)\\
SVI-02 & 100 & 0.51(0) & 2.67(0.05) & 44.65(1.24) & 20.7(0) & 47.75(2.65) & 977.42(103.22) & 0.78(0.04) & 0.7(0.03) & 0.67(0.03)\\
SVI-005 & 100 & 2.1(1.15) & 1.68(0.11) & 23.41(0.68) & 17.01(0.06) & 32.39(0.75) & 853.64(117.58) & 0.77(0.04) & 0.59(0.05) & 0.56(0.05)\\
POET & 100 & 0.36(0.02) & 9.53(0.38) & 1016.78(79.71) & 50.22(3.24) & 285.41(30.47) & 25786.85(1079.02) & 0.89(0.02) & 0.87(0.02) & 0.86(0.03)\\
ECM & 100 & \multicolumn{1}{c}{\longdash[5]} & \multicolumn{1}{c}{\longdash[5]} & \multicolumn{1}{c|}{\longdash[5]} & \multicolumn{1}{c}{\longdash[5]} & \multicolumn{1}{c}{\longdash[5]} & \multicolumn{1}{c|}{\longdash[5]} & \multicolumn{1}{c}{\longdash[5]} & \multicolumn{1}{c}{\longdash[5]} & \multicolumn{1}{c}{\longdash[5]} \\
ADVI & 100 & 219.76(199.4) & \multicolumn{1}{c}{\longdash[5]} & \multicolumn{1}{c|}{\longdash[5]} & 180.56(21.37) & \multicolumn{1}{c}{\longdash[5]} & \multicolumn{1}{c|}{\longdash[5]} & 0.81(0.16) & \multicolumn{1}{c}{\longdash[5]} & \multicolumn{1}{c}{\longdash[5]}\\
\specialrule{\cmidrulewidth}{0pt}{0.5pt}
GS & 500 & 156.12(1.97) & 735.09(10.25) & \multicolumn{1}{c|}{\longdash[5]} & 293.63(0.16) & 4133.36(1.01) & \multicolumn{1}{c|}{\longdash[5]} & 0.98(0.01) & 0.95(0.02) & \multicolumn{1}{c}{\longdash[5]}\\
CAVI & 500 & 6.79(0.47) & 35.69(0.5) & 765.9(60.91) & 37.11(0.06) & 46.95(0.33) & 609.63(6.19) & 0.96(0.01) & 0.93(0.01) & 0.91(0.02)\\
SVI-05 & 500 & 4.05(0.09) & 20.48(0.38) & 379.69(13.92) & 37.1(0) & 46.9(0) & 720.98(89.59) & 0.96(0.01) & 0.92(0.01) & 0.9(0.02)\\
SVI-02 & 500 & 1.98(0.05) & 9.94(0.21) & 172.84(5.76) & 32.4(0) & 46.9(0) & 647.32(55.43) & 0.95(0.01) & 0.92(0.01) & 0.9(0.01)\\
SVI-005 & 500 & 0.96(0.05) & 4.62(0.11) & 69.38(2.51) & 24.82(0.13) & 46.58(1.57) & 662.62(89.92) & 0.9(0.01) & 0.88(0.01) & 0.87(0.01)\\
POET & 500 & 1.15(0.09) & 20.75(0.5) & 2948.15(65.48) & 103.61(0.04) & 1195.91(0.04) & 102273.19(1163.95) & 0.97(0.01) & 0.97(0.01) & 0.97(0.01)\\
ECM & 500 & 3.11(0.06) & \multicolumn{1}{c}{\longdash[5]} & \multicolumn{1}{c|}{\longdash[5]} & 37.39(0.03) & \multicolumn{1}{c}{\longdash[5]} & \multicolumn{1}{c|}{\longdash[5]} & 0.97(0.01) & \multicolumn{1}{c}{\longdash[5]} & \multicolumn{1}{c}{\longdash[5]}\\
ADVI & 500 & 983.34(958.17) & \multicolumn{1}{c}{\longdash[5]}& \multicolumn{1}{c|}{\longdash[5]} & 177.36(29.92) & \multicolumn{1}{c}{\longdash[5]} & \multicolumn{1}{c|}{\longdash[5]} & 0.94(0.12) & \multicolumn{1}{c}{\longdash[5]} & \multicolumn{1}{c}{\longdash[5]}\\
\specialrule{\cmidrulewidth}{0pt}{0.5pt}
GS & 1000 & 183.18(2.68) & 835.14(14.12) & \multicolumn{1}{c|}{\longdash[5]} & 396.78(12.6) & 4276.11(1.8) & \multicolumn{1}{c|}{\longdash[5]} & 0.99(0) & 0.98(0.01) & \multicolumn{1}{c}{\longdash[5]}\\
CAVI & 1000 & 15.9(1.09) & 71.75(0.81) & 1541.9(78.94) & 36.71(0.06) & 59.86(4.26) & 783.69(107.97) & 0.98(0.01) & 0.96(0.01) & 0.96(0.01)\\
SVI-05 & 1000 & 7.19(0.25) & 39.69(0.77) & 858.27(48.3) & 36.7(0) & 61.1(0) & 784.1(97.46) & 0.97(0.01) & 0.96(0.01) & 0.95(0.01)\\
SVI-02 & 1000 & 3.4(0.09) & 18.54(0.35) & 382.5(21.08) & 36.7(0) & 60.78(2.28) & 765.89(92.06) & 0.97(0) & 0.95(0.01) & 0.94(0.01)\\
SVI-005 & 1000 & 1.55(0.06) & 8.24(0.19) & 147.21(8.26) & 33.57(0.17) & 60.82(1.97) & 811.15(124.89) & 0.94(0.01) & 0.93(0.01) & 0.93(0.01)\\
POET & 1000 & 3.35(0.06) & 35.84(1.02) & \multicolumn{1}{c|}{\longdash[5]} & 203(0) & 2385.5(0.08) & \multicolumn{1}{c|}{\longdash[5]} & 0.98(0) & 0.98(0) & \multicolumn{1}{c}{\longdash[5]}\\
ECM & 1000 & 3.18(0.07) & 74.35(3.23) & \multicolumn{1}{c|}{\longdash[5]} & 36.99(0.03) & 61.34(0.3) & \multicolumn{1}{c|}{\longdash[5]}& 0.98(0) & 0.98(0) & \multicolumn{1}{c}{\longdash[5]}\\
ADVI & 1000 & 1197.34(689.55) & \multicolumn{1}{c}{\longdash[5]} & \multicolumn{1}{c|}{\longdash[5]} & 180.18(21.42) & \multicolumn{1}{c}{\longdash[5]} & \multicolumn{1}{c|}{\longdash[5]} & 0.94(0.14) & \multicolumn{1}{c}{\longdash[5]} & \multicolumn{1}{c}{\longdash[5]}\\
\bottomrule
\end{tabular}}
\end{table}

\begin{table}[h!]
\caption{\it \small  Estimation accuracy: Frobenius norm, L1 Norm, reported as mean(sd) across 50 simulation replicates for simulation study of \S 4.1 with sparsity level of 80\% in true Factor Loading matrices. Note: we were unable to run GS, ADVI, and POET for some scenarios due to computation time and memory limitations, as the maximum allocation of 24 hours of run time and 128Gb of RAM for each iteration was insufficient. Additionally, we are unable to run ECM in scenarios with $P \geq N$. This is indicated with \longdash[5] in the table.}\label{single_study_extended_sims}
\centering
\tiny
\begin{tabular}{lr|lll|lll}
\multicolumn{2}{c}{} &
    \multicolumn{3}{c}{Frobenius Norm} &
    \multicolumn{3}{c}{L1 Norm} \\
method & N & P=100 & P=500 & P=5000 & P=100 & P=500 & P=5000 \\
\specialrule{\cmidrulewidth}{0pt}{0pt}
GS & 100 & 5.91(1) & 41.46(11) & \multicolumn{1}{c|}{\longdash[5]} & 10.95(2.52) & 93.21(25.83) & \multicolumn{1}{c}{\longdash[5]}\\
CAVI & 100 & 7.41(1) & 39.55(2) & 413.71(19) & 13.86(1.95) & 80.57(8.5) & 949.05(96.75)\\
SVI-05 & 100 & 8.36(1) & 44.45(3) & 462.86(22) & 15.67(2.32) & 93.32(10.03) & 1097.01(114.95)\\
SVI-02 & 100 & 10.05(1) & 55.46(4) & 577.63(33) & 19.2(2.87) & 121.8(14.31) & 1499.15(169.23)\\
SVI-005 & 100 & 24.13(3) & 128.1(16) & 1362.88(169) & 52.18(7.89) & 301.81(57.26) & 3797.11(701.43)\\
POET & 100 & 7.22(0) & 32.51(2) & 316.21(16) & 12.59(1.91) & 60.99(8.36) & 670.27(84.45)\\
ECM & 100 & \multicolumn{1}{c}{\longdash[5]} & \multicolumn{1}{c}{\longdash[5]} & \multicolumn{1}{c|}{\longdash[5]} & \multicolumn{1}{c}{\longdash[5]} & \multicolumn{1}{c}{\longdash[5]} & \multicolumn{1}{c}{\longdash[5]} \\
ADVI & 100 & 271.65(565) & \multicolumn{1}{c}{\longdash[5]} & \multicolumn{1}{c|}{\longdash[5]} & 548.63(1233.29) & \multicolumn{1}{c}{\longdash[5]} & \multicolumn{1}{c}{\longdash[5]}\\
\specialrule{\cmidrulewidth}{0pt}{0pt}
GS & 500 & 2.32(0) & 15.56(3) & \multicolumn{1}{c|}{\longdash[5]} & 4.28(0.63) & 33.42(6.93) & \multicolumn{1}{c}{\longdash[5]}\\
CAVI & 500 & 2.95(0) & 17.23(1) & 186.37(12) & 5.36(0.78) & 38.63(5.01) & 428.28(38.44)\\
SVI-05 & 500 & 3.3(0) & 19.05(1) & 206.32(11) & 5.93(0.78) & 42.86(5.12) & 469.65(39.31)\\
SVI-02 & 500 & 3.81(0) & 21.4(1) & 229.37(10) & 6.87(1.08) & 45.65(4.45) & 505.51(44.46)\\
SVI-005 & 500 & 6.3(0) & 36.81(2) & 401.08(17) & 11.86(2.15) & 78.94(7.73) & 961.17(90.65)\\
POET & 500 & 3.3(0) & 14.11(1) & 140.89(8) & 5.75(0.94) & 28.64(3.87) & 297.03(39.3)\\
ECM & 500 & 2.68(0) & \multicolumn{1}{c}{\longdash[5]} & \multicolumn{1}{c|}{\longdash[5]} & 4.78(0.68) & \multicolumn{1}{c}{\longdash[5]} & \multicolumn{1}{c}{\longdash[5]}\\
ADVI & 500 & 37.49(35) & \multicolumn{1}{c}{\longdash[5]} & \multicolumn{1}{c|}{\longdash[5]} & 69.76(64.01) & \multicolumn{1}{c}{\longdash[5]} & \multicolumn{1}{c}{\longdash[5]}\\
\specialrule{\cmidrulewidth}{0pt}{0pt}
GS & 1000 & 1.63(0) & 9.72(1) &  \multicolumn{1}{c|}{\longdash[5]} & 3.08(0.34) & 20.23(3.5) &  \multicolumn{1}{c}{\longdash[5]}\\
CAVI & 1000 & 2.01(0) & 11.65(1) & 130.53(9) & 3.77(0.48) & 22.96(2.79) & 304.01(29.78)\\
SVI-05 & 1000 & 2.33(0) & 12.88(1) & 143.39(8) & 4.29(0.48) & 25.51(2.83) & 332.33(27.96)\\
SVI-02 & 1000 & 2.68(0) & 14.46(1) & 157.03(7) & 5.06(0.58) & 28.35(2.75) & 347.37(23.77)\\
SVI-005 & 1000 & 4.22(0) & 23(1) & 246.34(8) & 8.2(1.12) & 47.1(5.1) & 553.55(48.99)\\
POET & 1000 & 2.68(0) & 9.94(1) &  \multicolumn{1}{c|}{\longdash[5]} & 4.71(0.59) & 19.36(2.14) &  \multicolumn{1}{c}{\longdash[5]}\\
ECM & 1000 & 1.91(0) & 10.36(1) &  \multicolumn{1}{c|}{\longdash[5]} & 3.49(0.37) & 24.82(5.23) &  \multicolumn{1}{c}{\longdash[5]}\\
ADVI & 1000 & 61.77(291) &  \multicolumn{1}{c}{\longdash[5]} &  \multicolumn{1}{c|}{\longdash[5]} & 81.84(292.2) &  \multicolumn{1}{c}{\longdash[5]} &  \multicolumn{1}{c}{\longdash[5]}\\
\bottomrule
\end{tabular}
\end{table}

\begin{figure}[h!]
    \centering
    \includegraphics[width=\textwidth]{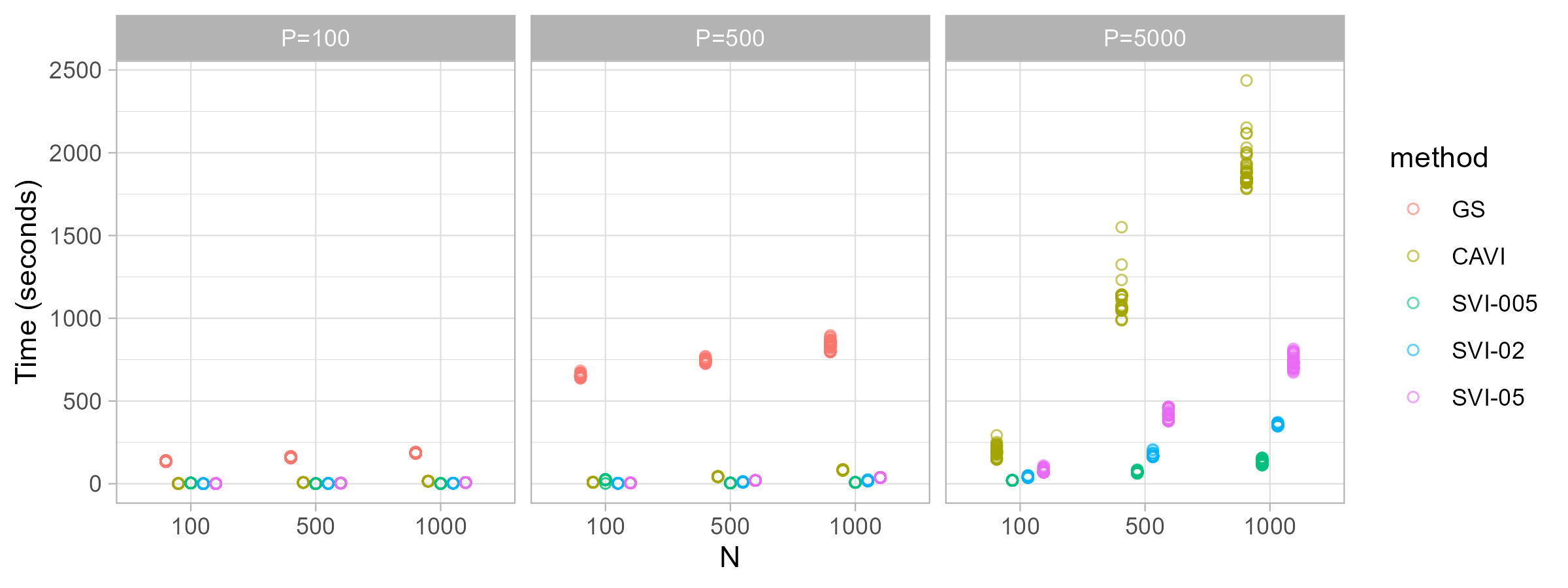}
    \caption{Visualization of results of \S 4.1 for GS, CAVI, and SVI algorithms. Computation time in seconds of GS, CAVI, and SVI Algorithms for single-study simulations across 50 replicates. Note: we were unable to run GS for some scenarios due to computation time and memory limitations, as the maximum allocation of 24 hours of run time and 128Gb of RAM for each iteration was insufficient.}
    \label{fig:time_single_study}
\end{figure}

\begin{figure}[h!]
    \centering
    \includegraphics[width=\textwidth]{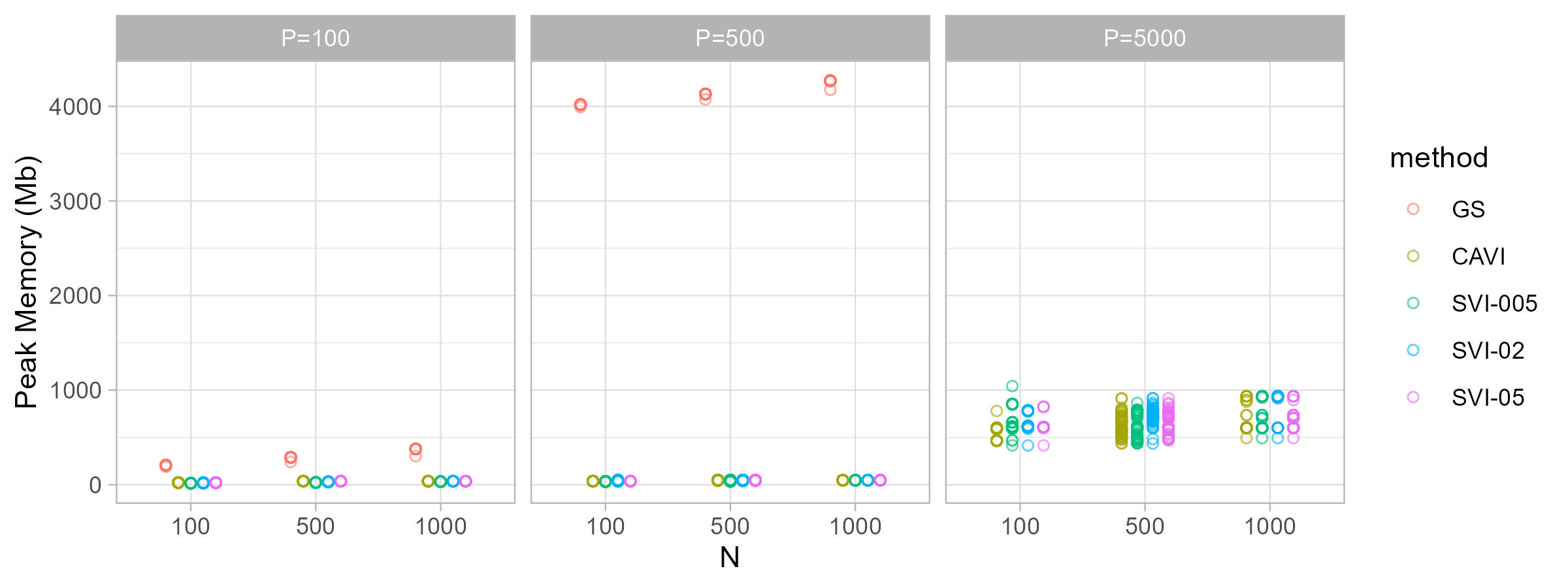}
    \caption{Visualization of results of \S 4.1 for GS, CAVI, and SVI algorithms. Peak RAM usage in Mb of GS, CAVI, and SVI Algorithms for single-study simulations across 50 replicates. Note: we were unable to run GS for some scenarios due to computation time and memory limitations, as the maximum allocation of 24 hours of run time and 128Gb of RAM for each iteration was insufficient.}
    \label{fig:memory_single_study}
\end{figure}

\begin{figure}[h!]
    \centering
    \includegraphics[width=\textwidth]{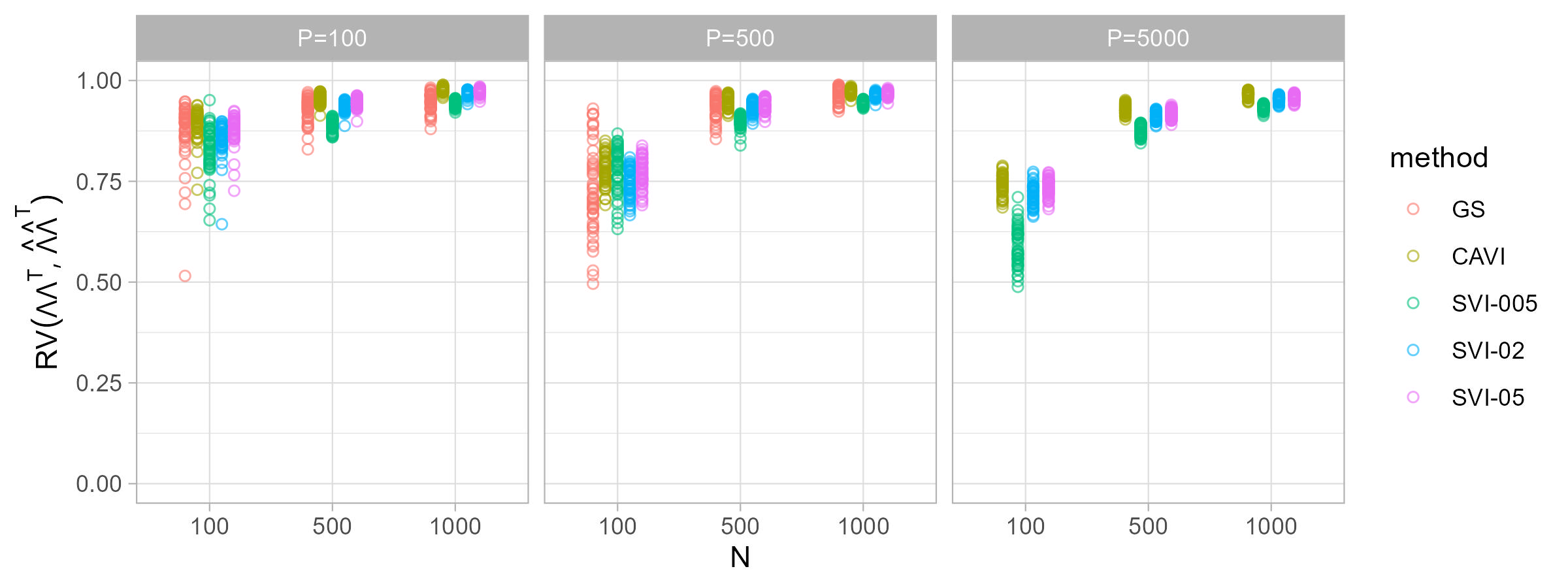}
    \caption{Visualization of results of \S 4.1 for GS, CAVI, and SVI algorithms. RV coefficient between estimated $\boldsymbol{\Lambda}\boldsymbol{\Lambda}^\intercal$ and the truth for single-study simulation study using GS, CAVI, and SVI Algorithms for single-study simulations across 50 replicates. Note: we were unable to run GS for some scenarios due to computation time and memory limitations, as the maximum allocation of 24 hours of run time and 128Gb of RAM for each iteration was insufficient.}
    \label{fig:lambda_rv_single_study}
\end{figure}

\begin{figure}[h!]
    \centering
    \includegraphics[width=\textwidth]{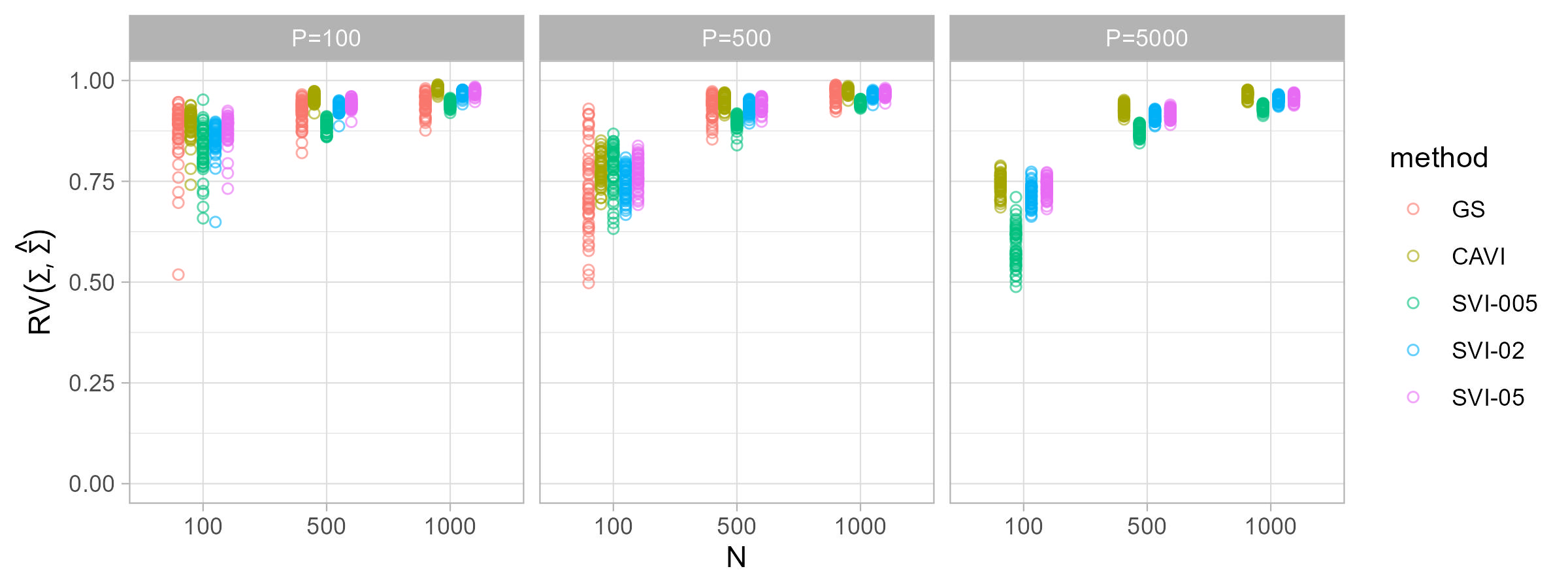}
    \caption{Visualization of results of \S 4.1 for GS, CAVI, and SVI algorithms. RV coefficient between estimated $\boldsymbol{\Sigma}=\boldsymbol{\Lambda}\boldsymbol{\Lambda}^\intercal + \boldsymbol{\Psi}$ and the truth using GS, CAVI, and SVI Algorithms for single-study simulations across 50 replicates. Note: we were unable to run GS for some scenarios due to computation time and memory limitations, as the maximum allocation of 24 hours of run time and 128Gb of RAM for each iteration was insufficient.}
    \label{fig:sigma_rv_single_study}
\end{figure}
 
\newpage

\FloatBarrier
 
\subsection{Multi-study Factor Analysis}

\begin{table}[ht!]
\tiny
\centering
\caption{\it \small Additional results of \S 4.2.  Average estimation accuracy, reported as $\lVert \boldsymbol{\Sigma}_s - \hat{\boldsymbol{\Sigma}}_s \rVert_F$  mean(sd),
 for 50 multi-study simulation replicates. Note: we were unable to run GS and ADVI for some scenarios due to computation time and memory limitations, as the maximum allocation of 24 hours of run time and 128Gb of RAM for each iteration was insufficient. Additionally, we are unable to run ECM in scenarios with $P \geq N$. This is indicated with \longdash[5] in the table.}\label{table-multistudy-F}
\begin{tabular}{lc|rrr|rrr}
\multicolumn{2}{c}{} &
\multicolumn{3}{c}{S = 5} & \multicolumn{3}{c}{S = 10}\\
Method & $\text{N}_s$ & P= 100 & P=500 & P=5000 & P=100 & P=500 & P=5000 \\
\cline{1-8}
GS & 100 & 9.75(0.9) & 48.8(7) & \multicolumn{1}{c|}{\longdash[5]} & 8.87(1) & 45.9(6.5) & \multicolumn{1}{c}{\longdash[5]}\\
CAVI & 100 & 12.8(1.1) & 60.4(5) & 588(42) & 11.4(1.3) & 56.4(4.4) & 558(41)\\
SVI-05 & 100 & 14.1(1.1) & 62.7(5.1) & 595(43) & 12.8(1.5) & 58.8(4.9) & 566(44)\\
SVI-02 & 100 & 14.9(1) & 67.8(4.9) & 652(42) & 13.6(1.5) & 64.3(4.9) & 628(44)\\
SVI-005 & 100 & 22.6(2.2) & 128(12) & 1320(110) & 21.6(2.3) & 124(12) & 1270(120)\\
ECM & 100 & \multicolumn{1}{c}{\longdash[5]} & \multicolumn{1}{c}{\longdash[5]} & \multicolumn{1}{c|}{\longdash[5]} & \multicolumn{1}{c}{\longdash[5]} & \multicolumn{1}{c}{\longdash[5]} & \multicolumn{1}{c}{\longdash[5]} \\
ADVI & 100 & 174(96) & \multicolumn{1}{c}{\longdash[5]} & \multicolumn{1}{c|}{\longdash[5]} & 166(30) & \multicolumn{1}{c}{\longdash[5]} & \multicolumn{1}{c}{\longdash[5]}\\
\specialrule{\cmidrulewidth}{0pt}{0.5pt}
GS & 500 & 4.43(0.36) & 20.3(1.9) & \multicolumn{1}{c|}{\longdash[5]} & 4.22(0.4) & 19.3(1.7) & \multicolumn{1}{c}{\longdash[5]}\\
CAVI & 500 & 8.93(1.2) & 40(7.8) & 356(76) & 7.92(0.95) & 32.7(5.5) & 303(66)\\
SVI-05 & 500 & 10.3(1.3) & 43.2(7.6) & 371(76) & 9.81(1.4) & 36.4(6.4) & 318(68)\\
SVI-02 & 500 & 9.93(1.2) & 44.3(7.3) & 386(72) & 9.64(1.4) & 37.8(6.1) & 335(64)\\
SVI-005 & 500 & 9.49(1) & 48.3(5.9) & 486(53) & 9.24(1) & 44.4(5.1) & 444(50)\\
ECM & 500 & 17.7(1) & \multicolumn{1}{c}{\longdash[5]} & \multicolumn{1}{c|}{\longdash[5]} & 14.6(0.77) & \multicolumn{1}{c}{\longdash[5]} & \multicolumn{1}{c}{\longdash[5]}\\
ADVI & 500 & 111(40) & \multicolumn{1}{c}{\longdash[5]} & \multicolumn{1}{c|}{\longdash[5]} & 104(21) & \multicolumn{1}{c}{\longdash[5]} & \multicolumn{1}{c}{\longdash[5]}\\
\specialrule{\cmidrulewidth}{0pt}{0.5pt}
GS & 1000 & 3.16(0.27) & 14.5(1) & \multicolumn{1}{c|}{\longdash[5]} & 3.12(0.3) & 13.8(1) & \multicolumn{1}{c}{\longdash[5]}\\
CAVI & 1000 & 8.02(1.8) & 37.6(4.6) & 318(85) & 7.03(0.47) & 30.6(6.2) & 259(67)\\
SVI-05 & 1000 & 8.88(1.6) & 40.6(4.8) & 335(86) & 8.51(0.89) & 34.1(7) & 279(72)\\
SVI-02 & 1000 & 8.48(1.6) & 40.8(4.7) & 342(82) & 8.08(0.79) & 34.9(6.8) & 287(72)\\
SVI-005 & 1000 & 7.76(1.3) & 40.4(4) & 390(69) & 7.42(0.6) & 35.6(5.8) & 343(60)\\
ECM & 1000 & 16.1(0.46) & 76.1(1.5) & \multicolumn{1}{c|}{\longdash[5]} & 17.3(0.71) & \multicolumn{1}{c}{\longdash[5]} & \multicolumn{1}{c}{\longdash[5]}\\
ADVI & 1000 & 92.4(42) & \multicolumn{1}{c}{\longdash[5]} & \multicolumn{1}{c|}{\longdash[5]} & 120(14) & \multicolumn{1}{c}{\longdash[5]} & \multicolumn{1}{c}{\longdash[5]}\\
\bottomrule
\end{tabular}
\end{table}

\begin{table}[ht!]
\tiny
\centering
\caption{\it \small Additional results of \S 4.2. Average estimation accuracy, reported as $\lVert \boldsymbol{\Phi} - \hat{\boldsymbol{\Phi}} \rVert_F$  mean(sd),
 for 50 multi-study simulation replicates. Note: we were unable to run GS and ADVI for some scenarios due to computation time and memory limitations, as the maximum allocation of 24 hours of run time and 128Gb of RAM for each iteration was insufficient. Additionally, we are unable to run ECM in scenarios with $P \geq N$. This is indicated with \longdash[5] in the table.}\label{table-multistudy-F}
\begin{tabular}{lc|rrr|rrr}
\multicolumn{2}{c}{} &
\multicolumn{3}{c}{S = 5} & \multicolumn{3}{c}{S = 10}\\
Method & $\text{N}_s$ & P= 100 & P=500 & P=5000 & P=100 & P=500 & P=5000 \\
\cline{1-8}
GS & 100 & 3.74(0.36) & 17(2.4) & \multicolumn{1}{c|}{\longdash[5]} & 2.29(0.18) & 11.5(1.1) & \multicolumn{1}{c}{\longdash[5]}\\
CAVI & 100 & 4.46(0.49) & 21.6(1.3) & 229(15) & 2.7(0.22) & 13.8(1.1) & 150(13)\\
SVI-05 & 100 & 5.67(0.43) & 23.7(1.1) & 237(12) & 3.82(0.21) & 16.1(0.96) & 163(12)\\
SVI-02 & 100 & 5.99(0.43) & 25.6(1) & 258(11) & 4(0.2) & 17.2(0.84) & 175(10)\\
SVI-005 & 100 & 7.07(0.45) & 35.7(1.4) & 368(16) & 4.81(0.2) & 23(0.82) & 238(7.8)\\
ECM & 100 & \multicolumn{1}{c}{\longdash[5]} & \multicolumn{1}{c}{\longdash[5]} & \multicolumn{1}{c|}{\longdash[5]} & \multicolumn{1}{c}{\longdash[5]} & \multicolumn{1}{c}{\longdash[5]} & \multicolumn{1}{c}{\longdash[5]} \\
ADVI & 100 & 76.6(27) & \multicolumn{1}{c}{\longdash[5]} & \multicolumn{1}{c|}{\longdash[5]} & 56.1(8.1) & \multicolumn{1}{c}{\longdash[5]} & \multicolumn{1}{c}{\longdash[5]}\\
\specialrule{\cmidrulewidth}{0pt}{0.5pt}
GS & 500 & 1.77(0.23) & 7.54(0.75) & \multicolumn{1}{c|}{\longdash[5]} & 1.03(0.092) & 5.41(0.49) & \multicolumn{1}{c}{\longdash[5]}\\
CAVI & 500 & 2.65(0.23) & 13.9(0.68) & 141(7.2) & 1.56(0.13) & 7.42(0.45) & 81.1(3.5)\\
SVI-05 & 500 & 4(0.24) & 16.5(0.74) & 150(6.7) & 2.9(0.12) & 9.64(0.4) & 87.8(3.9)\\
SVI-02 & 500 & 3.87(0.24) & 17.1(0.72) & 157(6.4) & 2.83(0.12) & 10.1(0.38) & 93(3.7)\\
SVI-005 & 500 & 3.7(0.27) & 18.5(0.68) & 190(6) & 2.59(0.12) & 11.5(0.36) & 116(2.5)\\
ECM & 500 & 19.2(0.53) & \multicolumn{1}{c}{\longdash[5]} & \multicolumn{1}{c|}{\longdash[5]} & 13.7(0.48) & \multicolumn{1}{c}{\longdash[5]} & \multicolumn{1}{c}{\longdash[5]}\\
ADVI & 500 & 68.6(27) & \multicolumn{1}{c}{\longdash[5]} & \multicolumn{1}{c|}{\longdash[5]} & 47.2(10) & \multicolumn{1}{c}{\longdash[5]} & \multicolumn{1}{c}{\longdash[5]}\\
\specialrule{\cmidrulewidth}{0pt}{0.5pt}
GS & 1000 & 1.21(0.12) & 5.6(0.44) & \multicolumn{1}{c|}{\longdash[5]} & 0.846(0.088) & 3.83(0.35) & \multicolumn{1}{c}{\longdash[5]}\\
CAVI & 1000 & 2.3(0.18) & 12.5(0.43) & 124(3.7) & 1.3(0.13) & 6.66(0.21) & 66.2(3.2)\\
SVI-05 & 1000 & 3.42(0.2) & 15(0.48) & 134(3.7) & 2.38(0.11) & 8.81(0.26) & 75.5(2.9)\\
SVI-02 & 1000 & 3.23(0.2) & 15.2(0.48) & 138(3.4) & 2.21(0.12) & 9.07(0.24) & 78.3(2.8)\\
SVI-005 & 1000 & 2.89(0.22) & 15.6(0.45) & 157(3.1) & 1.85(0.13) & 9.36(0.26) & 91.8(2.2)\\
ECM & 1000 & 18.6(0.69) & 72.6(4.5) & \multicolumn{1}{c|}{\longdash[5]} & 15.8(1.9) & \multicolumn{1}{c}{\longdash[5]} & \multicolumn{1}{c}{\longdash[5]}\\
ADVI & 1000 & 62.8(29) & \multicolumn{1}{c}{\longdash[5]} & \multicolumn{1}{c|}{\longdash[5]} & 80.3(11) & \multicolumn{1}{c}{\longdash[5]} & \multicolumn{1}{c}{\longdash[5]}\\
\bottomrule
\end{tabular}
\end{table}

\begin{table}[ht!]
\tiny
\centering
\caption{\it \small Additional results of \S 4.2. Average estimation accuracy, reported as $\lVert \boldsymbol{\Lambda}_s - \hat{\boldsymbol{\Lambda}}_s \rVert_F$  mean(sd),
 for 50 multi-study simulation replicates. Note: we were unable to run GS and ADVI for some scenarios due to computation time and memory limitations, as the maximum allocation of 24 hours of run time and 128Gb of RAM for each iteration was insufficient. Additionally, we are unable to run ECM in scenarios with $P \geq N$. This is indicated with \longdash[5] in the table.}\label{table-multistudy-F}
\begin{tabular}{lc|rrr|rrr}
\multicolumn{2}{c}{} &
\multicolumn{3}{c}{S = 5} & \multicolumn{3}{c}{S = 10}\\
Method & $\text{N}_s$ & P= 100 & P=500 & P=5000 & P=100 & P=500 & P=5000 \\
\cline{1-8}
GS & 100 & 8.99(0.97) & 45.6(7.4) & \multicolumn{1}{c|}{\longdash[5]} & 8.53(1.1) & 44.4(6.8) & \multicolumn{1}{c}{\longdash[5]}\\
CAVI & 100 & 12.4(1.3) & 57.9(7.2) & 558(71) & 11.3(1.3) & 55.1(4.9) & 542(51)\\
SVI-05 & 100 & 14(1.5) & 60.4(7.7) & 563(72) & 12.8(1.8) & 57.5(5.9) & 547(55)\\
SVI-02 & 100 & 14.6(1.5) & 65.1(7.2) & 616(68) & 13.5(1.7) & 62.8(5.8) & 608(53)\\
SVI-005 & 100 & 22.2(2.4) & 125(13) & 1280(120) & 21.4(2.4) & 122(12) & 1250(120)\\
ECM & 100 & \multicolumn{1}{c}{\longdash[5]} & \multicolumn{1}{c}{\longdash[5]} & \multicolumn{1}{c|}{\longdash[5]} & \multicolumn{1}{c}{\longdash[5]} & \multicolumn{1}{c}{\longdash[5]} & \multicolumn{1}{c}{\longdash[5]} \\
ADVI & 100 & 152(95) & \multicolumn{1}{c}{\longdash[5]} & \multicolumn{1}{c|}{\longdash[5]} & 154(33) & \multicolumn{1}{c}{\longdash[5]} & \multicolumn{1}{c}{\longdash[5]}\\
\specialrule{\cmidrulewidth}{0pt}{0.5pt}
GS & 500 & 4.04(0.38) & 18.8(2) & \multicolumn{1}{c|}{\longdash[5]} & 4.08(0.41) & 18.5(1.8) & \multicolumn{1}{c}{\longdash[5]}\\
CAVI & 500 & 9.23(1.5) & 41(11) & 354(110) & 8.09(1) & 32.8(6.2) & 301(78)\\
SVI-05 & 500 & 10.9(1.7) & 44.2(11) & 369(110) & 10.2(1.7) & 36.5(7.5) & 314(80)\\
SVI-02 & 500 & 10.4(1.6) & 45.2(11) & 382(110) & 10(1.6) & 37.8(7.2) & 330(76)\\
SVI-005 & 500 & 9.83(1.3) & 48.7(9.1) & 474(87) & 9.5(1.2) & 44.1(6) & 436(60)\\
ECM & 500 & 14.7(0.76) & \multicolumn{1}{c}{\longdash[5]} & \multicolumn{1}{c|}{\longdash[5]}& 13.2(0.89) & \multicolumn{1}{c}{\longdash[5]} & \multicolumn{1}{c}{\longdash[5]}\\
ADVI & 500 & 80.7(30) & \multicolumn{1}{c}{\longdash[5]} & \multicolumn{1}{c|}{\longdash[5]} & 89.7(21) & \multicolumn{1}{c}{\longdash[5]} & \multicolumn{1}{c}{\longdash[5]}\\
\specialrule{\cmidrulewidth}{0pt}{0.5pt}
GS & 1000 & 2.91(0.28) & 13.4(1.1) & \multicolumn{1}{c|}{\longdash[5]} & 2.99(0.31) & 13.2(1) & \multicolumn{1}{c}{\longdash[5]}\\
CAVI & 1000 & 8.46(2.3) & 39.4(6.5) & 323(120) & 7.16(0.51) & 31.2(7) & 260(77)\\
SVI-05 & 1000 & 9.53(2.4) & 42.4(7.1) & 339(120) & 8.85(1.1) & 34.7(8.1) & 279(83)\\
SVI-02 & 1000 & 9.05(2.3) & 42.5(6.9) & 346(120) & 8.37(0.95) & 35.4(7.9) & 287(83)\\
SVI-005 & 1000 & 8.21(1.9) & 42(6.2) & 389(100) & 7.61(0.69) & 36(6.9) & 339(71)\\
ECM & 1000 & 14.3(0.77) & 64(2.4) & \multicolumn{1}{c|}{\longdash[5]} & 12.7(1.3) &\multicolumn{1}{c}{\longdash[5]} & \multicolumn{1}{c}{\longdash[5]}\\
ADVI & 1000 & 60.7(28) & \multicolumn{1}{c}{\longdash[5]} & \multicolumn{1}{c|}{\longdash[5]} & 81.4(13) & \multicolumn{1}{c}{\longdash[5]} & \multicolumn{1}{c}{\longdash[5]}\\
\bottomrule
\end{tabular}
\end{table}

\begin{table}[ht!]
\tiny
\centering
\caption{\it \small Additional results of \S 4.2. Average estimation accuracy, reported as $\lVert \boldsymbol{\Sigma}_s - \hat{\boldsymbol{\Sigma}}_s \rVert_1$  mean(sd),
 for 50 multi-study simulation replicates. Note: we were unable to run GS and ADVI for some scenarios due to computation time and memory limitations, as the maximum allocation of 24 hours of run time and 128Gb of RAM for each iteration was insufficient. Additionally, we are unable to run ECM in scenarios with $P \geq N$. This is indicated with \longdash[5] in the table.}\label{table-multistudy-F}
\begin{tabular}{lc|rrr|rrr}
\multicolumn{2}{c}{} &
\multicolumn{3}{c}{S = 5} & \multicolumn{3}{c}{S = 10}\\
Method & $\text{N}_s$ & P= 100 & P=500 & P=5000 & P=100 & P=500 & P=5000 \\
\cline{1-8}
GS & 100 & 15.2(2.3) & 89.9(18) & \multicolumn{1}{c|}{\longdash[5]} & 14.2(2.5) & 84.4(18) & \multicolumn{1}{c}{\longdash[5]}\\
CAVI & 100 & 19.9(2.3) & 107(16) & 1180(140) & 18.3(2.7) & 99.1(11) & 1150(140)\\
SVI-05 & 100 & 21.9(2.7) & 110(16) & 1190(160) & 20.1(3.3) & 102(12) & 1150(150)\\
SVI-02 & 100 & 22.6(2.6) & 118(16) & 1290(170) & 21.1(3.3) & 112(15) & 1290(160)\\
SVI-005 & 100 & 37.8(6.3) & 251(39) & 3070(440) & 36.1(6.2) & 245(40) & 2960(450)\\
ECM & 100 & \multicolumn{1}{c}{\longdash[5]} & \multicolumn{1}{c}{\longdash[5]} & \multicolumn{1}{c|}{\longdash[5]} & \multicolumn{1}{c}{\longdash[5]} & \multicolumn{1}{c}{\longdash[5]} & \multicolumn{1}{c}{\longdash[5]} \\
ADVI & 100 & 267(180) & \multicolumn{1}{c}{\longdash[5]} & \multicolumn{1}{c|}{\longdash[5]} & 257(55) & \multicolumn{1}{c}{\longdash[5]} & \multicolumn{1}{c}{\longdash[5]}\\
\specialrule{\cmidrulewidth}{0pt}{0.5pt}
GS & 500 & 6.77(1) & 34.5(5) & \multicolumn{1}{c|}{\longdash[5]} & 6.58(0.99) & 33.3(4.9) & \multicolumn{1}{c}{\longdash[5]}\\
CAVI & 500 & 14.6(2.3) & 78.2(23) & 800(260) & 12.8(1.6) & 62.1(17) & 662(220)\\
SVI-05 & 500 & 16.8(2.8) & 84.2(22) & 830(260) & 15.4(2.5) & 69.3(19) & 690(230)\\
SVI-02 & 500 & 16(2.5) & 84.7(21) & 844(250) & 15.1(2.3) & 70.3(18) & 712(220)\\
SVI-005 & 500 & 14.6(2.2) & 87.5(15) & 974(170) & 14.2(1.9) & 78.7(13) & 900(150)\\
ECM & 500 & 27.4(1.3) & \multicolumn{1}{c}{\longdash[5]} & \multicolumn{1}{c|}{\longdash[5]} & 23.7(3.1) & \multicolumn{1}{c}{\longdash[5]} & \multicolumn{1}{c}{\longdash[5]}\\
ADVI & 500 & 161(60) & \multicolumn{1}{c}{\longdash[5]} & \multicolumn{1}{c|}{\longdash[5]} & 158(35) & \multicolumn{1}{c}{\longdash[5]} & \multicolumn{1}{c}{\longdash[5]}\\
\specialrule{\cmidrulewidth}{0pt}{0.5pt}
GS & 1000 & 4.63(0.57) & 24.1(3.1) & \multicolumn{1}{c|}{\longdash[5]} & 4.76(0.74) & 23.2(3) & \multicolumn{1}{c}{\longdash[5]}\\
CAVI & 1000 & 13.6(3.4) & 74.3(14) & 756(250) & 11.6(1.4) & 58.6(17) & 585(200)\\
SVI-05 & 1000 & 14.8(3) & 81.2(14) & 797(250) & 14(2.3) & 65.7(18) & 627(210)\\
SVI-02 & 1000 & 14.1(2.9) & 80.9(14) & 801(250) & 13.3(2.1) & 66.3(18) & 633(220)\\
SVI-005 & 1000 & 12.5(2.7) & 78.5(12) & 854(230) & 11.8(1.6) & 65.2(15) & 705(180)\\
ECM & 1000 & 23.3(1.2) & 138(9.5) & \multicolumn{1}{c|}{\longdash[5]} & 27.9(2.2) & \multicolumn{1}{c}{\longdash[5]} & \multicolumn{1}{c}{\longdash[5]}\\
ADVI & 1000 & 137(63) & \multicolumn{1}{c}{\longdash[5]} & \multicolumn{1}{c|}{\longdash[5]} & 187(26) & \multicolumn{1}{c}{\longdash[5]} & \multicolumn{1}{c}{\longdash[5]}\\
\bottomrule
\end{tabular}
\end{table}

\begin{table}[ht!]
\tiny
\centering
\caption{\it \small Additional results of \S 4.2. Average estimation accuracy, reported as $\lVert \boldsymbol{\Phi} - \hat{\boldsymbol{\Phi}} \rVert_1$  mean(sd),
 for 50 multi-study simulation replicates. Note: we were unable to run GS for scenarios with $P=5000$ due to memory limitations, as the allocated 128Gb of RAM for each algorithm was insufficient. This is indicated with \longdash[5] in the table.}\label{table-multistudy-F}
\begin{tabular}{lc|rrr|rrr}
\multicolumn{2}{c}{} &
\multicolumn{3}{c}{S = 5} & \multicolumn{3}{c}{S = 10}\\
Method & $\text{N}_s$ & P= 100 & P=500 & P=5000 & P=100 & P=500 & P=5000 \\
\cline{1-8}
GS & 100 & 5.92(0.98) & 29.6(5) & \multicolumn{1}{c|}{\longdash[5]} & 3.84(0.63) & 19.8(2.8) & \multicolumn{1}{c}{\longdash[5]}\\
CAVI & 100 & 7.14(1.3) & 38.2(4.1) & 460(40) & 4.55(0.78) & 24.4(3.1) & 320(48)\\
SVI-05 & 100 & 9.73(1.6) & 42.9(5.5) & 536(78) & 6.96(0.97) & 29.8(3.7) & 368(52)\\
SVI-02 & 100 & 10.1(1.6) & 45.3(5.3) & 564(75) & 7.07(0.98) & 31(3.8) & 388(60)\\
SVI-005 & 100 & 11.1(1.4) & 60.5(6) & 725(77) & 7.77(0.99) & 36.9(3.6) & 464(41)\\
ECM & 100 & \multicolumn{1}{c}{\longdash[5]} & \multicolumn{1}{c}{\longdash[5]} & \multicolumn{1}{c|}{\longdash[5]} & \multicolumn{1}{c}{\longdash[5]} & \multicolumn{1}{c}{\longdash[5]} & \multicolumn{1}{c}{\longdash[5]} \\
ADVI & 100 & 118(42) & \multicolumn{1}{c}{\longdash[5]} & \multicolumn{1}{c|}{\longdash[5]} & 106(18) & \multicolumn{1}{c}{\longdash[5]} & \multicolumn{1}{c}{\longdash[5]}\\
\specialrule{\cmidrulewidth}{0pt}{0.5pt}
GS & 500 & 2.88(0.56) & 13.5(2) & \multicolumn{1}{c|}{\longdash[5]} & 1.66(0.26) & 9.62(1.4) & \multicolumn{1}{c}{\longdash[5]}\\
CAVI & 500 & 5.35(0.88) & 33.8(4.6) & 367(28) & 2.81(0.41) & 14.9(1.3) & 192(14)\\
SVI-05 & 500 & 9.51(1.7) & 41.2(5.9) & 403(41) & 5.9(0.7) & 22.4(2.2) & 214(19)\\
SVI-02 & 500 & 9(1.6) & 41.6(5.9) & 408(41) & 5.7(0.7) & 22.6(2.3) & 218(22)\\
SVI-005 & 500 & 8.01(1.5) & 42.9(6.3) & 440(36) & 4.99(0.6) & 23.2(2) & 243(24)\\
ECM & 500 & 32.1(1.6) & \multicolumn{1}{c}{\longdash[5]} & \multicolumn{1}{c|}{\longdash[5]} & 24.6(1.3) & \multicolumn{1}{c}{\longdash[5]} & \multicolumn{1}{c}{\longdash[5]}\\
ADVI & 500 & 114(45) & \multicolumn{1}{c}{\longdash[5]} & \multicolumn{1}{c|}{\longdash[5]} & 82.8(19) & \multicolumn{1}{c}{\longdash[5]} & \multicolumn{1}{c}{\longdash[5]}\\
\specialrule{\cmidrulewidth}{0pt}{0.5pt}
GS & 1000 & 1.91(0.28) & 9.75(1.4) & \multicolumn{1}{c|}{\longdash[5]} & 1.39(0.22) & 6.75(1) & \multicolumn{1}{c}{\longdash[5]}\\
CAVI & 1000 & 4.51(0.4) & 34.3(2.5) & 359(25) & 2.17(0.3) & 14.9(1.1) & 175(12)\\
SVI-05 & 1000 & 6.93(0.62) & 42.2(3.1) & 401(47) & 4.77(0.69) & 20.9(1.5) & 206(23)\\
SVI-02 & 1000 & 6.51(0.64) & 42.1(3.1) & 402(50) & 4.32(0.66) & 21(1.6) & 208(23)\\
SVI-005 & 1000 & 5.69(0.6) & 42.3(3.2) & 420(62) & 3.42(0.55) & 20.6(1.3) & 217(18)\\
ECM & 1000 & 29.4(1.4) & 134(12) & \multicolumn{1}{c|}{\longdash[5]} & 28.9(4.1) & \multicolumn{1}{c}{\longdash[5]} & \multicolumn{1}{c}{\longdash[5]}\\
ADVI & 1000 & 102(48) & \multicolumn{1}{c}{\longdash[5]} & \multicolumn{1}{c|}{\longdash[5]} & 144(23) & \multicolumn{1}{c}{\longdash[5]} & \multicolumn{1}{c}{\longdash[5]}\\
\bottomrule
\end{tabular}
\end{table}

\begin{table}[ht!]
\tiny
\centering
\caption{\it \small Additional results of \S 4.2.  Average estimation accuracy, reported as $\lVert \boldsymbol{\Lambda}_s - \hat{\boldsymbol{\Lambda}}_s \rVert_{L1}$  mean(sd),
 for 50 multi-study simulation replicates. Note: we were unable to run GS and ADVI for some scenarios due to computation time and memory limitations, as the maximum allocation of 24 hours of run time and 128Gb of RAM for each iteration was insufficient. Additionally, we are unable to run ECM in scenarios with $P \geq N$. This is indicated with \longdash[5] in the table.}\label{table-multistudy-F}
\begin{tabular}{lc|rrr|rrr}
\multicolumn{2}{c}{} &
\multicolumn{3}{c}{S = 5} & \multicolumn{3}{c}{S = 10}\\
Method & $\text{N}_s$ & P= 100 & P=500 & P=5000 & P=100 & P=500 & P=5000 \\
\cline{1-8}
GS & 100 & 14.8(2.3) & 88(19) & \multicolumn{1}{c|}{\longdash[5]} & 14(2.6) & 83.6(18) & \multicolumn{1}{c}{\longdash[5]}\\
CAVI & 100 & 20(2.7) & 108(21) & 1200(230) & 18.4(2.9) & 98.9(12) & 1150(160)\\
SVI-05 & 100 & 23(3.5) & 112(23) & 1210(240) & 20.8(3.9) & 103(14) & 1150(170)\\
SVI-02 & 100 & 23.5(3.3) & 120(22) & 1310(230) & 21.6(3.8) & 112(16) & 1290(180)\\
SVI-005 & 100 & 37.9(6.2) & 250(39) & 3060(440) & 36.3(6.3) & 244(40) & 2960(450)\\
ECM & 100 & \multicolumn{1}{c}{\longdash[5]} & \multicolumn{1}{c}{\longdash[5]} & \multicolumn{1}{c|}{\longdash[5]} & \multicolumn{1}{c}{\longdash[5]} & \multicolumn{1}{c}{\longdash[5]} & \multicolumn{1}{c}{\longdash[5]} \\
ADVI & 100 & 256(180) & \multicolumn{1}{c}{\longdash[5]} & \multicolumn{1}{c|}{\longdash[5]} & 250(57) & \multicolumn{1}{c}{\longdash[5]} & \multicolumn{1}{c}{\longdash[5]}\\
\specialrule{\cmidrulewidth}{0pt}{0.5pt}
GS & 500 & 6.44(1.1) & 33.6(5.1) & \multicolumn{1}{c|}{\longdash[5]} & 6.49(1) & 32.9(4.9) & \multicolumn{1}{c}{\longdash[5]}\\
CAVI & 500 & 16.2(2.8) & 88.2(31) & 836(380) & 13.6(1.8) & 64.6(20) & 674(260)\\
SVI-05 & 500 & 19(3.6) & 94.8(31) & 873(380) & 16.9(2.9) & 72.8(23) & 706(270)\\
SVI-02 & 500 & 18.1(3.3) & 95(30) & 878(380) & 16.4(2.7) & 73.5(22) & 723(260)\\
SVI-005 & 500 & 16.4(2.9) & 93.6(26) & 1000(290) & 15(2.2) & 80.2(16) & 902(180)\\
ECM & 500 & 21.3(2.7) & \multicolumn{1}{c}{\longdash[5]} & \multicolumn{1}{c|}{\longdash[5]} & 20.8(2.6) & \multicolumn{1}{c}{\longdash[5]} & \multicolumn{1}{c}{\longdash[5]}\\
ADVI & 500 & 126(47) & \multicolumn{1}{c}{\longdash[5]} & \multicolumn{1}{c|}{\longdash[5]} & 142(34) & \multicolumn{1}{c}{\longdash[5]} & \multicolumn{1}{c}{\longdash[5]}\\
\specialrule{\cmidrulewidth}{0pt}{0.5pt}
GS & 1000 & 4.46(0.57) & 23.4(3.2) & \multicolumn{1}{c|}{\longdash[5]} & 4.66(0.75) & 22.9(3) & \multicolumn{1}{c}{\longdash[5]}\\
CAVI & 1000 & 15.2(4.2) & 86.6(21) & 797(380) & 12.2(1.6) & 62.6(20) & 607(240)\\
SVI-05 & 1000 & 17.2(4.3) & 93(22) & 834(380) & 15.4(2.8) & 70.8(23) & 653(260)\\
SVI-02 & 1000 & 16.3(4) & 92.1(22) & 835(380) & 14.4(2.5) & 71(23) & 654(260)\\
SVI-005 & 1000 & 14.5(3.7) & 86.6(20) & 870(340) & 12.6(1.9) & 68.3(19) & 710(220)\\
ECM & 1000 & 20(1.3) & 116(8.8) & \multicolumn{1}{c|}{\longdash[5]} & 21.3(3.2) & \multicolumn{1}{c}{\longdash[5]} & \multicolumn{1}{c}{\longdash[5]}\\
ADVI & 1000 & 94(45) & \multicolumn{1}{c}{\longdash[5]} & \multicolumn{1}{c|}{\longdash[5]} & 129(22) & \multicolumn{1}{c}{\longdash[5]} & \multicolumn{1}{c}{\longdash[5]}\\
\bottomrule
\end{tabular}
\end{table}

\FloatBarrier
 
\begin{figure}[h!]
    \centering
    \includegraphics[width=\textwidth]{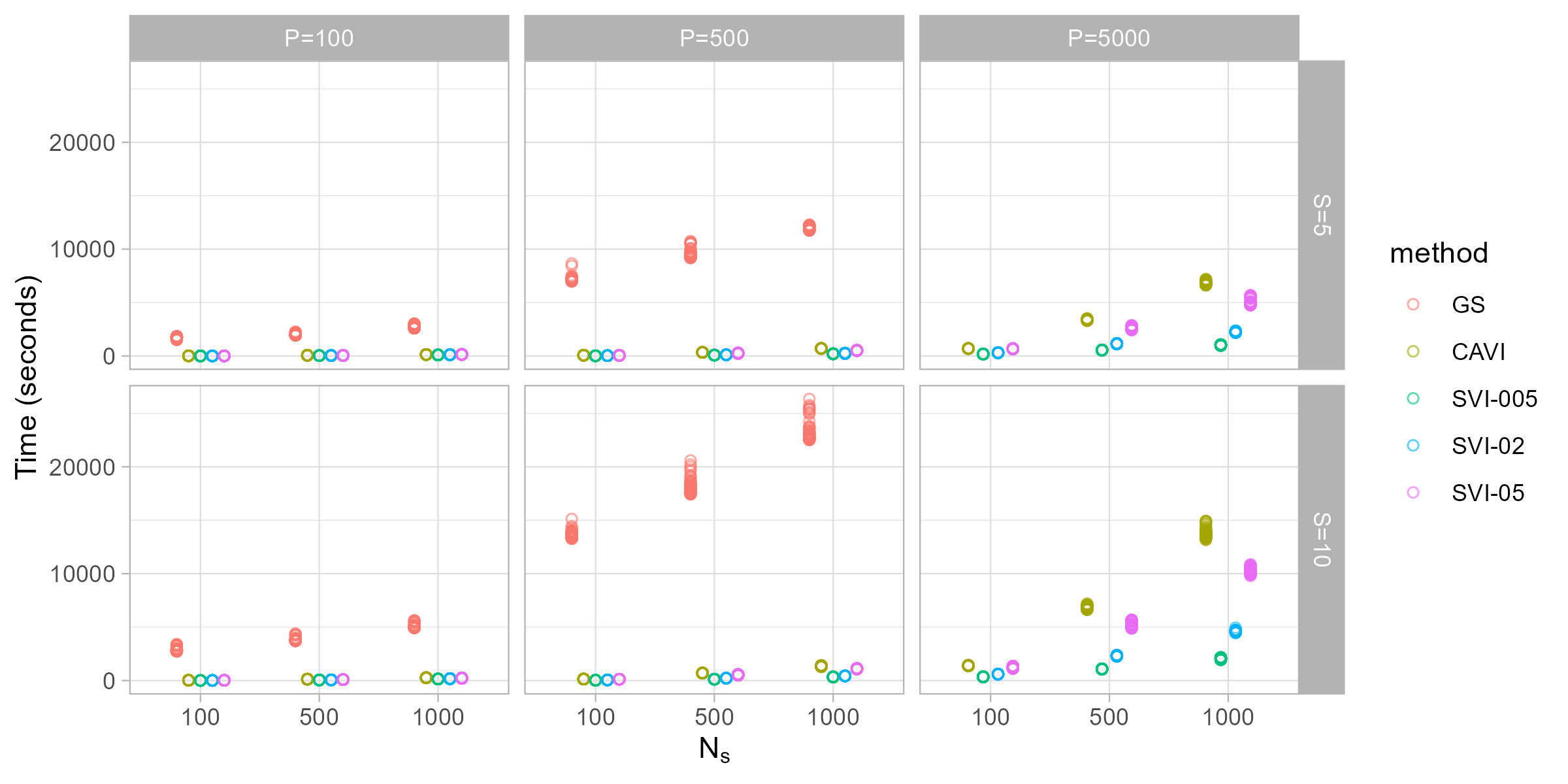}
    \caption{Visualization of results of \S 4.2 for GS, CAVI, and SVI algorithms. Computation time in seconds of GS, CAVI, and SVI Algorithms for multi-study simulations across 50 replicates.  Note: we were unable to run GS for some scenarios due to computation time and memory limitations, as the maximum allocation of 24 hours of run time and 128Gb of RAM for each iteration was insufficient.}
    \label{fig:time_multi_study}
\end{figure}

\begin{figure}[h!]
    \centering
    \includegraphics[width=\textwidth]{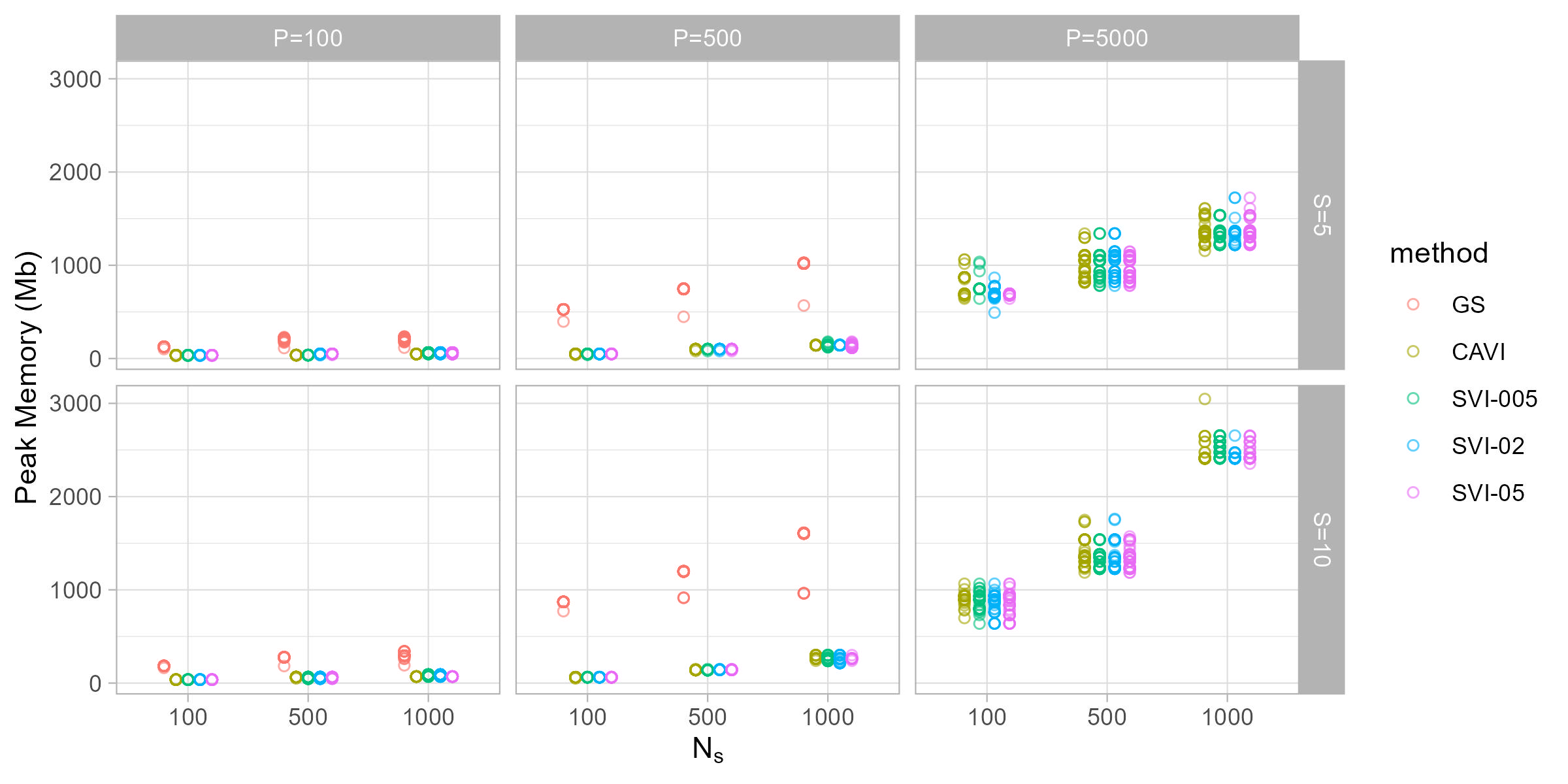}
    \caption{Visualization of results of \S 4.2 for GS, CAVI, and SVI algorithms. Peak RAM usage in Mb of GS, CAVI, and SVI Algorithms for multi-study simulations across 50 replicates. Note: we were unable to run GS for some scenarios due to computation time and memory limitations, as the maximum allocation of 24 hours of run time and 128Gb of RAM for each iteration was insufficient.}
    \label{fig:memory_multi_study}
\end{figure}

\begin{figure}[h!]
    \centering
    \includegraphics[width=\textwidth]{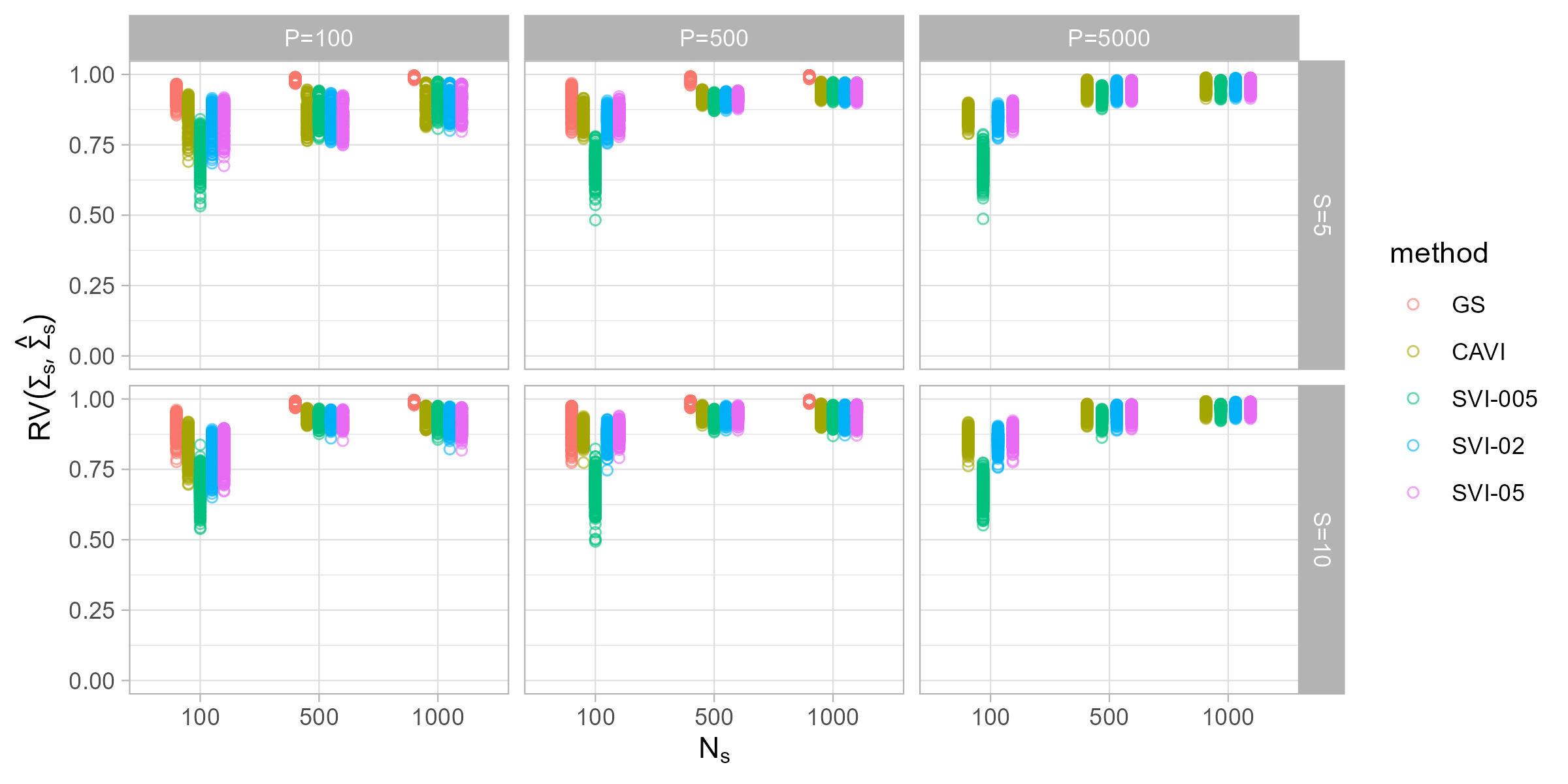}
    \caption{Visualization of results of \S 4.2 for GS, CAVI, and SVI algorithms. RV coefficient between estimated $\boldsymbol{\Sigma}=\boldsymbol{\Phi}\boldsymbol{\Phi}^\intercal + \boldsymbol{\Lambda}_s\boldsymbol{\Lambda}_s^\intercal + \boldsymbol{\Psi}_s$ and the truth using GS, CAVI, and SVI Algorithms for single-study simulations across 50 replicates.  Note: we were unable to run GS for some scenarios due to computation time and memory limitations, as the maximum allocation of 24 hours of run time and 128Gb of RAM for each iteration was insufficient.}
    \label{fig:sigma_rv_multi_study}
\end{figure}

\begin{figure}[h!]
    \centering
    \includegraphics[width=\textwidth]{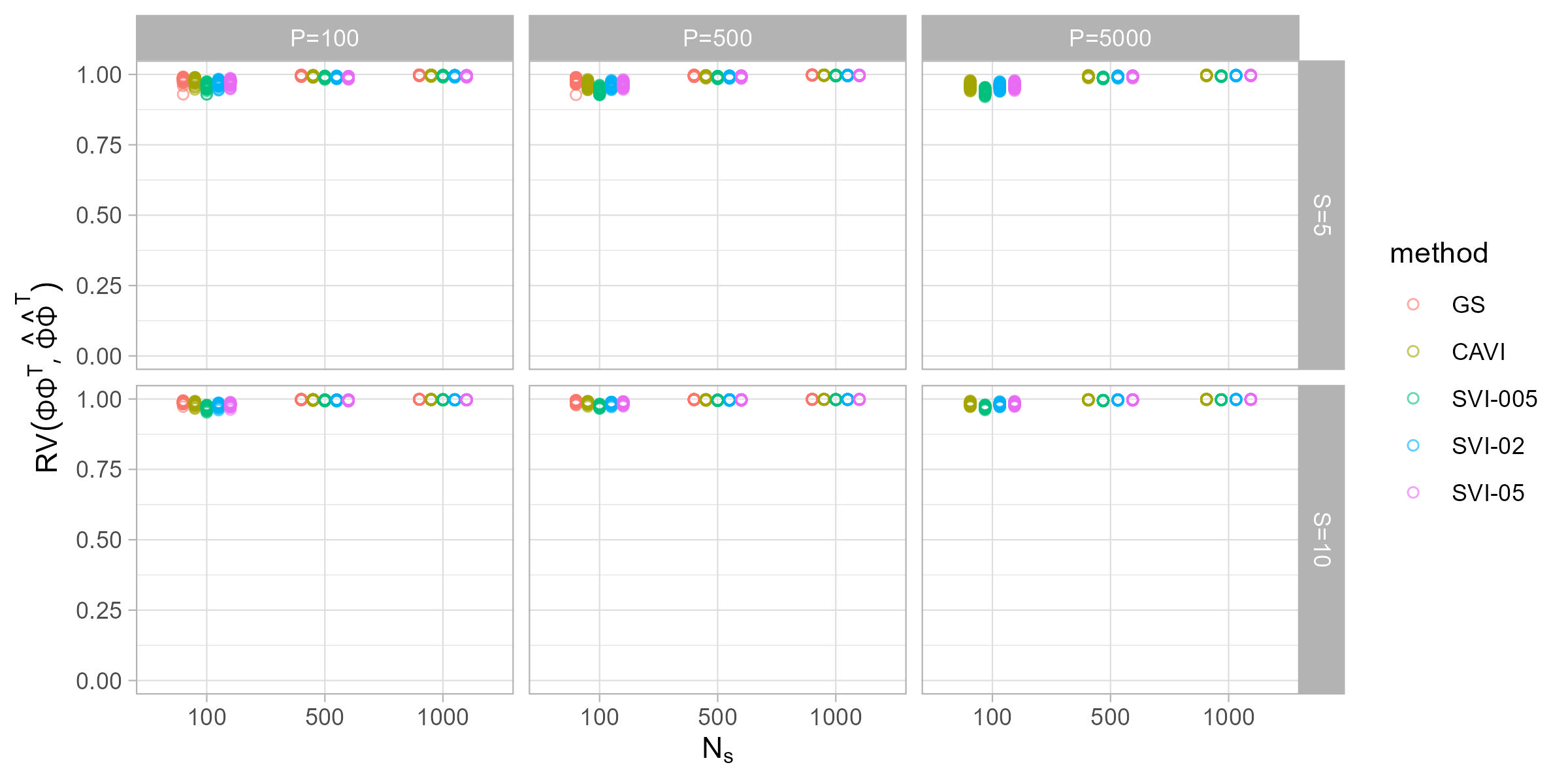}
    \caption{Visualization of results of \S 4.2 for GS, CAVI, and SVI algorithms. RV coefficient between estimated $\boldsymbol{\Phi}\boldsymbol{\Phi}^\intercal$ and the truth using GS, CAVI, and SVI Algorithms for single-study simulations across 50 replicates. Note: we were unable to run GS for some scenarios due to computation time and memory limitations, as the maximum allocation of 24 hours of run time and 128Gb of RAM for each iteration was insufficient.}
    \label{fig:phi_rv_multi_study}
\end{figure}

\begin{figure}[h!]
    \centering
    \includegraphics[width=\textwidth]{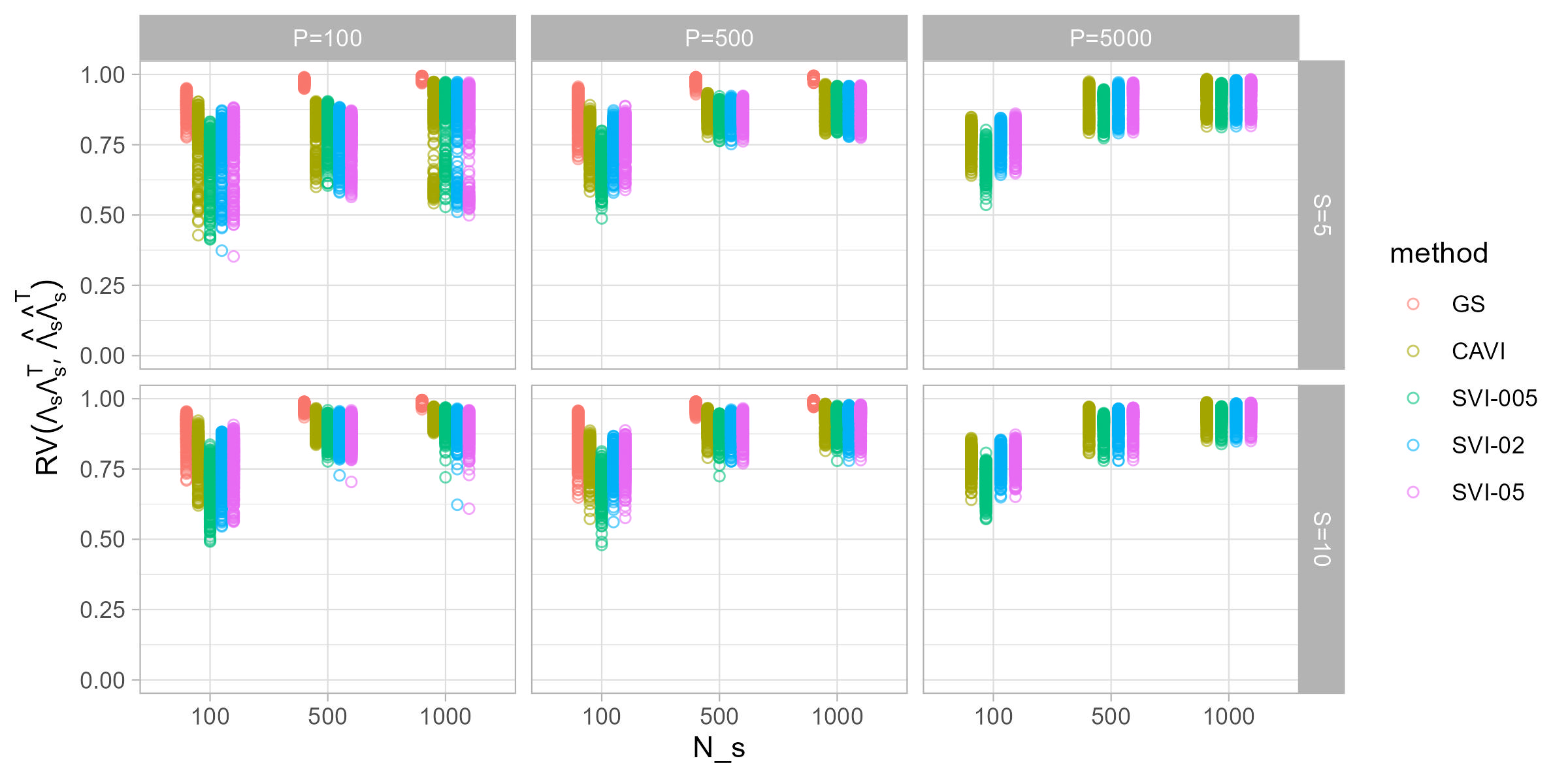}
    \caption{Visualization of results of \S 4.2 for GS, CAVI, and SVI algorithms. RV coefficient between estimated $ \boldsymbol{\Lambda}_s\boldsymbol{\Lambda}_s^\intercal$ and the truth using GS, CAVI, and SVI Algorithms for single-study simulations across 50 replicates. Note: we were unable to run GS for some scenarios due to computation time and memory limitations, as the maximum allocation of 24 hours of run time and 128Gb of RAM for each iteration was insufficient.}
    \label{fig:lambda_rv_multi_study}
\end{figure}
\FloatBarrier

\newpage

\section{Prediction estimates and uncertainty quantification}\label{sec:prediction}

 In this section, we compare our VB algorithms' prediction accuracy and coverage for (single-study) FA with the GS proposed in \cite{bhattacharya_sparse_2011}. Following \cite{bhattacharya_sparse_2011}, we simulate $\boldsymbol{\Lambda}$ and $\boldsymbol{\Psi}$ as in \S 4.1 of the main paper with $P=100, 500$ and a sparsity level of 67\%, fixing two random entries of the first row of $\boldsymbol{\Lambda}$ equal to 1 and -1. The remaining entries of the first row of $\boldsymbol{\Lambda}$ are equal to 0. 
 
 We partition the data as $\mathbf{x}_i = \left(y_i, \mathbf{z}_i^\intercal \right)^\intercal$, where  $y_i \in \mathbb R$ is a response and $\mathbf{z}_i \in \mathbb{R}^{P-1}$ are predictors. 
 Given a training set $(\mathbf{x}_1,\ldots, \mathbf{x}_N)$ of size $N$,
 the aim is to predict $M \geq 1$ future values of the response $(y_{N+1}, \ldots, y_{N+M})$ conditional on observed predictors, $(\mathbf{z}_{N+1}, \ldots, \mathbf{z}_{N+M})$ associated to them.   Thus, we generate $N= M =100$ training and testing samples. Each observation $\mathbf x_i,$ for $i=1,\ldots N+M,$ is sampled independently from $\mathcal{N}\left(\mathbf{0}, \boldsymbol{\Sigma}\right)$, where $\boldsymbol{\Sigma} = \boldsymbol{\Lambda}\boldsymbol{\Lambda}^\intercal + \boldsymbol{\Psi}$. By the properties of the multivariate normal distribution, we have:
\begin{equation}
    y_i \mid \mathbf{z}_i, \boldsymbol{\Sigma} \sim \mathcal{N}\left(\boldsymbol{\Sigma}_{yz}^\intercal\boldsymbol{\Sigma}_z^{-1} \mathbf{z}_i, \sigma^2_{y} -  \boldsymbol{\Sigma}_{yz}^\intercal \boldsymbol{\Sigma}_z^{-1}  \boldsymbol{\Sigma}_{yz}\right), \label{eq:pred_new_data}
\end{equation}
where $\boldsymbol{\Sigma}_z$ is the covariance matrix of $\mathbf{z}_i$, and $\boldsymbol{\Sigma}_{yz}$ is the covariance of $y_i$ and $\mathbf{z}_i$, and $\sigma_y^2$ is the variance of $y$. Note that these matrices partition the covariance of $\mathbf{x}_i$ as: 
\begin{equation}
    \Cov(\mathbf{x}_i) = \boldsymbol{\Sigma} = \begin{pmatrix}
            \sigma^2_y & \boldsymbol{\Sigma}_{yz}^\intercal \\
            \boldsymbol{\Sigma}_{yz} & \boldsymbol{\Sigma}_z
       \end{pmatrix}.
       \label{eq:cov_block}
\end{equation}
We can obtain samples from the predictive distribution of $y_i | \mathbf{z}_i,$ for each $i=N+1,\ldots, N+M$, using Monte Carlo integration. First, we (approximately) sample $\boldsymbol{\Sigma}$ from its posterior distribution $p(\boldsymbol{\Sigma}| \{\mathbf{x}_i\}_{i=1}^N),$ and then plug-in these samples in \eqref{eq:pred_new_data}. In other words, to produce $R$ samples from $y_i | \mathbf{z}_i$ we can use the following steps:

\begin{enumerate}
    \item Get a sample  $\{(\boldsymbol{\Phi}^{(r)}, \boldsymbol{\Psi}^{(r)})\}_{r=1}^R$  from the posterior $p(\boldsymbol{\Phi}, \boldsymbol{\Psi}| \{\mathbf{x}_i\}_{i=1}^N)$ by either:
    \begin{enumerate}
        \item[(i)] Using GS: Take $R$ samples from the MCMC chain of $(\boldsymbol{\Phi}, \boldsymbol{\Psi})$ after burn-in (see Section~{3} of~\cite{bhattacharya_sparse_2011} for the details of the sampling algorithm)
        \item[(ii)] Using a VB method: Draw $R$ independent samples from $$\boldsymbol{\Phi}^{(r)}\sim \prod_{p=1}^P q(\boldsymbol{\Lambda}_p; \boldsymbol{\mu}^{*\Lambda }_p, \boldsymbol{\Sigma}^{*\Lambda }_p)  \text{ and }
        \boldsymbol{\Psi}^{(r)}\sim \prod_{p=1}^P q(\boldsymbol{\psi}^{-2}_p; \alpha^{*\psi}_p, \beta^{*\psi}_p),$$
        for $r=1,\ldots,R.$
    \end{enumerate}
        The optimal variational parameters 
        $(\{\boldsymbol{\mu}^{*\Lambda }_p\}, \{\boldsymbol{\Sigma}^{*\Lambda }_p\},\{\alpha^{*\psi}_p\}, \{\beta^{*\psi}_p\})$ in (ii) can be obtained using the CAVI (Algorithm S1) or the SVI (Algorithm S2).
    \item For $r=1,\ldots, R,$ use the samples in step 1 to compute  $\boldsymbol{\Sigma}^{(r)} = \boldsymbol{\Lambda}^{(r)}[\boldsymbol{\Lambda}^{(r)}]^\intercal + \boldsymbol{\Psi}^{(r)}$ as well as $([\sigma^{(r)}]^2, \boldsymbol{\Sigma}^{(r)}_{yz},\boldsymbol{\Sigma}^{(r)}_{z})$ defined in \eqref{eq:cov_block}.
    \item Get $R$ predicted values for $y_i$ sampling from 
    $$
    \hat y^{(r)}_i \mid \mathbf{z}_i, \boldsymbol{\Sigma}^{(r)} \sim \mathcal{N}\left([\boldsymbol{\Sigma}^{(r)}_{yz}]^\intercal[\boldsymbol{\Sigma}^{(r)}_z]^{-1} \mathbf{z}_i, [\sigma^{(r)}_{y}]^2 -  [\boldsymbol{\Sigma}^{(r)}_{yz}]^\intercal [\boldsymbol{\Sigma}^{(r)}_z]^{-1}  \boldsymbol{\Sigma}^{(r)}_{yz}\right),
    $$
    for $r=1,\ldots, R,$ and $i=N+1, \ldots, N+M.$
\end{enumerate}

Using the samples $\{\hat y^{(r)}_i\}_{r=1}^R,$ for $i=N+1,\ldots,N+M$, we compare GS and our variational algorithms with the following four metrics: 
\begin{enumerate}

\item \textbf{Coverage:} The proportion of 95\% prediction intervals for  $\hat y_i$ in the test set that contains the true $y_i,$ 
$ M^{-1}\sum_{i =N+1}^{N+M}  \left[1\{ \hat y_i^{[0.025]} \leq y_i \leq \hat y_i^{[0.975]} \}\right].$ Here $\hat y_i^{[\alpha]}$ indicates the $\alpha$-level quantile of  $\{\hat y^{(r)}_i\}_{r = 1}^R.$

\item \textbf{Interval Length:} The average length of the 95\% prediction intervals for $\{\hat y_i\},$ $M^{-1}\sum_{i =N+1}^{N+M}  \left[\hat y_i^{[0.975]} - \hat y_i^{[0.025]}\right].$

\item \textbf{MSE:} The mean squared error defined as $M^{-1} \sum_{i = N+1}^{N+M} \left[ R^{-1} \sum_{r=1}^R (y_i -\hat y^{(r)}_i)^2\right].$
\end{enumerate}

We report the averages and standard deviations of metrics 1--3 with $R=1000$ (Table~\ref{pred-table}). Averages and standard deviations are computed, generating $50$ independent training/testing pairs. The GS and CAVI are comparable in terms of coverage, length of the prediction intervals, and MSE. The prediction intervals generated by SVI tend to be wider and with a coverage larger than the nominal level of $95\%.$ SVI also presents a larger MSE for the predicted values compared to CAVI and GS. This is not surprising since we are considering a setting with a sample size $(N=100)$ smaller or equal to the number of variables $P=100,500$.

\begin{table}[h!]
\caption{Summary of prediction simulations described in Section~\ref{sec:prediction}, mean (sd). Results are averaged across 50 independent training/testing set pairs, each set with a sample size of $100$. For each pair, we consider $1000$ samples from the predictive distribution of each observation in the testing set to compute the prediction interval and MSE.}\label{pred-table}
\centering
\begin{tabular}[t]{rl|lll}
P & Method & Coverage & Interval Length & MSE\\
\specialrule{\cmidrulewidth}{0pt}{0.5pt}
100 & GS & 0.95(0.03) & 3.52(0.23) & 1.61(0.18)\\
100 & CAVI & 0.95(0.03) & 3.53(0.23) & 1.62(0.19)\\
100 & SVI-005 & 0.92(0.04) & 3.68(0.34) & 1.98(0.25)\\
100 & SVI-02 & 0.96(0.02) & 4.05(0.24) & 1.95(0.28)\\
100 & SVI-05 & 0.97(0.02) & 4.09(0.18) & 1.94(0.21)\\
\specialrule{\cmidrulewidth}{0pt}{0.5pt}
500 & GS & 0.93(0.03) & 1.44(0.11) & 0.28(0.03)\\
500 & CAVI & 0.94(0.03) & 1.47(0.11) & 0.29(0.03)\\
500 & SVI-005 & 0.94(0.05) & 2.57(0.19) & 0.91(0.21)\\
500 & SVI-02 & 0.99(0.01) & 2.6(0.12) & 0.67(0.06)\\
500 & SVI-05 & 0.99(0.01) & 2.59(0.13) & 0.64(0.05)\\
\bottomrule
\end{tabular}
\end{table}

\FloatBarrier

\section{Shrinkage per Factor Loadings Column Index}

As discussed in \S 6 of the main text, we find that the shrinkage imposed by the gamma process shrinkage prior~\citep{bhattacharya_sparse_2011} affect the VB posterior distribution differently from the MCMC based one. To illustrate this point, we repeat the simulation study of the main text with a greater number of factors, fixing $J=8$, $J^*=8$, and $N=100$ for the single-study simulations and $K=8$, $K^*=10$, and $N_s=100$ for the multi-study simulations, keeping the dimensions of the study-specific factor loadings $J_s$ the same as before, $J_s=4$ and $J^*_s$ for all studies. Following \cite{bhattacharya_sparse_2011}, we use the proportion of estimated factor loadings near 0 to determine the effective number of factors suggested by the model. For this purpose we consider factor loadings in $(-0.01, 0.01)$ to be near zero. 

\begin{figure}[hp!]
    \centering
    \includegraphics[width=\textwidth]{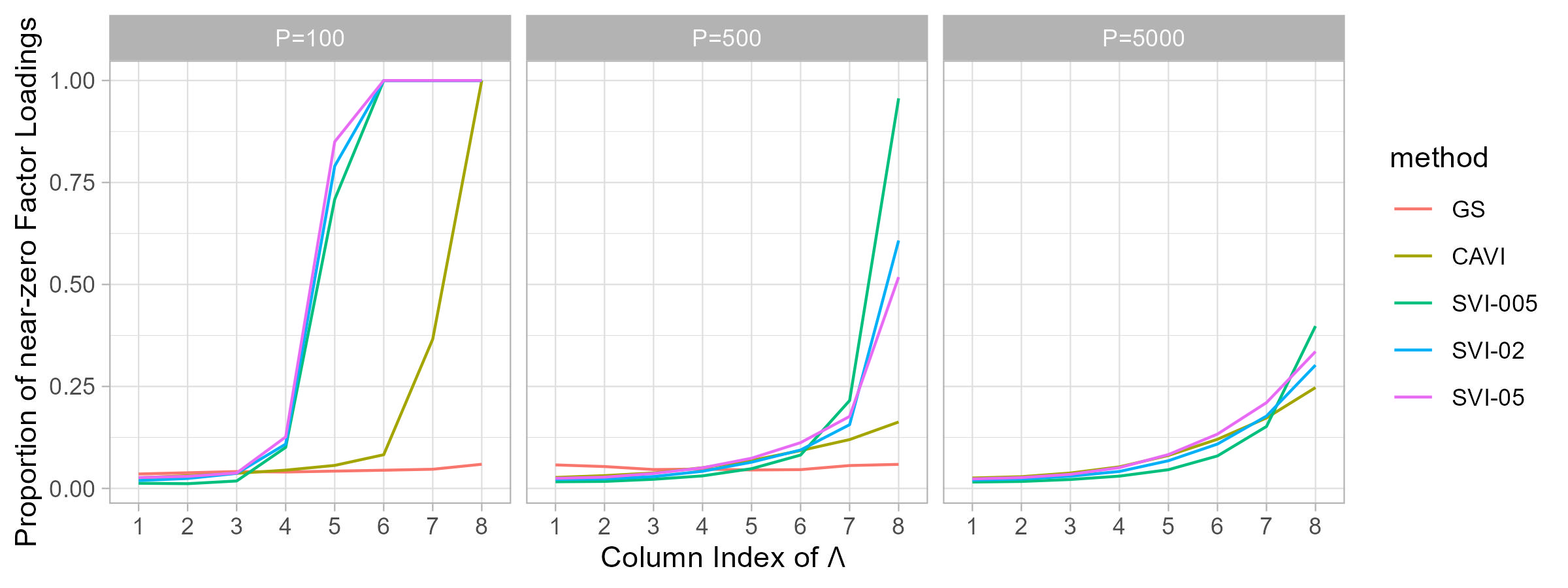}
    \caption{Average proportion of factor loadings $\boldsymbol{\Lambda}$ 
    in $(-0.01, 0.01)$ in single extended study simulations with $J=8$ factors. Note: we were unable to run GS for scenarios with P=5000 due to memory limitations, which we discuss in Section 4 of the main text.}
    \label{column_shrinkage_single_study}
\end{figure}

\begin{figure}[hp!]
    \centering
    \includegraphics[width=\textwidth]{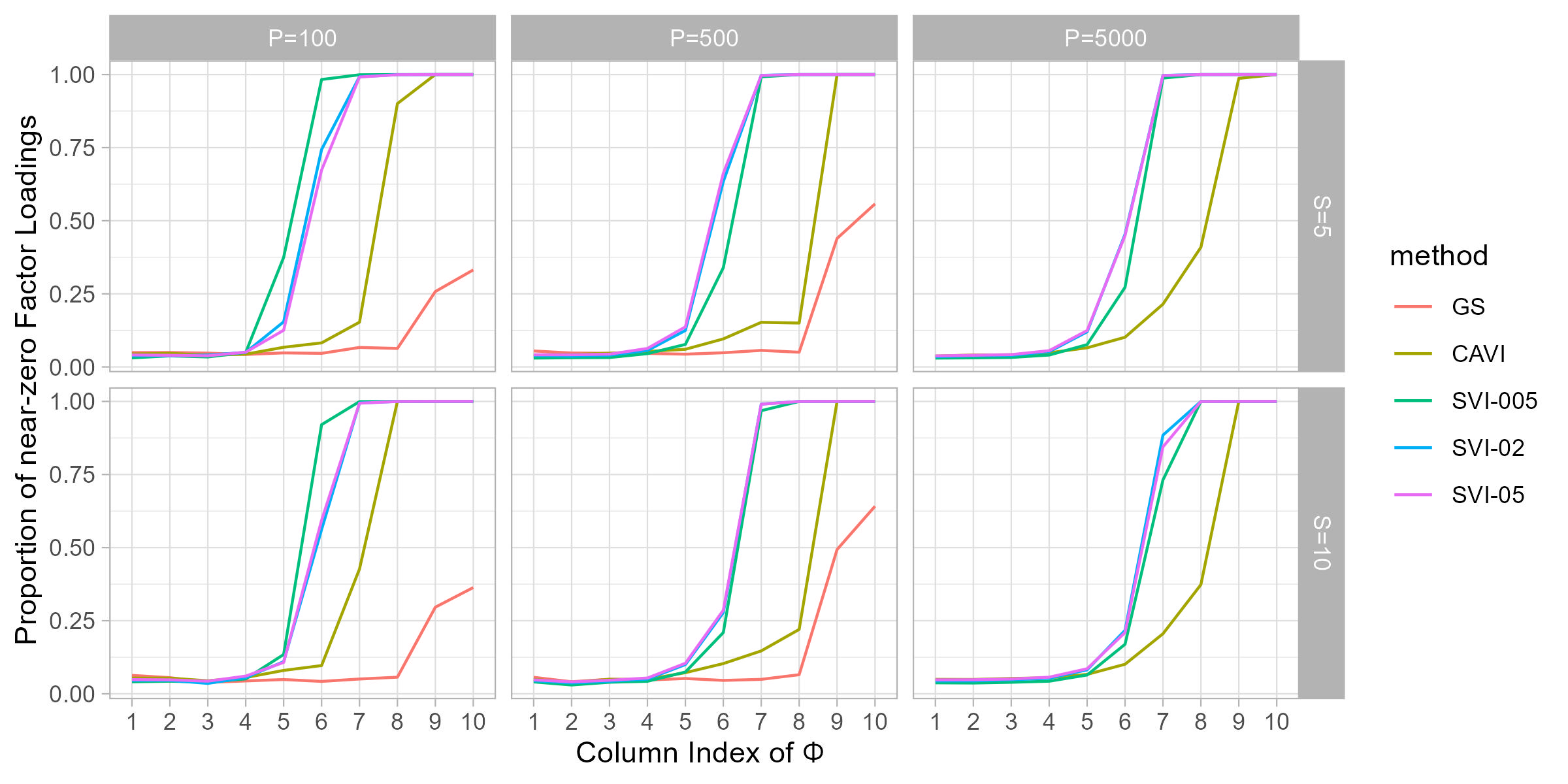}
    \caption{Average proportion of near-zero factor loadings in $\boldsymbol{\Phi}$ in multi-study simulations with $K=8$ factors. We increase the truncated number of factors to be $K^*=10$ account for the additional information provided by combining the multiple studies. Note: we were unable to run GS for scenarios with P=5000 due to memory limitations, which we discuss in Section 5 of the main text.}
    \label{column_shrinkage_multi_study}
\end{figure}

We can see that in both the single-study and multi-study simulations that the shrinkage of factor loadings is greater in estimates obtained by VB than it is for estimates obtained by MCMC. Figure~\ref{column_shrinkage_single_study} shows the proportion of factor loadings near zero for each column index. In the single-study simulations with $P=100$, the effective number of factors is  $8$ and it is correctly recovered by MCMC, while CAVI and SVI estimated on average $7$ and $6$ factors. CAVI and SVI correctly identify the number  factors with $P=500$ and $P=5000$. Note that  although the VB algorithms correctly identify the number of factors when $P=500,$ the proportion of near zero elements is higher that the one of the MCMC algorithm. We observe similar trends in the multi-study simulations as seen in \figurename~\ref{column_shrinkage_multi_study}. In all of the scenarios, the number of effective factors under GS is equal to $K^*=10$, although the number of zero elements spikes for factors 9 and 10. At the same time, the effective number of factors under CAVI was either $7$ or 8 and was between $5$ to $7$ under SVI. 

\newpage

\section{Derivation of the Optimal Variational Parameters}

\subsection{Factor Analysis}

We start by considering Bayesian FA in the single-study setting. The form of the posterior distribution is as follows: 
\begin{align}\label{posterior}
    p\left( \boldsymbol{\theta} | \mathbf{x} \right) \propto & \hspace{0.1cm} p\left(\mathbf{x}|\boldsymbol{\theta}\right)p\left(\boldsymbol{\theta}\right) \nonumber \\
    \propto & \left[ |\boldsymbol{\Psi}|^{-\frac{N}{2}} \prod_{i=1}^N \exp \left\{-\frac{1}{2} \left(\mathbf{x}_i - \boldsymbol{\Lambda}\mathbf{l}_i\right)^\intercal \boldsymbol{\Psi}^{-1} \left(\mathbf{x}_i - \boldsymbol{\Lambda}\mathbf{l}_i\right)\right\}\right] \times \\
    & \left[ \prod_{i=1}^N \exp\left\{ -\frac{1}{2} \mathbf{l}_i^\intercal\mathbf{l}_i\right\}\right]
    \left[ \prod_{p=1}^P \prod_{k=1}^{J^*} \left(\omega_{pj}\tau_j\right)^{1/2} \exp \left\{ -\frac{1}{2} \omega_{pj}\tau_j \lambda_{pj}^2 \right\}\right] \times \nonumber \\
    & \left[ \prod_{p=1}^P \prod_{j=1}^{J^*} \omega_{pj}^{\frac{\nu}{2}-1} \exp \left\{-\frac{\nu}{2}\omega_{pj}\right\}\right] \left[ \prod_{l=1}^{J^*} \delta_l^{a_l - 1} \exp\left\{-\delta_l\right\} \right] \times \nonumber \\
    & \left[ \prod_{p=1}^P \left(\psi_{p}^{-2}\right)^{a^\psi -1} \exp\left\{-b^\psi \psi_p^{-2}\right\}\right] \nonumber
\end{align}

To approximate $p\left( \boldsymbol{\theta} | \boldsymbol{x} \right)$ we take a Variational Bayes approach considering approximations $q(\boldsymbol{\theta}; \boldsymbol{\varphi})$.
Each element of $\boldsymbol{\theta}$, $\theta_j$, is assigned its own variational factor that is characterized by the variational parameter vector $\boldsymbol{\varphi}_j$. In the following sections, we derive closed-form expressions for the varitional parameters which optimize $ELBO(q(\boldsymbol{\theta})).$

In the following we will use $C$ that do not jointly depend on the data and the parameters.
\subsubsection{Optimal Variational Parameters for factor scores}

\begin{align*}
    \log p (\mathbf{l}_i | - ) &= -\frac{1}{2}\left( \mathbf{x}_i - \boldsymbol{\Lambda}\boldsymbol{l}_i\right)^\intercal \boldsymbol{\Psi}^{-1} \left( \mathbf{x}_i - \boldsymbol{\Lambda}\boldsymbol{l}_i\right) -\frac{1}{2}\mathbf{l}_i^\intercal\mathbf{l}_i + C \\
    &= -\frac{1}{2} \mathbf{x}_i^\intercal\boldsymbol{\Psi}^{-1}\mathbf{x}_i + \mathbf{x}_i^\intercal \boldsymbol{\Psi}^{-1}\boldsymbol{\Lambda}\mathbf{l}_i -\frac{1}{2} \mathbf{l}_i^\intercal \boldsymbol{\Phi}^\intercal\boldsymbol{\Psi}^{-1}\boldsymbol{\Phi}\mathbf{l}_i -\frac{1}{2}\mathbf{l}_i^\intercal\mathbf{l}_i + C \\
    &= -\frac{1}{2}\mathbf{l}_i^\intercal \left( \boldsymbol{I}_{J^*} + \boldsymbol{\Lambda}^\intercal\boldsymbol{\Psi}^{-1}\boldsymbol{\Phi}\right)\mathbf{l}_i + \left( \boldsymbol{\Lambda}^\intercal \boldsymbol{\Psi}^{-1} \mathbf{x}_i \right)^\intercal \mathbf{l}_i + C \\
    \mathbb{E}_q \left[\log p (\mathbf{l}_i | - ) \right]
    &= -\frac{1}{2}\mathbf{l}_i^\intercal \left( \boldsymbol{I}_{J^*} + \mathbb{E}\left[\boldsymbol{\Lambda}^\intercal\boldsymbol{\Psi}^{-1}\boldsymbol{\Phi}\right]\right)\mathbf{l}_i + \left( \mathbb{E}\left[\boldsymbol{\Lambda}\right]^\intercal \mathbb{E}\left[\boldsymbol{\Psi}^{-1}\right] \mathbf{x}_i \right)^\intercal \mathbf{l}_i + C
\end{align*}
where $\mathbb{E}\left[\boldsymbol{\Lambda}^\intercal\boldsymbol{\Psi}^{-1}\boldsymbol{\Lambda}\right]=\sum_{p=1}^P \mathbb{E}\left[\psi_{p}^{-2}\right]\mathbb{E}\left[\boldsymbol{\Lambda}_p\boldsymbol{\Lambda}_p^\intercal\right]$. This is the log of the kernel of a multivariate normal density with parameters:
\begin{align*}
    \boldsymbol{\Sigma}_i^l &= \left( \boldsymbol{I}_{J^*} + \mathbb{E}\left[\boldsymbol{\Lambda}^\intercal\boldsymbol{\Psi}^{-1}\boldsymbol{\Lambda}\right]\right)^{-1} \\
    &= \left( \boldsymbol{I}_{J^*} + \sum_{p=1}^P \frac{\alpha_{p}^\psi}{\beta_{p}^\psi} \left(\boldsymbol{\mu}_p^\Lambda\left[\boldsymbol{\mu}_p^\Lambda\right]^\intercal+ \boldsymbol{\Sigma}_p^\Lambda\right)\right)^{-1} \\
    \boldsymbol{\mu}_i^l &= \left( \boldsymbol{I}_{J^*} + \mathbb{E}\left[\boldsymbol{\Lambda}^\intercal\boldsymbol{\Psi}^{-1}\boldsymbol{\Lambda}\right]\right)^{-1}\left( \mathbb{E}\left[\boldsymbol{\Phi}\right]^\intercal \mathbb{E}\left[\boldsymbol{\Psi}^{-1}\right] \mathbf{x}_i \right) \\
    &= \boldsymbol{\Sigma}_i^l \left(\boldsymbol{\mu}_{1}^\Lambda,\ldots,\boldsymbol{\mu}_{P}^\Lambda\right) \text{diag}\left\{\frac{\alpha_{1}^\psi}{\beta_{1}^\psi},\ldots,\frac{\alpha_{P}^\psi}{\beta_{P}^\psi}\right\} \mathbf{x}_i\\
\end{align*}
which implies that the variational factor which maximizes the ELBO with respect to $q(\mathbf{l}_i)$, conditional on the remaining variational factors, is $q(\mathbf{l}_i; \boldsymbol{\varphi}^*)=\mathcal{N}\left(\mathbf{l}_i; \mathbf{\mu}_i^l, \boldsymbol{\Sigma}_i^l \right)$.

\subsubsection{Optimal Variational Parameters for Factor Loadings}
\begin{align*}
    \log p (\boldsymbol{\Lambda}_p|-) &= -\frac{1}{2}\sum_{j=1}^{J^*} \omega_{pj}\tau_j \lambda_{pj}^2 +  -\frac{1}{2} \sum_{i=1}^n  \left( \mathbf{x}_i - \boldsymbol{\Lambda}\mathbf{l}_i \right)^\intercal\boldsymbol{\Psi}^{-1} \left( \mathbf{x}_i - \boldsymbol{\Lambda}\mathbf{l}_i \right) + C \\ 
    &= -\frac{1}{2} \boldsymbol{\Lambda}_p^\intercal \boldsymbol{D}_p \boldsymbol{\Lambda}_p -\frac{1}{2} \sum_{i=1}^n \psi_p^{-2}\left( x_{ip} - \boldsymbol{\Lambda}_p^\intercal\mathbf{l}_i \right)^\intercal \left( x_{ip} - \boldsymbol{\Lambda}_p^\intercal\mathbf{l}_i \right) + C \\
    &= -\frac{1}{2} \boldsymbol{\Lambda}_p^\intercal \boldsymbol{D}_p \boldsymbol{\Lambda}_p -\frac{1}{2} \psi_p^{-2} \sum_{i=1}^n x_{ip}^2 -2x_{ip} \mathbf{l}_i^\intercal\boldsymbol{\Lambda}_p + \boldsymbol{\Lambda}_p^\intercal\mathbf{l}_i\mathbf{l}_i^\intercal\boldsymbol{\Lambda}_p + C\\
    &= -\frac{1}{2} \boldsymbol{\Lambda}_p^\intercal \boldsymbol{D}_p \boldsymbol{\Lambda}_p + -\frac{1}{2}\psi_p^{-2} \boldsymbol{\Lambda}_p^\intercal\left(\sum_{i=1}^n \mathbf{l}_i\mathbf{l}_i^\intercal \right)\boldsymbol{\Lambda}_p + \psi_p^{-2}\left(\sum_{i=1}^n x_{ip}\mathbf{l}_i^\intercal \right) \boldsymbol{\Lambda}_p + C \\
    &= -\frac{1}{2} \boldsymbol{\Lambda}_p^\intercal \boldsymbol{D}_p \boldsymbol{\Lambda}_p -\frac{1}{2} \psi_p^{-2} \boldsymbol{\Lambda}_p^\intercal\left(   \mathbf{l}^\intercal \mathbf{l} \right)\boldsymbol{\Lambda}_p + \left(  \mathbf{l}^\intercal \mathbf{x}_p \right)^\intercal \boldsymbol{\Lambda}_p + C \\
    &= -\frac{1}{2} \boldsymbol{\Lambda}_p^\intercal \left(\boldsymbol{D}_p + \psi_p^{-2} \mathbf{l}^\intercal \mathbf{l}\right) \boldsymbol{\Lambda}_p + \left( \psi_p^{-2} \mathbf{l}^\intercal \mathbf{x}_p \right)^\intercal \boldsymbol{\Lambda}_p + C
\end{align*}
where $\boldsymbol{D}_p=\text{diag}(\omega_{p1}\tau_1,\ldots,\omega_{pJ^*}\tau_{J^*})$, and $\boldsymbol{\Lambda}_p^\intercal$ is row $p$ of $\boldsymbol{\Lambda}$. Next, we take expectations with respect to $q$:
\begin{align*}
    \mathbb{E}_{-\boldsymbol{\Lambda}_p}\left[  \log p (\boldsymbol{\Lambda}_p|-) \right] &= -\frac{1}{2} \boldsymbol{\Lambda}_p^\intercal \left(\mathbb{E}\left[\boldsymbol{D}_p\right] + \mathbb{E}\left[\psi_p^{-2}\right] \mathbb{E}\left[\mathbf{l}^\intercal \mathbf{l}\right]\right) \boldsymbol{\Lambda}_p + \left( \mathbb{E}\left[\psi_p^{-2}\right] \mathbb{E}\left[\mathbf{l}\right]^\intercal \mathbf{x}_p \right)^\intercal \boldsymbol{\Lambda}_p + C
\end{align*}
where $\mathbb{E}\left[\mathbf{l}^\intercal \mathbf{l}\right]=\sum_{i=1}^n \mathbb{E}\left[\mathbf{l}_i\mathbf{l}_i^\intercal\right]$ and $\mathbb{E}\left[\boldsymbol{D}_p\right]=\text{diag}\left(\mathbb{E}\left[\omega_{p1}\right]\mathbb{E}\left[\delta_1\right],\ldots, \mathbb{E}\left[\omega_{pJ^*}\right]\prod_{l=1}^{J^*}\mathbb{E}\left[\delta_l\right]\right)$. This is the log of the kernel of a multivariate normal density with parameters:
\begin{align*}
    \boldsymbol{\Sigma}_p^\Lambda &= \left(\mathbb{E}\left[\boldsymbol{D}_p\right] + \mathbb{E}\left[\psi_p^{-2}\right] \mathbb{E}\left[\mathbf{l}^\intercal \mathbf{l}\right]\right)^{-1} \\
    &= \left(\text{diag}\left\{\frac{\alpha_{p1}^\omega}{\beta_{p1}^\omega}\frac{\alpha_{1}^\delta}{\beta_{1}^\delta},\ldots,\frac{\alpha_{pJ^*}^\omega}{\beta_{pJ^*}^\omega}\prod_{j=1}^{J^*}\frac{\alpha_{j}^\delta}{\beta_{j}^\delta}\right\} + \frac{\alpha_{p}^\psi}{\beta_{p}^\psi}\sum_{i=1}^n\left(\boldsymbol{\mu}_{i}^l\left[\boldsymbol{\mu}_{i}^l\right]^\intercal + \boldsymbol{\Sigma}_i^l \right) \right)^{-1} \\
    \boldsymbol{\mu}_p^\Lambda &= \mathbb{E}\left[\psi_p^{-2}\right]\left(\mathbb{E}\left[\boldsymbol{D}_p\right] + \mathbb{E}\left[\psi_p^{-2}\right] \mathbb{E}\left[\mathbf{l}^\intercal \mathbf{l}\right]\right)^{-1}  \mathbb{E}\left[\mathbf{l}\right]^\intercal \mathbf{x}_p \\
    &= \frac{\alpha_{p}^\psi}{\beta_{p}^\psi}\boldsymbol{\Sigma}_p^\Lambda \left[\boldsymbol{\mu}_{i}^l\right]^\intercal\mathbf{x}_p
\end{align*}
which implies that the variational factor which maximimzes the ELBO with respect to $q(\boldsymbol{\Lambda}_p)$, conditional on the remaining factors, is $q(\boldsymbol{\Lambda}_p;\boldsymbol{\varphi}^*)=\mathcal{N}\left(\boldsymbol{\Lambda}_p;\boldsymbol{\mu}_p^\Lambda, \boldsymbol{\Sigma}_p^\Lambda\right)$.

\subsubsection{Optimal Variational Parameters for local shrinkage terms}
\begin{align*}
    \log p(\omega_{pj}|-) &= \frac{1}{2}\log \omega_{pj} - \frac{1}{2} \omega_{pj}\tau_j\lambda_{pj}^2 + \left(\frac{\nu}{2} - 1\right)\log \omega_{pj} - \frac{\nu}{2} \omega_{pj} + C\\
    &= \left(\frac{\nu + 1}{2}-1\right)\log \omega_{pj} - \left(\frac{\nu + \tau_k\lambda_{pj}^2}{2}\right)\omega_{pj} + C \\
    \mathbb{E}_{-\omega_{pj}}\left[\log p(\omega_{pj}|-)\right] &= \left(\frac{\nu + 1}{2}-1\right)\log \omega_{pj} - \left(\frac{\nu + \mathbb{E}\left[\tau_j\right]\mathbb{E}\left[\lambda_{pj}^2\right]}{2}\right)\omega_{pj} + C \\
\end{align*}
which is the log of the kernel of a gamma density with shape and rate parameters:
\begin{align*}
    \alpha_{pj}^\omega &= \frac{\nu+1}{2} \\
    \beta_{pj}^\omega &= \frac{\nu + \mathbb{E}\left[\tau_j\right]\mathbb{E}\left[\lambda_{pj}^2\right]}{2} \\
    &= \frac{\nu + \left(\prod_{l=1}^j \frac{\alpha_{l}^\delta}{\beta_{l}^\delta}\right)\left(\left[\mu_{pj}^\lambda\right]^2 + \left[\boldsymbol{\Sigma}_{p}^\Lambda\right]_{j,j}\right)}{2}
\end{align*}
This implies that the variational factor which maximizes the ELBO with respect to $q(\omega_{pj})$, conditional on the remaining factors, is $q(\omega_{pj};\boldsymbol{\varphi}^*)=\Gamma(\omega_{pj}; \alpha_{pj}^\omega, \beta_{pj}^\omega)$.

\subsubsection{Optimal Variational Parameters for global shrinkage terms}
Recall that $\tau_j=\prod_{j=1}^l\delta_l$. Also, we use the fact that $\delta_l$ only appears in the prior for $\lambda_{pj}$ in columns greater than or equal to $l$.

\begin{align*}
    \log p (\delta_l|-) &= \left(\sum_{p=1}^P\sum_{j=l}^{J^*}\frac{1}{2}\log\left(\omega_{pl}\tau_l\right) -\frac{1}{2}\omega_{pl}\tau_l\lambda_{pl}^2\right) + (a_l - 1) \log \delta_l - \delta_l + C\\
    &= \left(\sum_{p=1}^P \sum_{j=l}^{J^*} \frac{1}{2}\log \delta_l - \frac{1}{2}\omega_{pl}\tau_l\phi_{pl}^2\right) +  (a_l - 1) \log \delta_l - \delta_l + C \\
    &= \left(\sum_{p=1}^{P}\sum_{j=l}^{J^*} \frac{1}{2}\right) \log \delta_l -\frac{1}{2}\left(\sum_{p=1}^P\sum_{j=l}^{J^*} \omega_{pj}\tau_{j}\lambda_{pj}^2\right) + (a_l - 1) \log \delta_l - \delta_l + C\\
    &= \left( \alpha_l -1 +\sum_{p=1}^P\sum_{j=l}^{J^*} \frac{1}{2} \right)\log \delta_l - \left(1 + \frac{1}{2}\sum_{p=1}^P\sum_{j=l}^{J^*}\omega_{pj}\tau_j\lambda_{pj}^2 \right)\delta_l + C \\
    &= \left(a_l + \frac{P(J^*-l+1)}{2} - 1\right)\log \delta_l - \left(1 + \frac{1}{2}\sum_{p=1}^P\sum_{j=1}^{J^*}\omega_{pj}\left(\prod_{1\leq r \leq l, r \neq l} \delta_r\right)\lambda_{pj}^2 \right)\delta_l + C\\
    \mathbb{E}\left[ \log p (\delta_l|-)\right] &= \left(a_l + \frac{P(J^*-l+1)}{2} - 1\right)\log \delta_l - \left(1 + \frac{1}{2}\sum_{p=1}^P\sum_{j=1}^{J^*}\mathbb{E}[\omega_{pj}]\left(\prod_{1\leq r \leq l, r \neq l} \mathbb{E}\left[\delta_r\right]\right)\mathbb{E}[\lambda_{pj}^2] \right)\delta_l + C
\end{align*}
which is the log of the kernel of a gamma density with shape and rate parameters: 
\begin{align*}
    \alpha_l^\delta &= a_l + \frac{P(J^*-l+1)}{2} \\
    \beta_l^\delta &= 1 + \frac{1}{2}\left(\sum_{p=1}^P\sum_{j=1}^{J^*}\mathbb{E}[\omega_{pj}]\left(\prod_{1\leq r \leq l, r \neq l} \mathbb{E}\left[\delta_r\right]\right)\mathbb{E}[\lambda_{pj}^2]\right) \\
    &= 1 + \frac{1}{2}\left(\sum_{p=1}^P\sum_{j=l}^{J^*} \frac{\alpha_{pj}^\omega}{\beta_{pj}}\left(\prod_{1\leq r \leq l, r\neq l}\frac{\alpha_{r}^\delta}{\beta_{r}^\delta}\right)\left(\left[\mu^\Lambda_{pj}\right]^2 + \left[\boldsymbol{\Sigma}_p^\Phi\right]_{j,j}\right)\right)
\end{align*}
This implies that the variational factor with maximizes the ELBO with respect to $q(\delta_l)$, conditional on the remaining factors, is $q(\delta_l;\boldsymbol{\varphi}^*)=\Gamma(\delta_l;\alpha_l^\delta, \beta_l^\delta)$.

\subsubsection{Optimal Variational Parameters for idiosyncratic errors}
\begin{align*}
    p(\psi_p^{-2}|-) &\propto \left(\psi_p^{-2}\right)^{a^\psi - 1} \exp\left\{-b^\psi \psi_p^{-2}\right\} |\Psi|^{-n/2}\prod_{i=1}^n \exp\left\{-\frac{1}{2}\left(x_i - \boldsymbol{\Lambda}\mathbf{l}_i\right)^\intercal\boldsymbol{\Psi}^{-1}\left(x_i - \boldsymbol{\Lambda}\mathbf{l}_i\right)\right\} \\
    &\propto \left(\psi_p^{-2}\right)^{a^\psi - 1} \exp\left\{-b^\psi \psi_p^{-2}\right\} \left(\psi_p^{-2}\right)^{n/2}\exp\left\{-\left(\frac{1}{2}\sum_{i=1}^n\left(x_{ip} - \boldsymbol{\Lambda}_p^\intercal \mathbf{l}_i\right)^2\right)\psi_p^{-2}\right\} \\
    &= \left( \psi_p^{-2} \right)^{(a^\psi + n/2 - 1)}\exp\left\{-\left(b^\psi + \frac{1}{2}\sum_{i=1}^n\left(x_{ip} - \boldsymbol{\Lambda}_p^\intercal \mathbf{l}_i\right)^2\right)\psi_p^{-2}\right\} \\
    \log  p(\psi_p^{-2}|-) &= (a^\psi + n/2 - 1) \log \psi_p^{-2} -\left(b^\psi + \frac{1}{2}\sum_{i=1}^n\left(x_{ip} - \boldsymbol{\Lambda}_p^\intercal \mathbf{l}_i\right)^2\right)\psi_p^{-2} + C \\
    \mathbb{E}\left[\log  p(\psi_p^{-2}|-)\right] &= (a^\psi + n/2 - 1) \log \psi_p^{-2} -\left(b^\psi + \frac{1}{2}\sum_{i=1}^n\mathbb{E}\left[\left(x_{ip} - \boldsymbol{\Lambda}_p^\intercal \mathbf{l}_i\right)^2\right]\right)\psi_p^{-2} + C
\end{align*}
This is the log of the kernel of a gamma density with shape and rate parameters:
\begin{align*}
    \alpha_p^\psi &= a^\psi + n/2 \\
    \beta_p^\psi &= b^\psi + \frac{1}{2}\sum_{i=1}^n\mathbb{E}\left[\left(x_{ip} - \boldsymbol{\Lambda}_p^\intercal \mathbf{l}_i\right)^2\right]
\end{align*}
This implies that the variational factor which maximizes the ELBO with respect to $q(\psi_p^{-2})$, conditional on the remaining factos, is $q(\psi_p^{-2};\boldsymbol{\varphi}^*)=\Gamma(\psi_p^{-2};\alpha_p^\psi, \beta_p^\psi)$.

The expectation within $\beta_p^\psi$ can be expanded as follows:
\begin{align*}
    \mathbb{E}\left[\left(x_{ip} - \boldsymbol{\Lambda}_p^\intercal \mathbf{l}_i\right)^2\right] &= \mathbb{E}\left[x_{ip}^2 -2 x_{ip} \boldsymbol{\Lambda}_p^\intercal\mathbf{l}_i + \left(\boldsymbol{\Lambda}_p^\intercal\mathbf{l}_i\right)^2 \right]\\
    &= x_{ip}^2 - 2 x_{ip}\mathbb{E}[\boldsymbol{\Lambda}_p]^\intercal\mathbb{E}[\mathbf{l}_i] + \mathbb{E}\left[ \boldsymbol{\Lambda}_p^\intercal\mathbf{l}_i\mathbf{l}_i^\intercal\boldsymbol{\Lambda}_p\right] \\
    &= x_{ip}^2 - 2 x_{ip}\mathbb{E}[\boldsymbol{\Lambda}_p]^\intercal\mathbb{E}[\mathbf{l}_i] + \mathbb{E}\left[\boldsymbol{\Lambda}_p^\intercal \mathbb{E}\left[ \mathbf{l}_i \mathbf{l}_i^\intercal \right]\boldsymbol{\Lambda}_p \right] \\
    &= x_{ip}^2 - 2 x_{ip}\mathbb{E}[\boldsymbol{\Lambda}_p]^\intercal\mathbb{E}[\mathbf{l}_i] + \mathbb{E}\left[\boldsymbol{\Lambda}_p^\intercal \left( \boldsymbol{\Sigma}_i^l + \boldsymbol{\mu}_i^l \left[\boldsymbol{\mu}_i^l\right]^\intercal \right)\boldsymbol{\Lambda}_p \right] \\
    &= x_{ip}^2 - 2 x_{ip}\left[\boldsymbol{\mu}_p^\Lambda\right]^\intercal\boldsymbol{\mu}_i^l + \text{Tr}\left\{ \boldsymbol{\Sigma}_p^\Lambda\left( \boldsymbol{\Sigma}_i^l + \boldsymbol{\mu}_i^l \left[\boldsymbol{\mu}_i^l\right]^\intercal\right)  \right\} + \left[\boldsymbol{\mu}_p^\Lambda\right]^\intercal \left( \boldsymbol{\Sigma}_i^l + \boldsymbol{\mu}_i^l \left[\boldsymbol{\mu}_i^l\right]^\intercal\right) \boldsymbol{\mu}_p^\Lambda \\
    &= (x_{ip}-\left[\boldsymbol{\mu}_p^\Lambda\right]^\intercal\boldsymbol{\mu}_i^l)^2  + \text{Tr}\left\{ \boldsymbol{\Sigma}_p^\Lambda \left( \boldsymbol{\Sigma}_i^l + \boldsymbol{\mu}_i^l \left[\boldsymbol{\mu}_i^l\right]^\intercal\right)  \right\} + \left[\boldsymbol{\mu}_p^\Lambda\right]^\intercal \boldsymbol{\Sigma}_i^l \boldsymbol{\mu}_p^\Lambda \\
    \sum_{i=1}^n \mathbb{E}\left[\left(x_{ip} - \boldsymbol{\Lambda}_p^\intercal \mathbf{l}_i\right)^2\right] &= \sum_{i=1}^n (x_{ip}-\left[\boldsymbol{\mu}_p^\Lambda\right]^\intercal\boldsymbol{\mu}_i^l)^2 + \text{Tr}\left\{  \boldsymbol{\Sigma}_p^\Lambda \left( \sum_{i=1}^n \boldsymbol{\Sigma}_i^l + \boldsymbol{\mu}_i^l \left[\boldsymbol{\mu}_i^l\right]^\intercal\right) \right\} + \left[\boldsymbol{\mu}_p^\Lambda\right]^\intercal \left( \sum_{i=1}^n \boldsymbol{\Sigma}_i^l \right)\boldsymbol{\mu}_p^\Lambda
\end{align*}

\newpage

\subsection{Bayesian Multi-Study Factor Analysis}
We now consider Bayesian Multi-Study FA. In this scenario, the posterior distribution is expressed as:
\begin{align}\label{posterior-multistudy}
    p\left( \boldsymbol{\theta} | \boldsymbol{X} \right) \propto & \hspace{0.1cm} p\left(\boldsymbol{X}|\boldsymbol{\theta}\right)p\left(\boldsymbol{\theta}\right) \nonumber \\
    = & \left[ \prod_{s=1}^S |\boldsymbol{\Psi}_s|^{-\frac{N_s}{2}} \prod_{i=1}^{N_s} \exp \left\{-\frac{1}{2} \left(\mathbf{x}_{si} - \boldsymbol{\Phi}\boldsymbol{f}_{si} - \boldsymbol{\Lambda}_s\mathbf{l}_{si}\right)^\intercal \boldsymbol{\Psi}_s^{-1} \left(\mathbf{x}_{si} - \boldsymbol{\Phi}\boldsymbol{f}_{si}- \boldsymbol{\Lambda}_s\mathbf{l}_{si}\right)\right\}\right] \times \\
    & \left[ \prod_{s=1}^S \prod_{i=1}^{N_s} \exp\left\{ -\frac{1}{2} \boldsymbol{f}_{si}^\intercal\boldsymbol{f}_{si}\right\}\right] \left[ \prod_{s=1}^S \prod_{i=1}^{N_s} \exp\left\{ -\frac{1}{2} \mathbf{l}_{si}^\intercal\mathbf{l}_{si}\right\}\right] \times \nonumber \\
    &\left[ \prod_{p=1}^P \prod_{k=1}^{K^*} \left(\omega_{pk}\tau_k\right)^{1/2} \exp \left\{ -\frac{1}{2} \omega_{pk}\tau_k \phi_{pk}^2 \right\}\right]
    \left[ \prod_{s=1}^S \prod_{p=1}^P \prod_{j=1}^{J^*_s} \left(\omega_{spj}\tau_{sj}\right)^{1/2} \exp \left\{ -\frac{1}{2} \omega_{spj}\tau_{sj} \lambda_{spj}^2 \right\}\right]\times \nonumber \\
    & \left[ \prod_{p=1}^P \prod_{k=1}^{K^*} \omega_{pk}^{\nu/2-1} \exp \left\{-\frac{\nu}{2}\omega_{pk}\right\}\right] \left[ \prod_{l=1}^{K^*} \delta_l^{a_l - 1} \exp\left\{-\delta_l\right\} \right] \times \nonumber \\
    & \left[\prod_{s=1}^S \prod_{p=1}^P \prod_{j=1}^{J^*_s} \omega_{spj}^{\nu_s/2-1}\exp\left\{-\frac{\nu_s}{2}\omega_{spj}\right\} \right] \left[ \prod_{s=1}^S \prod_{l=1}^{J^*_s} \delta_{sl}^{a_{sl} - 1} \exp\left\{-\delta_{sl}\right\}\right] \times \nonumber \\
    & \left[ \prod_{s=1}^S \prod_{p=1}^P \left(\psi_{sp}^{-2}\right)^{a^\psi -1} \exp\left\{-b^\psi \psi_{sp}^{-2}\right\}\right] \nonumber
\end{align}

To approximate $p\left( \boldsymbol{\theta} | \boldsymbol{X} \right)$ we take a Variational Bayes approach considering approximations $q(\boldsymbol{\theta}; \boldsymbol{\varphi})$.
Each element of $\boldsymbol{\theta}$, $\theta_j$, is assigned its own variational factor that is characterized by the variational parameter vector $\boldsymbol{\varphi}_j$. In the following sections, we derive closed-form expressions for the varitional parameters which optimize $ELBO(q(\boldsymbol{\theta})).$

In the following we will use $C$ that do not jointly depend on the data and the parameters.

\subsubsection{Optimal Variational Factors for common factor scores}
\begin{align*}
    \log p (\boldsymbol{f}_{si} | - ) &= -\frac{1}{2}\left( \mathbf{x}_{si} - \boldsymbol{\Phi}\boldsymbol{f}_{si} - \boldsymbol{\Lambda}_s\mathbf{l}_{si}\right)^\intercal \boldsymbol{\Psi}_s^{-1} \left( \mathbf{x}_{si} - \boldsymbol{\Phi}\boldsymbol{f}_{si}- \boldsymbol{\Lambda}_s\mathbf{l}_{si}\right) -\frac{1}{2}\boldsymbol{f}_{si}^\intercal\boldsymbol{f}_{si} + C \\
    &= -\frac{1}{2} \mathbf{x}_{si}^\intercal\boldsymbol{\Psi}^{-1}\mathbf{x}_{si} + \mathbf{x}_{si}^\intercal \boldsymbol{\Psi}_s^{-1}\boldsymbol{\Phi}\boldsymbol{f}_{si} + \mathbf{x}_{si}^\intercal \boldsymbol{\Psi}_s^{-1}\boldsymbol{\Lambda}_s\mathbf{l}_{si} - \boldsymbol{f}_{si}^\intercal\boldsymbol{\Phi}_p^\intercal\boldsymbol{\Psi}_s^{-1}\boldsymbol{\Lambda}_s\mathbf{l}_{si} - \\
    &\qquad \frac{1}{2} \boldsymbol{f}_{si}\boldsymbol{\Phi}^\intercal\boldsymbol{\Psi}_s^{-1}\boldsymbol{\Phi}\boldsymbol{f}_{si} - \frac{1}{2} \mathbf{l}_{si}\boldsymbol{\Lambda}_s^\intercal\boldsymbol{\Psi}_s^{-1}\boldsymbol{\Lambda}_s\mathbf{l}_{si} -\frac{1}{2}\boldsymbol{f}_{si}^\intercal\boldsymbol{f}_{si} + C \\
    &= -\frac{1}{2}\boldsymbol{f}_{si}^\intercal \left( \boldsymbol{I}_k + \boldsymbol{\Phi}^\intercal\boldsymbol{\Psi}_s^{-1}\boldsymbol{\Phi}\right)\boldsymbol{f}_{si} + \left( \boldsymbol{\Phi}^\intercal \boldsymbol{\Psi}_s^{-1} \left(\mathbf{x}_{si} - \boldsymbol{\Lambda}_s\mathbf{l}_{si}\right) \right)^\intercal \boldsymbol{f}_{si} + C \\
    \mathbb{E}_{-f_{si}} \left[\log p (\boldsymbol{f}_{si} | - ) \right]&= -\frac{1}{2}\boldsymbol{f}_i^\intercal \left( \boldsymbol{I}_k + \mathbb{E}\left[\boldsymbol{\Phi}^\intercal\boldsymbol{\Psi}_s^{-1}\boldsymbol{\Phi}\right]\right)\boldsymbol{f}_{si} + \left( \mathbb{E}\left[\boldsymbol{\Phi}\right]^\intercal \mathbb{E}\left[\boldsymbol{\Psi}_s^{-1}\right] \left(\mathbf{x}_{si} - \mathbb{E}\left[\boldsymbol{\Lambda}_s\right]\mathbb{E}\left[\mathbf{l}_{si}\right]\right) \right)^\intercal \boldsymbol{f}_{si} + \mathbb{E}[C]
\end{align*}
where $\mathbb{E}\left[\boldsymbol{\Phi}^\intercal\boldsymbol{\Psi}_s^{-1}\boldsymbol{\Phi}\right]=\sum_{p=1}^P \mathbb{E}\left[\psi_{sp}^{-2}\right]\mathbb{E}\left[\boldsymbol{\Phi}_p\boldsymbol{\Phi}_p^\intercal\right]$. This is the log of the kernel of a multivariate normal density with parameters:
\begin{align*}
    \boldsymbol{\Sigma}_{si}^f &= \left( \boldsymbol{I}_k + \mathbb{E}\left[\boldsymbol{\Phi}^\intercal\boldsymbol{\Psi}_s^{-1}\boldsymbol{\Phi}\right]\right)^{-1} \\
    &= \left(\boldsymbol{I}_k + \sum_{p=1}^P\frac{\alpha_{sp}^\psi}{\beta_{sp}^\psi}
    \left(\boldsymbol{\mu}_p^\Phi\left[\boldsymbol{\mu}_p^\Phi\right]^\intercal + \boldsymbol{\Sigma}_p^\Phi\right)\right)^{-1} \\
    \boldsymbol{\mu}_{si}^f
    &= \left( \boldsymbol{I}_k + \mathbb{E}\left[\boldsymbol{\Phi}^\intercal\boldsymbol{\Psi}_s^{-1}\boldsymbol{\Phi}\right]\right)^{-1} \mathbb{E}\left[\boldsymbol{\Phi}\right]^\intercal \mathbb{E}\left[\boldsymbol{\Psi}_s^{-1}\right] \left(\mathbf{x}_{si} - \mathbb{E}\left[\boldsymbol{\Lambda}_s\right]\mathbb{E}\left[\mathbf{l}_{si}\right]\right) \\
    &= \boldsymbol{\Sigma}_{si}^f \left(\boldsymbol{\Phi}_1,\ldots,\boldsymbol{\Phi}_P\right)\text{diag}\left\{\frac{\alpha_{s1}^\psi}{\beta_{s1}^\psi},\ldots,\frac{\alpha_{sP}^\psi}{\beta_{sP}^\psi}\right\}\left(\mathbf{x}_{si} - \left(\boldsymbol{\Lambda}_{s1},\ldots,\boldsymbol{\Lambda}_{sP}\right)^\intercal\boldsymbol{\mu}_{si}^l\right) \\
\end{align*}
which implies that the variational factor which maximizes the ELBO with respect to $q(\boldsymbol{f}_{si})$, conditional on the remaining variational factors, is $q(\boldsymbol{f}_{si}; \boldsymbol{\varphi}^*)=\mathcal{N}\left(\boldsymbol{f}_{si}; \boldsymbol{\mu}_{si}^f, \boldsymbol{\Sigma}_{si}^f \right)$.

\subsubsection{Optimal Variational Factors for study-specific factor scores}
\begin{align*}
    \log (\mathbf{l}_{si}|-) &=  -\frac{1}{2}\left( \mathbf{x}_{si} - \boldsymbol{\Phi}\mathbf{l}_{si} - \boldsymbol{\Lambda}_s\mathbf{l}_{si}\right)^\intercal \boldsymbol{\Psi}_s^{-1} \left( \mathbf{x}_{si} - \boldsymbol{\Phi}\boldsymbol{f}_{si}- \boldsymbol{\Lambda}_s\mathbf{l}_{si}\right) -\frac{1}{2}\mathbf{l}_{si}^\intercal\mathbf{l}_{si} + C\\
    &=  -\frac{1}{2} \mathbf{x}_{si}^\intercal\boldsymbol{\Psi}^{-1}\mathbf{x}_{si} + \mathbf{x}_{si}^\intercal \boldsymbol{\Psi}_s^{-1}\boldsymbol{\Phi}\boldsymbol{f}_{si} + \mathbf{x}_{si}^\intercal \boldsymbol{\Psi}_s^{-1}\boldsymbol{\Lambda}_s\mathbf{l}_{si} - \boldsymbol{f}_{si}^\intercal\boldsymbol{\Phi}_p^\intercal\boldsymbol{\Psi}_s^{-1}\boldsymbol{\Lambda}_s\mathbf{l}_{si} - \\
    &\qquad \frac{1}{2} \boldsymbol{f}_{si}\boldsymbol{\Phi}^\intercal\boldsymbol{\Psi}_s^{-1}\boldsymbol{\Phi}\boldsymbol{f}_{si} - \frac{1}{2} \mathbf{l}_{si}\boldsymbol{\Lambda}_s^\intercal\boldsymbol{\Psi}_s^{-1}\boldsymbol{\Lambda}_s\mathbf{l}_{si} -\frac{1}{2}\mathbf{l}_{si}^\intercal\mathbf{l}_{si} + C \\
    &= -\frac{1}{2}\mathbf{l}_{si}^\intercal \left( \boldsymbol{I}_{J_s} + \boldsymbol{\Lambda}_s^\intercal\boldsymbol{\Psi}_s^{-1}\boldsymbol{\Lambda}_s\right)\mathbf{l}_{si} + \left( \boldsymbol{\Lambda}_s^\intercal \boldsymbol{\Psi}_s^{-1} \left(\mathbf{x}_{si} - \boldsymbol{\Phi}\boldsymbol{f}_{si}\right) \right)^\intercal \mathbf{l}_{si} + C \\
    \mathbb{E}_{-l_{si}} \left[\log p (\mathbf{l}_{si} | - ) \right]&= -\frac{1}{2}\mathbf{l}_i^\intercal \left( \boldsymbol{I}_{J_s} + \mathbb{E}\left[\boldsymbol{\Lambda}_s^\intercal\boldsymbol{\Psi}_s^{-1}\boldsymbol{\Lambda}_s\right]\right)\mathbf{l}_{si} + \left( \mathbb{E}\left[\boldsymbol{\Lambda}_s\right]^\intercal \mathbb{E}\left[\boldsymbol{\Psi}_s^{-1}\right] \left(\mathbf{x}_{si} - \mathbb{E}\left[\boldsymbol{\Phi}\right]\mathbb{E}\left[\boldsymbol{f}_{si}\right]\right) \right)^\intercal \mathbf{l}_{si} + \mathbb{E}[C]
\end{align*}
where $\mathbb{E}\left[\boldsymbol{\Lambda}_s^\intercal\boldsymbol{\Psi}_s^{-1}\boldsymbol{\Lambda}_s\right]=\sum_{p=1}^P \mathbb{E}[\psi_{ps}^{-2}]\mathbb{E}\left[\boldsymbol{\Lambda}_{sp}\boldsymbol{\Lambda}_{sp}^\intercal\right]$. This is the log of the kernel of a multivariate normal distribution with parameters:
\begin{align*}
    \Sigma_{si}^l &= \left( \boldsymbol{I}_{J_s} + \mathbb{E}\left[\boldsymbol{\Lambda}_s^\intercal\boldsymbol{\Psi}_s^{-1}\boldsymbol{\Lambda}_s\right]\right)^{-1} \\
    &= \left(\boldsymbol{I}_{J_s} + \sum_{p=1}^P\frac{\alpha_{sp}^\psi}{\beta_{sp}^\psi}\left(\boldsymbol{\mu}_p^\Lambda\left[\boldsymbol{\mu}_{sp}^\Lambda \right]^\intercal + \boldsymbol{\Sigma}_{sp}^\Lambda\right)\right)^{-1} \\
    \mu_{si}^l &= \left( \boldsymbol{I}_{J_s} + \mathbb{E}\left[\boldsymbol{\Lambda}_s^\intercal\boldsymbol{\Psi}_s^{-1}\boldsymbol{\Lambda}_s\right]\right)^{-1} \mathbb{E}\left[\boldsymbol{\Lambda}_s\right]^\intercal \mathbb{E}\left[\boldsymbol{\Psi}_s^{-1}\right] \left(\mathbf{x}_{si} - \mathbb{E}\left[\boldsymbol{\Phi}\right]\mathbb{E}\left[\boldsymbol{f}_{si}\right]\right)\\
    &= \boldsymbol{\Sigma}_{si}^l \left(\boldsymbol{\mu}_{s1}^\Lambda,\ldots,\boldsymbol{\mu}_{sP}^\Lambda \right)\text{diag}\left\{\frac{\alpha_{s1}^\psi}{\beta_{s1}^\psi},\ldots,\frac{\alpha_{sP}^\psi}{\beta_{sP}^\psi}\right\}\left(\mathbf{x}_{si} - \left(\boldsymbol{\mu}_1^\Phi,\ldots,\boldsymbol{\mu}_P^\Phi\right)^\intercal \boldsymbol{\mu}_{si}^f\right)
\end{align*}
which implies that the the variational factor which maximizes the ELBO with respect to $q(\mathbf{l}_{si})$, conditional on the remaining factors, is $q(\mathbf{l}_{si};\boldsymbol{\varphi}^*)=\mathcal{N}\left(\mathbf{l}_{si};\boldsymbol{\mu}_{si}^l, \boldsymbol{\Sigma}_{si}^l \right)$.

\subsubsection{Optimal Variational Factors for common factor loadings}
\begin{align*}
    \small
    \log p(\boldsymbol{\Phi}_p|-) &= -\frac{1}{2}\left(\sum_{p=1}^P \sum_{k=1}^K \omega_{pk}\tau_{k}\phi_{pk}^2\right) - \frac{1}{2} \left(\sum_{s=1}^S \sum_{i=1}^{N_s} \left(\mathbf{x}_{si} - \boldsymbol{\Phi}\boldsymbol{f}_{si} -\boldsymbol{\Lambda}_s\mathbf{l}_{si} \right)^\intercal \boldsymbol{\Psi}_s \left(\mathbf{x}_{si} - \boldsymbol{\Phi}\boldsymbol{f}_{si} -\boldsymbol{\Lambda}_s\mathbf{l}_{si} \right)\right) + C \\
    &= -\frac{1}{2} \left(\sum_{k=1}^K \omega_{pk}\tau_k\phi_{pk}^2\right) -\frac{1}{2}\left(\sum_{s=1}^S\psi_{sp}^{-2}\sum_{i=1}^{N_s}\left(x_{isp} - \boldsymbol{\Phi}_p^\intercal\boldsymbol{f}_{si} - \boldsymbol{\Lambda}_{sp}^\intercal\mathbf{l}_{si}\right)^2\right) + C \\
    &= -\frac{1}{2}\boldsymbol{\Phi}^\intercal\boldsymbol{D}_p\boldsymbol{\Phi} - \frac{1}{2}\sum_{s=1}^S \psi_{sp}^{-2}\sum_{i=1}^{N_s} \left(x_{isp}^2 + \boldsymbol{\Phi}_p^\intercal\boldsymbol{f}_{si}\boldsymbol{f}_{si}^\intercal\boldsymbol{\Phi}_p + \boldsymbol{\Lambda}_{sp}^\intercal\mathbf{l}_{si}\mathbf{l}_{si}^\intercal\boldsymbol{\Lambda}_{sp} \right. \\
    & \left. \qquad\qquad \qquad\qquad\qquad\qquad\quad -2 x_{isp}\boldsymbol{\Phi}_p^\intercal\boldsymbol{f}_{si}-2x_{isp}\boldsymbol{\Lambda}_{sp}^\intercal\mathbf{l}_{si} + 2 \boldsymbol{\Lambda}_{sp}^\intercal\boldsymbol{\Phi}_p^\intercal\boldsymbol{f}_{si}\right) + C \\
    &= -\frac{1}{2}\boldsymbol{\Phi}^\intercal\boldsymbol{D}_p\boldsymbol{\Phi} - \frac{1}{2}\sum_{s=1}^S \psi_{sp}^{-2}\left(\sum_{i=1}^{N_s} \boldsymbol{\Phi}_p^\intercal\boldsymbol{f}_{si}\boldsymbol{f}_{si}^\intercal\boldsymbol{\Phi}_p -2 \left(\boldsymbol{f}_{si}\left(x_{isp}-\mathbf{l}_{si}\boldsymbol{\Lambda}_{sp}\right)\right)^\intercal\boldsymbol{\Phi}_p\right) + C \\
    &= -\frac{1}{2}\boldsymbol{\Phi}^\intercal\boldsymbol{D}_p\boldsymbol{\Phi} -\frac{1}{2}\sum_{s=1}^S\left(
    \boldsymbol{\Phi}_p^\intercal\left(\psi_{sp}^{-2}\sum_{i=1}^{N_s}\boldsymbol{f}_{si}\boldsymbol{f}_{si}^\intercal\right)\boldsymbol{\Phi}_p -2 \left(\psi_{sp}^{-2}\sum_{i=1}^{N_s}\boldsymbol{f}_{si}\left(x_{isp}-\mathbf{l}_{si}^\intercal\boldsymbol{\Lambda}_{sp}\right)\right)^\intercal\boldsymbol{\Phi}_p\right) + C \\
    &= -\frac{1}{2}\boldsymbol{\Phi}_p^\intercal\left(\boldsymbol{D}_p + \sum_{s=1}^S \psi_{sp}^{-2}\sum_{i=1}^{N_s}\boldsymbol{f}_{si}\boldsymbol{f}_{si}^\intercal\right)\boldsymbol{\Phi}_p + \left(\sum_{s=1}^S \psi_{sp}^{-2}\sum_{i=1}^{N_s}\boldsymbol{f}_{si}\left(x_{isp}-\mathbf{l}_{si}^\intercal\boldsymbol{\Lambda}_{sp}\right)\right)^\intercal\boldsymbol{\Phi}_p \\
    &= -\frac{1}{2}\boldsymbol{\Phi}_p^\intercal\left(\boldsymbol{D}_p + \sum_{s=1}^S \psi_{sp}^{-2}\boldsymbol{f}_s^\intercal\boldsymbol{f}_s\right)\boldsymbol{\Phi}_p + \left(\sum_{s=1}^S \psi_{sp}^{-2}\boldsymbol{f}_s^\intercal\left(\mathbf{x}_{sp}-\mathbf{l}_s\boldsymbol{\Lambda}_{sp}\right)\right)^\intercal\boldsymbol{\Phi}_p + C
\end{align*}
where $\boldsymbol{D}_p=\text{diag}\left(\omega_{p1}\tau_1,\ldots,\omega_{pK}\tau_K\right)$, $\boldsymbol{f}_s=(\boldsymbol{f}_{s1},\ldots,\boldsymbol{f}_{sN_s})^\intercal$, $\mathbf{l}_s=(\mathbf{l}_{s1},\ldots,\mathbf{l}_{sN_s})^\intercal$, $\mathbf{x}_{sp}=(x_{s1p},\ldots,x_{sN_sp})^\intercal$. Next, we apply expectations with respect to $q$:
\begin{align*}
    \mathbb{E}\left[\log p(\boldsymbol{\Phi}_p|-)\right] &= -\frac{1}{2}\boldsymbol{\Phi}_p^\intercal\left(\mathbb{E}\left[\boldsymbol{D}_p\right] + \sum_{s=1}^S \mathbb{E}\left[\psi_{sp}^{-2}\right]\mathbb{E}\left[\boldsymbol{f}_s^\intercal\boldsymbol{f}_s\right]\right)\boldsymbol{\Phi}_p + \\
    &\quad + \left(\sum_{s=1}^S \mathbb{E}\left[\psi_{sp}^{-2}\right] \mathbb{E}\left[\boldsymbol{f}_s\right]^\intercal\left(\mathbf{x}_{sp}-\mathbb{E}\left[\mathbf{l}_s\right]\mathbb{E}\left[\boldsymbol{\Lambda}_{sp}\right]\right)\right)^\intercal\boldsymbol{\Phi}_p + C
\end{align*}
which is the log of the kernel of a multivariate normal density with parameters:
\begin{align*}
    \boldsymbol{\Sigma}_p^\Phi &= \left(\mathbb{E}\left[\boldsymbol{D}_p\right] + \sum_{s=1}^S \mathbb{E}\left[\psi_{sp}^{-2}\right]\mathbb{E}\left[\boldsymbol{f}_s^\intercal\boldsymbol{f}_s\right]\right)^{-1} \\ 
    &= \left(\text{diag}\left\{\frac{\alpha_{1}^\omega}{\beta_1^\omega}\frac{\alpha_{p1}^\delta}{\beta_{p1}^\delta},\ldots,\frac{\alpha_{pK}^\omega}{\beta_{pK}^\omega}\prod_{l=1}^K\frac{\alpha_{pl}^\delta}{\beta_{pl}^\delta}\right\} + \sum_{s=1}^S \frac{\alpha_{sp}^\psi}{\beta_{sp}^\psi}\left(\sum_{i=1}^{N_s}\boldsymbol{\mu}_{si}^f\left[\boldsymbol{\mu}_{si}^f\right]^\intercal + \boldsymbol{\Sigma}_{si}^f\right)\right)^{-1}\\
    \boldsymbol{\mu}_p^\Phi &= \left(\mathbb{E}\left[\boldsymbol{D}_p\right] + \sum_{s=1}^S \mathbb{E}\left[\psi_{sp}^{-2}\right]\mathbb{E}\left[\boldsymbol{f}_s^\intercal\boldsymbol{f}_s\right]\right)^{-1}\left(\sum_{s=1}^S \mathbb{E}\left[\psi_{sp}^{-2}\right]\mathbb{E}\left[\boldsymbol{f}_s\right]^\intercal\left(\mathbf{x}_{sp}-\mathbb{E}\left[\mathbf{l}_s\right]\mathbb{E}\left[\boldsymbol{\Lambda}_{sp}\right]\right)\right)\\
    &= \boldsymbol{\Sigma}_p^\Phi \left(\sum_{s=1}^S \frac{\alpha_{sp}^\psi}{\beta_{sp}^\Psi}\left(\boldsymbol{f}_{s1},\ldots,\boldsymbol{f}_{sN_s}\right)\left(\mathbf{x}_{sp} - \left(\boldsymbol{\mu}_{s1}^l,\ldots,\boldsymbol{\mu}_{sN_s}^l\right)^\intercal\boldsymbol{\mu}_{sp}^\Lambda\right)\right)\\
\end{align*}
which implies that the factor which maximizes the ELBO with respect to $q(\boldsymbol{\Phi}_p)$, conditional on the remaining factors, is $q(\boldsymbol{\Phi}_p;\boldsymbol{\varphi}^*)=\mathcal{N}\left(\boldsymbol{\Phi}_p;\boldsymbol{\mu}_p, \boldsymbol{\Sigma}_p \right)$.

\subsubsection{Optimal Variational Factors for study-specific factor loadings}
\begin{align*}
    \small
    \log p\left( \boldsymbol{\Lambda}_{sp} | -  \right) &=  \frac{1}{2}  \left(\sum_{i=1}^{N_s} \left(\mathbf{x}_{si} - \boldsymbol{\Phi}\boldsymbol{f}_{si} - \boldsymbol{\Lambda}_s\mathbf{l}_{si}\right)^\intercal \boldsymbol{\Psi}_s^{-1}\left(\mathbf{x}_{si} - \boldsymbol{\Phi}\boldsymbol{f}_{si} - \boldsymbol{\Lambda}_s\mathbf{l}_{si}\right)\right) -  \frac{1}{2} \sum_{p=1}^P \sum_{j=1}^{J_s} \omega_{spj}\tau_{sj}\lambda_{spj}^2+ C\\ 
    &= - \frac{1}{2} \boldsymbol{\Lambda}_{sp}^\intercal \boldsymbol{D}_{sp}\boldsymbol{\Lambda}_{sp} -\frac{1}{2}\sum_{i=1}^{N_s} \psi_{sp}^{-2}\left( x_{isp} - \boldsymbol{\Phi}_p^\intercal\boldsymbol{f}_{si} - \boldsymbol{\Lambda}_{sp}^\intercal\mathbf{l}_{si} \right)^2 + C\\
    &= - \frac{1}{2} \boldsymbol{\Lambda}_{sp}^\intercal \boldsymbol{D}_{sp}\boldsymbol{\Lambda}_{sp} -\frac{1}{2}\psi_{sp}^{-2}\sum_{i=1}^{N_s} \left( x_{isp}^2 + \boldsymbol{\Phi}_p^\intercal\boldsymbol{f}_{si}\boldsymbol{f}_{si}^\intercal\boldsymbol{\Phi}_p + \boldsymbol{\Lambda}_{sp}^\intercal\mathbf{l}_{si}\mathbf{l}_{si}^\intercal\boldsymbol{\Lambda}_{sp} \right. \\
    & \left. \qquad\qquad\qquad\qquad\qquad\qquad\enspace -2x_{isp}\boldsymbol{\Phi}^\intercal\boldsymbol{f}_{si} -2 x_{isp}\boldsymbol{\Lambda}_{sp}^\intercal\mathbf{l}_{si} +2 \boldsymbol{\Phi}_p^\intercal\boldsymbol{f}_{si}\boldsymbol{\Lambda}_{sp}^\intercal\mathbf{l}_{si}\right) + C \\
    &= - \frac{1}{2} \boldsymbol{\Lambda}_{sp}^\intercal \boldsymbol{D}_{sp}\boldsymbol{\Lambda}_{sp} -\frac{1}{2}\psi_{sp}^{-2}\sum_{i=1}^{N_s} \boldsymbol{\Lambda}_{sp}^\intercal\mathbf{l}_{si}\mathbf{l}_{si}^\intercal\boldsymbol{\Lambda}_{sp} + \left( \mathbf{l}_{si}\left(x_{isp} - \boldsymbol{f}_{si}^\intercal\boldsymbol{\Phi}_p\right) \right)^\intercal\boldsymbol{\Lambda}_{sp} + C \\
    &= - \frac{1}{2} \boldsymbol{\Lambda}_{sp}^\intercal \boldsymbol{D}_{sp}\boldsymbol{\Lambda}_{sp} - \frac{1}{2}\boldsymbol{\Lambda}_{sp}^\intercal\left( \psi_{sp}^{-2}\sum_{i=1}^{N_s}\mathbf{l}_{si}\mathbf{l}_{si}^\intercal \right)\boldsymbol{\Lambda}_{sp} + \left(\sum_{i=1}^{N_s} \psi_{sp}^{-2} \mathbf{l}_{si}\left(x_{isp} - \boldsymbol{f}_{si}^\intercal\boldsymbol{\Phi}_p\right)\right)^\intercal\boldsymbol{\Lambda}_{sp} + C \\
    &= - \frac{1}{2} \boldsymbol{\Lambda}_{sp}^\intercal \boldsymbol{D}_{sp}\boldsymbol{\Lambda}_{sp} - \frac{1}{2}\boldsymbol{\Lambda}_{sp}^\intercal\left( \psi_{sp}^{-2}\mathbf{l}_s^\intercal\mathbf{l}_s \right)\boldsymbol{\Lambda}_{sp} + \left( \psi_{sp}^{-2}\mathbf{l}_s^\intercal \left(\mathbf{x}_{sp} - \boldsymbol{f}_s\boldsymbol{\Phi}_p\right) \right)^\intercal\boldsymbol{\Lambda}_{sp} + C \\
    &= - \frac{1}{2}\boldsymbol{\Lambda}_{sp}^\intercal\left( \boldsymbol{D}_{sp} + \psi_{sp}^{-2}\mathbf{l}_s^\intercal\mathbf{l}_s \right)\boldsymbol{\Lambda}_{sp} + \left( \psi_{sp}^{-2}\mathbf{l}_s^\intercal \left(\mathbf{x}_{sp} - \boldsymbol{f}_s\boldsymbol{\Phi}_p\right) \right)^\intercal\boldsymbol{\Lambda}_{sp} + C
\end{align*}
where $\boldsymbol{D}_{sp}=\text{diag}\left(\omega_{sp1}\tau_{s1},\ldots,\omega_{spJ_s}\tau_{sJ_s}\right)$, $\boldsymbol{f}_s=(\boldsymbol{f}_{s1},\ldots,\boldsymbol{f}_{sN_s})^\intercal$, $\mathbf{l}_s=(\mathbf{l}_{s1},\ldots,\mathbf{l}_{sN_s})^\intercal$, $\mathbf{x}_{sp}=(x_{s1p},\ldots,x_{sN_sp})^\intercal$. 

Next, we take the expectation with respect to $q$:
\begin{align*}
    \mathbb{E}\left[\log p(\boldsymbol{\Lambda}_{sp}|-)\right] &= - \frac{1}{2}\boldsymbol{\Lambda}_{sp}^\intercal\left( \mathbb{E}\left[\boldsymbol{D}_{sp}\right] + \mathbb{E}\left[\psi_{sp}^{-2}\right]\mathbb{E}\left[\mathbf{l}_s^\intercal\mathbf{l}_s\right] \right)\boldsymbol{\Lambda}_{sp} + \\
    &\quad \left( \mathbb{E}\left[\psi_{sp}^{-2}\right]\mathbb{E}\left[\mathbf{l}_s\right]^\intercal \left(\mathbf{x}_{sp} - \mathbb{E}\left[\boldsymbol{f}_s\right]\mathbb{E}\left[\boldsymbol{\Phi}_p\right]\right) \right)^\intercal\boldsymbol{\Lambda}_{sp} + \mathbb{E}[C]
\end{align*}

This is the log of the kernel of a multivariate normal density with parameters: 
\begin{align*}
    \boldsymbol{\Sigma}_{sp}^\Lambda &= \left( \mathbb{E}\left[\boldsymbol{D}_{sp}\right] + \mathbb{E}\left[\psi_{sp}^{-2}\right]\mathbb{E}\left[\mathbf{l}_s^\intercal\mathbf{l}_s\right] \right)^{-1} \\
    &= \left( \text{diag}\left\{\frac{\alpha_{sp1}^\omega}{\beta_{sp1}^\omega}\frac{\alpha_{s1}^\delta}{\beta_{s1}^\delta},\ldots,\frac{\alpha_{spJ_s}^\omega}{\beta_{spJ_s}^\omega}\prod_{l=1}^{J_s}\frac{\alpha_{sl}^\delta}{\beta_{sl}^\delta}\right\} + \frac{\alpha_{sp}^\psi}{\beta_{sp}^\psi}\sum_{i=1}^{N_s}\left(\boldsymbol{\mu}_{si}^l\left[\boldsymbol{\mu}_{si}^l\right]^\intercal + \boldsymbol{\Sigma}_{si}^l\right) \right)^{-1}\\
    \boldsymbol{\mu}_{sp}^\Lambda &= \mathbb{E}\left[\psi_{sp}^{-2}\right]\left( \mathbb{E}\left[\boldsymbol{D}_{sp}\right] + \mathbb{E}\left[\psi_{sp}^{-2}\right]\mathbb{E}\left[\mathbf{l}_s^\intercal\mathbf{l}_s\right] \right)^{-1}\mathbb{E}\left[\mathbf{l}_s\right]^\intercal\left(\mathbf{x}_{sp} - \mathbb{E}\left[\boldsymbol{f}_s\right]\mathbb{E}\left[\boldsymbol{\Phi}_p\right]\right)\\
    &= \frac{\alpha_{sp}^\psi}{\beta_{sp}^\psi}\boldsymbol{\Sigma}_{sp}^\Lambda \left(\boldsymbol{\mu}_{s1}^l,\ldots,\boldsymbol{\mu}_{sN_s}^l\right) \left( \mathbf{x}_{sp} - \left(\boldsymbol{\mu}_{s1}^f,\ldots,\boldsymbol{\mu}_{sN_s}^f\right)\boldsymbol{\mu}_p^\Phi \right)
\end{align*}
which implies that the factor which maximizes the ELBO with respect to $q(\boldsymbol{\Lambda}_{sp})$, conditional on the remaining factors, is $q(\boldsymbol{\Lambda}_{sp};\boldsymbol{\varphi}^*)=\mathcal{N}\left(\boldsymbol{\Lambda}_{sp};\boldsymbol{\mu}_{sp}^\Lambda, \boldsymbol{\Sigma}_{sp}^\Lambda\right)$.

\subsubsection{Optimal Variational Factors for local shrinkage of common factor loadings}
\begin{align*}
    \log p(\omega_{pk}|-) &= \frac{1}{2}\log \omega_{pk} - \frac{1}{2} \omega_{pk}\tau_k\phi_{pk}^2 + \left(\frac{\nu}{2} - 1\right)\log \omega_{pk} - \frac{\nu}{2} \omega_{pk} + C\\
    &= \left(\frac{\nu + 1}{2}-1\right)\log \omega_{pk} - \left(\frac{\nu + \tau_k\phi_{pk}^2}{2}\right)\omega_{pk} + C \\
    \mathbb{E}_{-\omega_{pk}}\left[\log p(\omega_{pk}|-)\right] &= \left(\frac{\nu + 1}{2}-1\right)\log \omega_{pk} - \left(\frac{\nu + \mathbb{E}\left[\tau_k\right]\mathbb{E}\left[\phi_{pk}^2\right]}{2}\right)\omega_{pk} + \mathbb{E}[C]
\end{align*}
which is the log of the kernel of a gamma density with shape and rate parameters:
\begin{align*}
    \alpha_{pk}^\omega &= \frac{\nu+1}{2} \\
    \beta_{pk}^\omega &= \frac{\nu + \mathbb{E}\left[\tau_k\right]\mathbb{E}\left[\phi_{pk}^2\right]}{2} \\
    &= \frac{\nu + \left(\prod_{l=1}^k\frac{\alpha_{k}^\delta}{\beta_k^\delta}\right)\left(\left(\left[\boldsymbol{\mu}_p^\Phi\right]_k\right)^2 + \left[\boldsymbol{\Sigma}_p^\Phi\right]_{k,k}\right)}{2}
\end{align*}
This implies that the variational factor which maximizes the ELBO with respect to $q(\omega_{pk})$, conditional on the remaining factors, is $q(\omega_{pk};\boldsymbol{\varphi}^*)=\Gamma(\omega_{pk}; \alpha_{pk}^\omega, \beta_{pk}^\omega)$.

\subsubsection{Optimal Variational Factors for local shrinkage of study-specific factor loadings}
\begin{align*}
    \log p(\omega_{spj}|-) &= \frac{1}{2}\log \omega_{spj} - \frac{1}{2} \omega_{spj}\tau_{sj}\lambda_{spj}^2 + \left(\frac{\nu_s}{2} - 1\right)\log \omega_{spj} - \frac{\nu_s}{2} \omega_{spj} + C\\
    &= \left(\frac{\nu_s + 1}{2}-1\right)\log \omega_{spj} - \left(\frac{\nu_s + \tau_{sj}\lambda_{spj}^2}{2}\right)\omega_{spj} + C \\
    \mathbb{E}_{-\omega_{spj}}\left[\log p(\omega_{spj}|-)\right] &= \left(\frac{\nu_s + 1}{2}-1\right)\log \omega_{spj} - \left(\frac{\nu_s + \mathbb{E}\left[\tau_{js}\right]\mathbb{E}\left[\lambda_{spj}^2\right]}{2}\right)\omega_{spj} + \mathbb{E}[C]
\end{align*}
which is the log of the kernel of a gamma density with shape and rate parameters:
\begin{align*}
    \alpha_{spj}^\omega &= \frac{\nu_s+1}{2} \\
    \beta_{pk}^\omega &= \frac{\nu_s + \mathbb{E}\left[\tau_{sj}\right]\mathbb{E}\left[\lambda_{spj}^2\right]}{2} \\
    &= \frac{\nu_s + \left(\prod_{l=1}^j \frac{\alpha_{spl}^\delta}{\beta_{spl}^\delta}\right)\left( \left[\mu_{spj}^\Lambda\right]^2 + \left[\boldsymbol{\Sigma}_{sp}^\Lambda\right]_{j,j}\right)}{2}
\end{align*}
This implies that the variational factor which maximizes the ELBO with respect to $q(\omega_{spj})$, conditional on the remaining factors, is $q(\omega_{spj};\boldsymbol{\varphi}^*)=\Gamma(\omega_{spj}; \alpha_{spj}^\omega, \beta_{spj}^\omega)$.

\subsubsection{Optimal Variational Factors for global shrinkage of common factor loadings}
\begin{align*}
    \small
    \log p (\delta_l|-) &= \left(\sum_{p=1}^P\sum_{k=l}^{K^*}\frac{1}{2}\log\left(\omega_{pk}\tau_k\right) -\frac{1}{2}\omega_{pk}\tau_k\phi_{pk}^2\right) + (a_l - 1) \log \delta_l - \delta_l + C\\
    &= \left(\sum_{p=1}^P \sum_{k=l}^{K^*} \frac{1}{2}\log \tau_k - \frac{1}{2}\omega_{pk}\tau_k\phi_{pk}^2\right) + (a_l - 1) \log \delta_l - \delta_l + C \\
    &= \left(\sum_{p=1}^{P}\sum_{k=l}^{K^*} \frac{1}{2}\right) \log \delta_l -\frac{1}{2}\left(\sum_{p=1}^P\sum_{k=l}^{K^*} \omega_{pk}\tau_{k}\phi_{pk}^2\right) + (a_l - 1) \log \delta_l - \delta_l + C\\
    &= \left( a_l -1 +\sum_{p=1}^P\sum_{k=l}^{K^*} \frac{1}{2} \right)\log \delta_l - \left(1 + \frac{1}{2}\sum_{p=1}^P\sum_{k=l}^{K^*}\omega_{pk}\left(\prod_{1\leq r \leq k, r\neq l} \delta_r\right)\phi_{pk}^2 \right)\delta_l + C \\
    &= \left(a_l + \frac{P(K^*-l+1)}{2} - 1\right)\log \delta_l - \left(1 + \frac{1}{2}\sum_{p=1}^P\sum_{k=l}^{K^*}\omega_{pk}\left(\prod_{1\leq r \leq k, r\neq l} \delta_r\right)\phi_{pk}^2 \right)\delta_l + C\\
    \mathbb{E}\left[ \log p (\delta_l|-)\right] &= \left(a_l + \frac{P(K^*-l+1)}{2} - 1\right)\log \delta_l \\
    & \quad - \left(1 + \frac{1}{2}\sum_{p=1}^P\sum_{k=l}^{K^*}\mathbb{E}[\omega_{pk}]\left(\prod_{1\leq r \leq k, r\neq l} \mathbb{E}\left[\delta_r\right]\right)\mathbb{E}[\phi_{pk}^2] \right)\delta_l + C
\end{align*}
which is the log of the kernel of a gamma density with shape and rate parameters: 
\begin{align*}
    \alpha_l^\delta &= a_l + \frac{P(K^*-l+1)}{2} \\
    \beta_l^\delta &= 1 + \frac{1}{2}\left(\sum_{p=1}^P\sum_{k=l}^{K^*}\mathbb{E}[\omega_{pk}]\left(\prod_{1\leq r \leq k, r\neq l} \mathbb{E}\left[\delta_r\right]\right)\mathbb{E}[\phi_{pk}^2]\right) \\
    &= 1 + \frac{1}{2}\left(\sum_{p=1}^P\sum_{k=l}^{K^*} \frac{\alpha_{pk}^\omega}{\beta_{pk}}\left(\prod_{1\leq r \leq l, r\neq l}\frac{\alpha_{r}^\delta}{\beta_{r}^\delta}\right)\left(\left[\boldsymbol{\mu}_{pk}^\Phi\right]^2 + \left[\boldsymbol{\Sigma}_p^\Phi\right]_{k,k}\right)\right)
\end{align*}
This implies that the variational factor with maximizes the ELBO with respect to $q(\delta_l)$, conditional on the remaining factors, is $q(\delta_l;\boldsymbol{\varphi}^*)=\Gamma(\delta_l;\alpha_l^\delta, \beta_l^\delta)$.

\subsubsection{Optimal Variational Factors for globbal shrinkage of study-specific factor loadings}
\begin{align*}
    \small
    \log p (\delta_{sl}|-) &= \left(\sum_{p=1}^P\sum_{j=l}^{J_s} \frac{1}{2}\log\left(\upsilon_{spj}\gamma_{sj}\right) - \frac{1}{2}\upsilon_{spj}\gamma_{sj}\lambda_{spj}^2\right)  + (a_{sl} - 1) \log \tau_{sl} - \tau_{sl} + C\\
    &= \left(\sum_{p=1}^P \sum_{j=l}^{J_s} \frac{1}{2}\log \delta_{sl} - \frac{1}{2}\delta_{sl}\omega_{spj}\left(\prod_{1\leq r \leq j, l \neq r} \delta_{sr}\right)\lambda_{spj}^2\right) + (a_{sl} - 1) \log \delta_{sl} - \delta_{sl} + C \\
    &= \left(a_{sl} - 1 + \sum_{p=1}^P\sum_{j=l}^{J^*_s}\frac{1}{2}\right)\log\delta_{sl} -\left(1 + \frac{1}{2}\sum_{p=1}^P\sum_{j=l}^{J^*_s}\omega_{spj}\left(\prod_{1\leq r \leq j, l \neq r} \delta_{sr}\right)\lambda_{spj}^2\right)\delta_{sl} + C\\
    &= \left(a_{sl} + \frac{P(J^*_s-l+1)}{2} - 1\right)\log \delta_{sl} - \left(1 + \frac{1}{2}\sum_{p=1}^P\sum_{j=l}^{J_s} \omega_{spl} \left(\prod_{1\leq r \leq j, l \neq r} \delta_{sr}\right) \lambda_{spj}^2 \right) \delta_{sl} + C\\
    \mathbb{E}\left[ \log p (\delta_{sl}|-)\right] &= \left(a_{sl} + \frac{P(J^*_s-l+1)}{2} - 1\right)\log \delta_{sj} \\
    & \quad - \left(1 + \frac{1}{2}\sum_{p=1}^P\sum_{l=j}^{J^*_s}\mathbb{E}[\omega_{spj}]\left(\prod_{1\leq r \leq j, l \neq r} \mathbb{E}\left[\delta_{sr}\right]\right)\mathbb{E}[\lambda_{spj}^2] \right)\delta_{sl} + C
\end{align*}
which is the log of the kernel of a gamma density with shape and rate parameters: 
\begin{align*}
    \alpha_{sl}^\delta &= a_{sl} + \frac{P(J^*_s-l+1)}{2} \\
    \beta_{sj}^\delta &= 1 + \frac{1}{2}\sum_{p=1}^P\sum_{j=l}^{J^*_s}\mathbb{E}[\omega_{spj}]\left(\prod_{1\leq r \leq j, l \neq r} \mathbb{E}\left[\delta_{sr}\right]\right)\mathbb{E}[\lambda_{spj}^2] \\
    &= 1 + \frac{1}{2} \left(\sum_{p=1}^P\sum_{j=l}^{J^*_s} \frac{\alpha_{spj}^\omega}{\beta_{spj}^\omega}\left(\prod_{l \leq r \leq J_s, r\neq j} \frac{\alpha_{sr}^\delta}{\beta_{sr}^\delta}\right)\left(\left[\boldsymbol{\mu}_{spj}^\Lambda\right]^2 + \left[\boldsymbol{\Sigma}_{sp}^\Lambda\right]_{j,j}\right)\right)
\end{align*}
This implies that the variational factor with maximizes the ELBO with respect to $q(\delta_k)$, conditional on the remaining factors, is $q(\delta_k;\boldsymbol{\varphi}^*)=\Gamma(\delta_k;\alpha_k^\delta, \beta_k^\delta)$.

\subsubsection{Optimal Variational Factors for idiosyncratic errors}
\begin{align*}
    p(\psi_{sp}^{-2}|-) &\propto \left(\psi_{sp}^{-2}\right)^{a^\psi - 1} \exp\left\{-b^\psi \psi_{sp}^{-2}\right\} |\Psi_s|^{-N_s/2}\times \\
    & \quad \prod_{i=1}^{N_s} \exp\left\{-\frac{1}{2}\left(\mathbf{x}_{si} - \boldsymbol{\Phi}\boldsymbol{f}_i - \boldsymbol{\Lambda}_s\mathbf{l}_{si}\right)^\intercal\boldsymbol{\Psi}^{-1}\left(\mathbf{x}_{si} - \boldsymbol{\Phi}\boldsymbol{f}_{si}- \boldsymbol{\Lambda}_s\mathbf{l}_{si}\right)\right\} \\
    &\propto \left(\psi_{sp}^{-2}\right)^{a^\psi - 1} \exp\left\{-b^\psi \psi_{sp}^{-2}\right\} \left(\psi_{sp}^{-2}\right)^{N_s/2}\exp\left\{-\left(\frac{1}{2}\sum_{i=1}^{N_s}\left(x_{sip} - \boldsymbol{\Phi}_p^\intercal \boldsymbol{f}_{si} - \boldsymbol{\Lambda}_{sp}^\intercal \mathbf{l}_{si}\right)^2\right)\psi_{sp}^{-2}\right\} \\
    &= \left( \psi_{sp}^{-2} \right)^{(a^\psi + N_s/2 - 1)}\exp\left\{-\left(b^\psi + \frac{1}{2}\sum_{i=1}^{N_s}\left(x_{sip} - \boldsymbol{\Phi}_p^\intercal \boldsymbol{f}_{si} - \boldsymbol{\Lambda}_{sp}^\intercal\mathbf{l}_{si}\right)^2\right)\psi_{sp}^{-2}\right\} \\
    \log  p(\psi_{sp}^{-2}|-) &= (a^\psi + N_s/2 - 1) \log \psi_{sp}^{-2} -\left(b^\psi + \frac{1}{2}\sum_{i=1}^{sp}\left(x_{sip} - \boldsymbol{\Phi}_p^\intercal \boldsymbol{f}_i\right)^2 - \boldsymbol{\Lambda}_{sp}^\intercal\mathbf{l}_{si}\right)\psi_{sp}^{-2} + C \\
    \mathbb{E}\left[\log  p(\psi_{sp}^{-2}|-)\right] &= (a^\psi + N_s/2 - 1) \log \psi_{sp}^{-2} -\left(b^\psi + \frac{1}{2}\sum_{i=1}^{N_s}\mathbb{E}\left[\left(x_{sip} - \boldsymbol{\Phi}_p^\intercal \boldsymbol{f}_i-\boldsymbol{\Lambda}_{sp}^\intercal\mathbf{l}_{si}\right)^2\right]\right)\psi_{sp}^{-2} + \mathbb{E}[C]
\end{align*}
This is the log of the kernel of a gamma density with shape and rate parameters:
\begin{align*}
    \alpha_{sp}^\psi &= a^\psi + N_s/2 \\
    \beta_{sp}^\psi &= b^\psi + \frac{1}{2}\sum_{i=1}^{N_s}\mathbb{E}\left[\left(x_{sip} - \boldsymbol{\Phi}_p^\intercal \boldsymbol{f}_{si} - \boldsymbol{\Lambda}_{sp}^\intercal\mathbf{l}_{si}\right)^2\right]
\end{align*}
This implies that the variational factor which maximizes the ELBO with respect to $q(\psi_p^{-2})$, conditional on the remaining factos, is $q(\psi_p^{-2};\boldsymbol{\varphi}^*)=\Gamma(\psi_p^{-2};\alpha_p^\psi, \beta_p^\psi)$.

The individual terms of the summed expectations within $\beta_p^\psi$ can be simplified as follows:
\small
\begin{align*}
    \mathbb{E}\left[\left(x_{sip} - \boldsymbol{\Phi}_p^\intercal \boldsymbol{f}_{si} - \Lambda_{sp}^\intercal\mathbf{l}_{si}\right)^2\right] = &\mathbb{E}\left[x_{sip}^2 -2 x_{sip} \boldsymbol{\Phi}_p^\intercal\boldsymbol{f}_{si} -2 x_{sip} \boldsymbol{\Lambda}_{sp}^\intercal \mathbf{l}_{si} + 2 \left(\boldsymbol{\Phi}_p\boldsymbol{f}_{si}\right)\left(\boldsymbol{\Lambda}_{sp}^\intercal\mathbf{l}_{si}\right) + \right.\\
    &\quad \left. \left(\boldsymbol{\Phi}_p^\intercal\boldsymbol{f}_{si}\right)^2 + \left(\boldsymbol{\Lambda}_{sp}^\intercal \mathbf{l}_{si}\right)^2\right]\\
    = &x_{ips}^2 -2 x_{ips} \left[\boldsymbol{\mu}^{\Phi}_{p}\right]^\intercal \boldsymbol{\mu}^f_{si} -2 x_{ips} \left[\boldsymbol{\mu}^{\Lambda}_{sp}\right]^\intercal \boldsymbol{\mu}^l_{si} + 2 \left(\left[\boldsymbol{\mu}^{\Phi}_{p}\right]^\intercal \boldsymbol{\mu}^f_{si} \right) \left(\left[\boldsymbol{\mu}^{\Lambda}_{sp}\right]^\intercal  \boldsymbol{\mu}^l_{si} \right) + \\
    &\quad \text{Tr}\left( \boldsymbol{\Sigma}^{\Phi}_{p}\left( \boldsymbol{\mu}^f_{si}\left[\boldsymbol{\mu}^f_{si}\right]^\intercal  + \boldsymbol{\Sigma}^f_{si}\right) \right) + \left[\boldsymbol{\mu}^{\Phi}_{p}\right]^\intercal \left( \boldsymbol{\mu}^f_{si} \left[\boldsymbol{\mu}^f_{si}\right]^\intercal  + \boldsymbol{\Sigma}^f_{si} \right)  \boldsymbol{\mu}^{\Phi}_{p} + \\
    & \quad \text{Tr}\left(\boldsymbol{\Sigma}^{\Lambda}_{sp} \left( \boldsymbol{\mu}^l_{si} \left[\boldsymbol{\mu}^l_{si}\right]^\intercal + \boldsymbol{\Sigma}^l_{si}\right) \right) + \left[\boldsymbol{\mu}^{\Lambda}_{sp}\right]^\intercal\left( \boldsymbol{\mu}^l_{si} \left[\boldsymbol{\mu}^l_{si}\right]^\intercal + \boldsymbol{\Sigma}^l_{si}\right)\boldsymbol{\mu}^{\Lambda}_{sp} \\
    = &\left( x_{sip} - \left[\boldsymbol{\mu}_p^\Phi\right]^\intercal\boldsymbol{\mu}_{si}^f - \left[\boldsymbol{\mu}_{sp}^\Lambda\right]^\intercal \boldsymbol{\mu}_{si}^l\right)^2 + \\
    & \quad \left[\boldsymbol{\mu}^{\Phi}_{p}\right]^\intercal \boldsymbol{\Sigma}^f_{si}  \boldsymbol{\mu}^{\Phi}_{p} + \left[\boldsymbol{\mu}^{\Lambda}_{sp}\right]^\intercal\boldsymbol{\Sigma}^l_{si}\boldsymbol{\mu}^{\Lambda}_{sp}   \\
    & \quad \text{Tr}\left( \boldsymbol{\Sigma}^{\Phi}_{p}\left( \boldsymbol{\mu}^f_{si}\left[\boldsymbol{\mu}^f_{si}\right]^\intercal  + \boldsymbol{\Sigma}^f_{si}\right) \right) + \text{Tr}\left(\boldsymbol{\Sigma}^{\Lambda}_{sp} \left( \boldsymbol{\mu}^l_{si} \left[\boldsymbol{\mu}^l_{si}\right]^\intercal + \boldsymbol{\Sigma}^l_{si}\right) \right) \\
    \sum_{i=1}^{N_s}\mathbb{E}\left[\left(x_{sip} - \boldsymbol{\Phi}_p^\intercal \boldsymbol{f}_{si} - \Lambda_{sp}^\intercal\mathbf{l}_{si}\right)^2\right] = & \sum_{i=1}^{N_s} \left( x_{sip} - \left[\boldsymbol{\mu}_p^\Phi\right]^\intercal\boldsymbol{\mu}_{si}^f - \left[\boldsymbol{\mu}_{sp}^\Lambda\right]^\intercal \boldsymbol{\mu}_{si}^l\right)^2 + \\
    & \quad \left[\boldsymbol{\mu}^{\Phi}_{p}\right]^\intercal \left( \sum_{i=1}^{N_s} \boldsymbol{\Sigma}^f_{si} \right)  \boldsymbol{\mu}^{\Phi}_{p} + \left[\boldsymbol{\mu}^{\Lambda}_{sp}\right]^\intercal\left(\sum_{i=1}^{N_s}\boldsymbol{\Sigma}^l_{si}\right)\boldsymbol{\mu}^{\Lambda}_{sp} +   \\
    & \quad \text{Tr}\left( \boldsymbol{\Sigma}^{\Phi}_{p}\left( \sum_{i=1}^{N_s}\boldsymbol{\mu}^f_{si}\left[\boldsymbol{\mu}^f_{si}\right]^\intercal + \boldsymbol{\Sigma}^f_{si}\right) \right) +  \text{Tr}\left(\boldsymbol{\Sigma}^{\Lambda}_{sp} \left( \sum_{i=1}^{N_s} \boldsymbol{\mu}^l_{si} \left[\boldsymbol{\mu}^l_{si}\right]^\intercal + \boldsymbol{\Sigma}^l_{si}\right) \right) \\
\end{align*}



\newpage
\bibliography{refs_trunc}
